\documentclass[12pt]{article}
\usepackage[utf8]{inputenc}
\usepackage{amsfonts,amssymb,amsmath, epsfig}
\usepackage{color,graphicx,graphics}
\usepackage{amsthm}
\usepackage{amsmath,amstext,amssymb,amsfonts, amscd}
\usepackage[lofdepth,lotdepth]{subfig}
\usepackage{xcolor}
\usepackage{hyperref}          
\usepackage{circledsteps}

\def\Box{\leavevmode\vbox{\hrule
     \hbox{\vrule\kern4pt\vbox{\kern4pt}%
           \vrule}\hrule}}
\def\blackbox{\leavevmode\vrule height 5pt width 4pt depth 0pt\relax}
\def\endproof{\null\hfill {$\blackbox$}\bigskip}

\def\Section#1{\setcounter{equation}{0} \section{#1} \markboth{#1}{#1}
   \leavevmode\par}
\def\paragraph#1{{\bf #1\ }}

\newtheorem{lemma}{Lemma}[section]  

\newtheorem{theorem}[lemma]{Theorem}

\newtheorem{corollary}[lemma]{Corollary}

\newtheorem{definition}[lemma]{Definition}

\newtheorem{proposition}[lemma]{Proposition}

\newtheorem{remark}{Remark}[section]

\title{Macroscopic limit of a Fokker-Planck model of swarming rigid bodies} 
\author{P. Degond$^{(1)}$, A. Frouvelle$^{(2)}$} 
\date{} 
\begin{document}

\maketitle

\vspace{0.5 cm}

\begin{center}

$^{(1)}$ Institut de Math\'ematiques de Toulouse ; UMR5219 \\
Universit\'e de Toulouse ; CNRS \\
UPS, F-31062 Toulouse Cedex 9, France\\
email: pierre.degond@math.univ-toulouse.fr

\bigskip

$^{(2)}$ CEREMADE, CNRS, Universit\'e Paris-Dauphine\\
Universit\'e PSL, 75016 Paris, France \\
email: frouvelle@ceremade.dauphine.fr

\end{center}

\vspace{0.5 cm}
\begin{abstract}

We consider self-propelled rigid-bodies interacting through local body-attitude alignment  modelled by stochastic differential equations. We derive a hydrodynamic model of this system at large spatio-temporal scales and particle numbers in any dimension $n \geq 3$. This goal was already achieved in dimension $n=3$, or in any dimension $n \geq 3$ for a different system involving jump processes. However, the present work corresponds to huge conceptual and technical gaps compared with earlier ones. The key difficulty is to determine an auxiliary but essential object, the generalized collision invariant. We achieve this aim by using the geometrical structure of the rotation group, namely, its maximal torus, Cartan subalgebra and Weyl group as well as other concepts of representation theory and Weyl's integration formula. The resulting hydrodynamic model appears as a hyperbolic system whose coefficients depend on the generalized collision invariant. 
\end{abstract}

\medskip
\noindent
{\bf Acknowledgements:} PD holds a visiting professor association with the Department of Mathematics, Imperial College London, UK. AF acknowledges support from the Project EFI ANR-17-CE40-0030 of the French National Research Agency. AF thanks the hospitality of the Laboratoire de Math\'ematiques et Applications (LMA, CNRS) in the Universit\'e de Poitiers, where part of this research was conducted.

\medskip
\noindent
{\bf Competing interests:} the authors declare none.

\medskip
\noindent
{\bf Key words: } particle model, macroscopic model, stochastic differential equation, Fokker-Planck operator, self-organized hydrodynamics, generalized collision invariant, rotation group representations, maximal torus, Weyl group.

\medskip
\noindent
{\bf AMS Subject classification: } 22E70, 35Q70, 35Q91, 35Q92, 60J60, 60J65, 65C65, 82C22, 82C40, 82C70, 92D50. 
\vskip 0.4cm

\setcounter{equation}{0}
\section{Introduction}
\label{sec_intro}

In this paper, we consider a system of self-propelled rigid-bodies interacting through local body-attitude alignment.  Such systems describe flocks of birds \cite{hildenbrandt2010self} for instance. The model consists of coupled stochastic differential equations describing the positions and body-attitudes of the agents. We aim to derive a hydrodynamic model of this system at large spatio-temporal scales and particle numbers. This goal has already been achieved in dimension $n=3$ \cite{degond2017new, Degond_etal_proc19, Degond_etal_MMS18} or in any dimension $n \geq 3$ for a different model where the stochastic differential equations are replaced by jump processes \cite{degond2021body}. In the present paper, we realize this objective for the original system of stochastic differential equations in any dimension $n \geq 3$. The resulting hydrodynamic model appears as a hyperbolic system of first-order partial differential equations for the particle density and mean body orientation. The model is formally identical with that of \cite{degond2021body} but for the expressions of its coefficients, whose determination is the main difficulty here. It has been shown in \cite{degond2022hyperbolicity} that this system is hyperbolic, which is a good indication of its (at least local) well-posedness. 

The passage from dimension $n=3$ to any dimension $n \geq 3$ involves huge conceptual and technical difficulties. One of them is the lack of an appropriate coordinate system of the $n$-dimensional rotation group $\mathrm{SO}_n$, by contrast to dimension $n=3$ where  the Rodrigues formula \cite{degond2017new} and the quaternion representation~\cite{Degond_etal_MMS18} are available. This difficulty was already encountered in \cite{degond2021body} and solved by the use of representation theory \cite{faraud2008Analysis, fulton2013representation} and Weyl's integration formula \cite{Simon}. Here, additional difficulties arise because an auxiliary but essential object, the generalized collision invariant (GCI)  which will be defined below, becomes highly non trivial. Indeed, the GCI is the object that leads to explicit formulas for the coefficients of the hydrodynamic model. In this paper, we will develop a completely new method to determine the GCI relying on the Cartan subalgebra and Weyl group of $\mathrm{SO}_n$ \cite{fulton2013representation}. While biological agents live in a 3-dimensional space, deriving models in arbitrary dimensions is useful to uncover underlying algebraic or geometric structures that would be otherwise hidden. This has been repeatedly used in physics where large dimensions (e.g. in the theory of glasses) \cite[Ch. 1]{parisi2020theory} or zero dimension (e.g. in the replica method) \cite{castellani2005spin} have provided invaluable information on real systems. In data science, data belonging to large-dimensional manifolds may also be encountered, which justifies the investigation of models in arbitrary dimensions. 

Collective dynamics can be observed in systems of interacting self-propelled agents such as locust swarms \cite{bazazi2008collective}, fish schools \cite{lopez2012behavioural} or bacterial colonies \cite{be2019statistical} and manifests itself by coordinated movements and patterns (see e.g. the review \cite{vicsek2012collective}). The system of interacting rigid-bodies which motivates the present study is only one among many examples of collective dynamics models. Other examples are the three-zone model \cite{aoki1982simulation, cao2020asymptotic}, the Cucker-Smale model \cite{aceves2019hydrodynamic, barbaro2016phase, barbaro2012phase, carrillo2010asymptotic, cucker2007emergent, ha2009simple, motsch2011new}, the Vicsek model \cite{chate2008collective, degond2013macroscopic, degond2015phase, frouvelle2012dynamics, toner1998flocks, vicsek1995novel} (the literature is huge and the proposed citations are for illustration purposes only). Other collective dynamics models involving rigid-body attitudes or geometrically complex objects can be found in \cite{cho2023continuum, fetecau2022emergent, golse2019mean, ha2017emergent, hildenbrandt2010self, sarlette2010coordinated, sarlette2009autonomous, sepulchre2010consensus}.

Collective dynamics can be studied at different scales, each described by a different class of models. At the finest level of description lie particle models which consist of systems of ordinary or stochastic differential equations  \cite{aoki1982simulation, cao2020asymptotic, chate2008collective, cucker2007emergent, lopez2012behavioural, motsch2011new, vicsek1995novel}. When the number of particles is large, a statistical description of the system is substituted, which leads to kinetic or mean-field models \cite{barbaro2016phase, bertin2006boltzmann, carrillo2010asymptotic, Degond_eal_JNLS20, degond2013macroscopic, figalli2018global, gamba2016global, griette2019kinetic}. At large spatio-temporal scales, the kinetic description can be approximated by fluid models, which describe the evolution of locally averaged quantities such as the density or mean orientation of the particles \cite{bertin2009hydrodynamic, bertozzi2015ring, degond2021bulk, degond2022hyperbolicity, toner1998flocks}. The rigorous passage from particle to kinetic models of collective dynamics has been investigated in \cite{bolley2012mean, briant2022cauchy, diez2020propagation, ha2009simple} while passing from kinetic to fluid models has been formally shown in \cite{degond2013macroscopic, degond2015phase, degond2008continuum, frouvelle2012continuum} and rigorously in \cite{jiang2016hydrodynamic}. Phase transitions in kinetic models have also received a great deal of attention \cite{Degond_eal_JNLS20, degond2013macroscopic, degond2015phase, frouvelle2021body, frouvelle2012dynamics}. 

The paper is organized as follows. After some preliminaries on rotation matrices, Section \ref{sec_prelim} describes the particle model and its associated kinetic model. The main result, which is the identification of the associated fluid model, is stated in Section \ref{sec_hydro}. The model has the same form as that derived in \cite{degond2021body}, but for the expression of its coefficients. In classical kinetic theory, fluid models are strongly related to the collision invariants of the corresponding kinetic model. However, in collective dynamics models, there are often not enough collision invariants. This has been remediated by the concept of generalized collision invariant (GCI) first introduced in \cite{degond2008continuum} for the Vicsek model. In Section \ref{sec_GCI}, the definition and first properties of the GCI are stated and proved. To make the GCI explicit, we need to investigate the geometry of $\mathrm{SO}_n$ in more detail. This is done in Section \ref{subsec:rotmore} where the notions of maximal torus, Cartan subalgebra and Weyl group are recalled. After these preparations, we derive an explicit system of equations for the GCI in section \ref{subsec_GCIId} and show its well-posedness. Once the GCI are known, we can proceed to the the derivation of the hydrodynamic model in Section \ref{subsec:Gamma}. This involves again the use of representation theory and the Weyl integration formula as the results from \cite{degond2021body} cannot be directly applied due to the different shapes of the GCI. Finally a conclusion is drawn in Section \ref{sec_conclu}. In Appendix~\ref{appsec_direct_strong_form}, an alternate derivation of the equations for the GCI is given: while Section \ref{subsec_GCIId} uses the variational form of these equations, Appendix \ref{appsec_direct_strong_form} uses their strong form. The two methods rely on different Lie algebra formulas and can be seen as cross validations of one another.

\setcounter{equation}{0}
\section{Microscopic model and scaling}
\label{sec_prelim}

\subsection{Preliminaries: rotations and the rotation group}
\label{subsec_rotations}

Before describing the model, we need to recall some facts about rotation matrices (see~\cite{degond2023radial} for more detail). Throughout this paper, the dimension $n$ will be supposed greater or equal to $3$. We denote by $\mathrm{M}_n$ the space of $n \times n$ matrices with real entries and by $\mathrm{SO}_n$ the subset of $\mathrm{M}_n$ consisting of rotation matrices:
$$ \mathrm{SO}_n = \big\{ A \in \mathrm{M}_n \quad | \quad A A^T = A^T A = \mathrm{I} \quad \mathrm{and} \quad \mathrm{det} A = 1 \big\}, $$
where $A^T$ is the transpose of $A$, $\mathrm{I}$ is the identity matrix of $\mathrm{M}_n$ and $\mathrm{det}$ stands for the determinant. For $x$, $y \in {\mathbb R}^n$, we denote by $x \cdot y$ and $|x|$ the euclidean inner product and norm. Likewise, we define a Euclidean inner product on $\mathrm{M}_n$ as follows: 
\begin{equation}
M \cdot N = \frac{1}{2} \mathrm{Tr} (M^T N) = \frac{1}{2} \sum_{i,j} M_{ij}N_{ij}, \quad \forall M, \, N \in \mathrm{M}_n, 
\label{eq:rot_inner_product}
\end{equation}
where $\mathrm{Tr}$ is the trace. We note the factor $\frac{1}{2}$ which differs from the conventional definition of the Frobenius inner product, but which is convenient when dealing with rotation matrices. We use the same symbol $\cdot$ for vector and matrix inner products as the context easily waives the ambiguity. The set $\mathrm{SO}_n$ is a compact Lie group, i.e. it is a group for matrix multiplication and an embedded manifold in $\mathrm{M}_n$ for which group multiplication and inversion are $C^\infty$, and it is compact. Let $A \in \mathrm{SO}_n$. We denote by $T_A$ the tangent space to $\mathrm{SO}_n$ at $A$. The tangent space $T_{\mathrm{I}}$ at the identity is the Lie algebra $\mathfrak{so}_n$ of skew-symmetric matrices with real entries endowed with the Lie bracket $[X,Y] = XY-YX$, $\forall X, \, Y \in  \mathfrak{so}_n$. Let $A \in \mathrm{SO}_n$. Then, 
\begin{equation}
T_A = A \, \mathfrak{so}_n = \mathfrak{so}_n \, A. 
\label{eq:tangent_at_A}
\end{equation}
For $M \in \mathrm{M}_n$, we denote by $P_{T_A} M$ its orthogonal projection onto $T_A$. It is written: 
$$ P_{T_A} M = A \frac{A^T M - M^T A}{2} = \frac{M A^T - A M^T}{2} A. $$

A Riemannian structure on $\mathrm{SO}_n$ is induced by the Euclidean structure of $\mathrm{M}_n$ following from \eqref{eq:rot_inner_product}. This Riemannian metric is given by defining the inner product of two elements $M$, $N$ of $T_A$ by $M \cdot N$. Given that there are $P$ and $Q$ in $\mathfrak{so}_n$ such that $M=AP$, $N = AQ$ and that $A$ is orthogonal, we have $M \cdot N = P \cdot Q$. As any Lie-group is orientable, this Riemannian structure gives rise to a Riemannian volume form and measure~$\omega$. This Riemannian measure is left-invariant by group translation \cite[Lemma 2.1]{degond2023radial} and is thus equal to the normalized Haar measure on $\mathrm{SO}_n$ up to a multiplicative constant. We recall that on compact Lie groups, the Haar measure is also right invariant and invariant by group inversion. On Riemannian manifolds, the gradient $\nabla$, divergence $\nabla \cdot$ and Laplacian~$\Delta$ operators can be defined. Given a smooth map $f$: $\mathrm{SO}_n \to {\mathbb R}$, the gradient $\nabla f(A) \in T_A$ is defined by 
\begin{equation}
\nabla f (A) \cdot X = df_A(X), \quad \forall X \in T_A, 
\label{eq:rot_gradient_def}
\end{equation}
where $df_A$ is the derivative of $f$ at $A$ and is a linear map $T_A \to {\mathbb R}$, and $df_A(X)$ is the image of $X$ by $df_A$. The divergence of a smooth vector field $\phi$ on $\mathrm{SO}_n$ (i.e. a map $\mathrm{SO}_n \to \mathrm{M}_n$ such that $\phi(A) \in T_A$, $\forall A \in \mathrm{SO}_n$) is defined by duality by 
$$ \int_{\mathrm{SO}_n} \nabla \cdot \varphi \, f \, dA = - \int_{\mathrm{SO}_n} \varphi \cdot \nabla f \, dA, \quad \forall f \in C^{\infty}(\mathrm{SO}_n), $$
where $C^{\infty}(\mathrm{SO}_n)$ denotes the space of smooth maps $\mathrm{SO}_n \to {\mathbb R}$ and where we have denoted the Haar measure by $dA$. The Laplacian of a smooth map $f$: $\mathrm{SO}_n \to {\mathbb R}$ is defined by $\Delta f = \nabla \cdot (\nabla f)$. 

It is not easy to find a convenient coordinate system on $\mathrm{SO}_n$ to express the divergence and Laplace operators, so we will rather use an alternate expression which uses the matrix exponential $\exp$: $\mathfrak{so}_n \to \mathrm{SO}_n$. Let $X \in \mathfrak{so}_n$. Then, $\varrho(X)$ denotes the map $C^{\infty}(\mathrm{SO}_n) \to C^{\infty}(\mathrm{SO}_n)$ such that 
\begin{equation} 
\big( \varrho(X)(f) \big)(A) = \frac{d}{dt} \big(f(A e^{tX}) \big)|_{t=0}, \quad \forall f \in C^{\infty}(\mathrm{SO}_n), \quad \forall A \in \mathrm{SO}_n. 
\label{eq:rot_def_rho}
\end{equation}
We note that 
\begin{equation} 
\big( \varrho(X)(f) \big)(A) = df_A(AX) = \nabla f(A) \cdot (AX). 
\label{eq:rot_operator_rho}
\end{equation}

Let $F_{ij}$ be the matrix with entries 
\begin{equation}
(F_{ij})_{k\ell} = \delta_{ik} \delta_{j \ell} - \delta_{i \ell} \delta_{jk}. 
\label{eq:rot_def_Fij}
\end{equation}
We note that $(F_{ij})_{1 \leq i < j \leq n}$ forms an orthonormal basis of $\mathfrak{so}_n$ for the inner product \eqref{eq:rot_inner_product}. Then, we have \cite[Lemma 2.2]{degond2023radial} 
\begin{equation}
(\Delta f)(A) = \sum_{1 \leq i < j \leq n} \big( \varrho(F_{ij})^2 f \big) (A). 
\label{eq:rot_Laplacian}
\end{equation}
The expression remains valid if the basis $(F_{ij})_{1 \leq i < j \leq n}$ is replaced by another orthonormal basis of $\mathfrak{so}_n$. 

Finally, let $M \in \mathrm{M}_n^+$ where $\mathrm{M}_n^+$ is the subset of $\mathrm{M}_n$ consisting of matrices with positive determinant. There exists a unique pair $(A, S) \in \mathrm{SO}_n \times {\mathcal S}^+_n$ where ${\mathcal S}^+_n$ denotes the cone of symmetric positive-definite matrices, such that $M = A S$. The pair $(A, S)$ is the polar decomposition of $M$ and we define a map ${\mathcal P}$: $\mathrm{M}_n^+ \to \mathrm{SO}_n$, $M \mapsto A$. We note that ${\mathcal P}(M) = (M M^T)^{-1/2} M$. We recall that a positive definite matrix $S$ can be written $S = U D U^T$ where $U \in \mathrm{SO}_n$ and $D = \mathrm{diag}(d_1, \ldots, d_n)$ is the diagonal matrix with diagonal elements $d_1, \ldots, d_n$ with $d_i >0$, $\forall i=1, \ldots, n$. Then, $S^{-1/2} = U D^{-1/2} U^T$ with $D^{-1/2} = \mathrm{diag}(d_1^{-1/2}, \ldots, d_n^{-1/2})$.


\subsection{The particle model}
\label{subsec_particle}

We consider a system of $N$ agents moving in an $n$-dimensional space ${\mathbb R}^n$. They have  positions $(X_k(t))_{k=1}^N$ with $X_k(t) \in {\mathbb R}^n$ at time $t$. All agents are identical rigid bodies. An agent's attitude can be described by a moving direct orthonormal frame $(\omega_1^k(t), \ldots, \omega_n^k(t))$ referred to as the agent's local body frame or body attitude. Let ${\mathbb R}^n$ be endowed with a reference direct orthonormal frame $({\mathbf e}_1, \ldots, {\mathbf e}_n)$. We denote by $A_k(t)$ the unique rotation which maps $({\mathbf e}_1, \ldots, {\mathbf e}_n)$ to $(\omega_1^k(t), \ldots, \omega_n^k(t))$, i.e. $\omega_j^k(t) = A_k(t) {\mathbf e}_j$. Therefore, the $k$-th agent's body attitude can be described equivalently by the local basis $(\omega_1^k(t), \ldots, \omega_n^k(t))$ or by the rotation $A_k(t)$.

The particle dynamics is as follows: particles move with speed $c_0$ in the direction of the first basis vector $\omega_1^k(t) = A_k(t) {\mathbf e}_1$ of the local body frame, hence leading to the equation 
\begin{equation}
d X_k= c_0  A_k(t) {\mathbf e}_1 \, dt . 
\label{eq:part_dXkdt}
\end{equation}
Body frames are subject to two processes. The first one tends to relax the particle's body frame to a target frame which represents the average of the body frames of the neighboring particles. The second one is diffusion noise. We first describe how the average of the body frames of the neighboring particles is computed. Define 
\begin{equation}
\tilde J_k(t) = \frac{1}{N \, R^n} \sum_{\ell=1}^N K \Big( \frac{|X_k(t) - X_\ell(t)|}{R} \Big) \, A_\ell(t), 
\label{eq:part_Jk_def}
\end{equation} 
where the sensing function $K$: $[0,\infty) \to [0,\infty)$ and the sensing radius $R >0$ are  given. Here we have assumed that the sensing function is radially symmetric for simplicity. Then, the body frame dynamics is as follows:
\begin{equation}
d A_k = \nu \, P_{T_{A_k}} \big( {\mathcal P}(\tilde J_k) \big) \, dt + \sqrt{2D} \, P_{T_{A_k}} \circ dW_t^k, 
\label{eq:part_dAkdt}
\end{equation}
where $\nu$ and $D$ are positive constants, $dW_t^k$ are independent Brownian motions on $\mathrm{M}_n$ and the symbol $\circ$ indicates that the stochastic differential equation is meant in the Stratonovich sense. Here, it is important to stress that these Brownian motions are defined using the metric induced by the inner product \eqref{eq:rot_inner_product}. In the first term, we note that the matrix $\tilde J_k$ is projected onto the rotation matrix issued from the polar decomposition by the map~${\mathcal P}$. For consistency, we need to assume that $\tilde J_k(t)$ remains in $M_n^+$, which will be true as long as the body frames of neighboring particles are close to each other. The rotation $\Gamma_k = {\mathcal P}(\tilde J_k(t))$ can be seen as the average body frame of the neighbors to particle~$k$. For \eqref{eq:part_dAkdt} to define a motion on $\mathrm{SO}_n$, the right-hand side must be a tangent vector to $\mathrm{SO}_n$ at $A_k(t)$. This is ensured by the projection operator $P_{T_{A_k}}$ which multiplies each term of the right-hand side of \eqref{eq:part_dAkdt}, and, for the second term, by the fact that the stochastic differential equation is taken in the Stratonovich sense. Indeed, according to \cite{hsu2002stochastic}, the second term of~\eqref{eq:part_dAkdt} generates a Brownian motion on $\mathrm{SO}_n$. 

Finally, the particle system must be supplemented with initial conditions specifying the values of $(X_k,A_k)(0)$. As discussed in \cite{degond2017new}, the first term of \eqref{eq:part_dAkdt} relaxes the $k$-th particle body frame to the neighbor's average body frame $\Gamma_k$ and models the agents' tendency to adopt the same body attitude as their neighbors. The second term of \eqref{eq:part_dAkdt} is an idiosyncratic noise term that models either errors in the agent's computation of the neighbor's average body frame or the agent's will to detach from the group and explore new environments.

\subsection{The mean-field model}
\label{subsec_mean-field}

When $N \to \infty$, the particle system can be approximated by a mean-field kinetic model. Denote by $f(x,A,t)$ the probability distribution of the particles, namely $f(x,A,t)$ is the probability density of the particles at $(x,A) \in {\mathbb R}^n \times \mathrm{SO}_n$ at time $t$. Then, provided that~$K$ satisfies 
$$ \int_{{\mathbb R}^n} K(|x|) \, dx = 1, \quad \int_{{\mathbb R}^n} K(|x|) \, |x|^2\, dx < \infty,$$
$f$ is the solution of the following system: 
\begin{eqnarray}
&&\hspace{-1cm}
\partial_t f + c_0 \,  A {\mathbf e}_1 \cdot \nabla_x f + \nu \nabla_A \cdot \big( P_{T_A} ({\mathcal P} (\tilde J_f)) f \big) - D \Delta_A f = 0, \label{eq:mf_kin} \\
&&\hspace{-1cm}
\tilde J_f(x,t) = \frac{1}{R^n} \int_{{\mathbb R}^n \times \mathrm{SO}_n} K \Big( \frac{|x-y|}{R} \Big) \, f(y, B, t) \, B \, dy \, dB, 
\label{eq:mf_J}
\end{eqnarray}
where again, in \eqref{eq:mf_J}, the integral over $\mathrm{SO}_n$ is taken with respect to the normalized Haar measure. The operators $\nabla_A \cdot$ and $\Delta_A$ are respectively the divergence and Laplacian on $\mathrm{SO}_n$ as defined in Section \ref{subsec_rotations}. The index $A$ is there to distinguish them from analog operators acting on the spatial variable $x$, which will be indicated (as in $\nabla_x f$) with an index $x$. A small remark may be worth making: according to \cite{hsu2002stochastic}, the stochastic process defined by the second term of \eqref{eq:part_dAkdt} has infinitesimal generator $D \tilde \Delta$ where for any $f \in C^\infty(\mathrm{SO}_n)$, 
$$ \tilde \Delta f(A) = \sum_{i,j=1}^n (P_{T_A} E_{ij} \cdot \nabla_A)^2 f(A), \qquad \forall A \in \mathrm{SO}_n, $$
and where $(E_{ij})_{i,j=1}^n$ is any orthogonal basis (for the inner product \eqref{eq:rot_inner_product}) of $\mathrm{M}_n$. For instance, we can take the matrices $E_{ij}$ with entries $(E_{ij})_{k \ell} = \sqrt{2} \delta_{ik} \delta_{j \ell}$. It is shown in \cite[Lemma 2.2]{degond2023radial} that $\tilde \Delta$ coincides with the Laplacian $\Delta$ defined by \eqref{eq:rot_Laplacian}. This gives a justification of the last term in \eqref{eq:mf_kin}. We note that \eqref{eq:mf_kin} is a nonlinear Fokker-Planck equation, where the nonlinearity arises in the third term. 

The proof of the convergence of the particle system \eqref{eq:part_dXkdt}, \eqref{eq:part_dAkdt} to the kinetic model \eqref{eq:mf_kin}, \eqref{eq:mf_J} is still open. The difficulty is in the presence of the projection operator~${\mathcal P}$ in~\eqref{eq:part_dAkdt} which requires to control that the determinant of $\tilde J_k$ remains positive. In the Vicsek case where a similar singular behavior is observed, a local-in-time convergence result is shown in \cite{briant2022cauchy}. This result supports the conjecture that System \eqref{eq:part_dXkdt}, \eqref{eq:part_dAkdt} converges to System \eqref{eq:mf_kin}, \eqref{eq:mf_J} in the limit $N \to \infty$ in small time. We will assume it.

\subsection{Scaling and statement of the problem}
\label{subsec_scaling}

Let $t_0$ be a time scale and define the spatial scale $x_0 = c_0 t_0$. We introduce the following dimensionless parameters: 
$$ \tilde D = D t_0, \qquad \tilde R = \frac{R}{x_0}, \qquad \kappa = \frac{\nu}{D}. $$
Then, we change variables and unknowns to dimensionless variables $x' = x/x_0$, $t'=t/t_0$ and unknowns $f'(x',A,t') = x_0^n f(x,A,t)$, $\tilde J'_{f'}(x',t') = x_0^n \tilde J_f(x,t)$. Inserting these changes into \eqref{eq:mf_kin}, \eqref{eq:mf_J} leads to (omitting the primes for simplicity): 
\begin{eqnarray}
&&\hspace{-1cm}
\partial_t f +   A {\mathbf e}_1 \cdot \nabla_x f + \tilde D \big[ \kappa \nabla_A \cdot \big( P_{T_A} ({\mathcal P} (\tilde J_f)) f \big) - \Delta_A f \big] = 0, \label{eq:sca_kin} \\
&&\hspace{-1cm}
\tilde J_f(x,t) = \frac{1}{\tilde R^n} \int_{{\mathbb R}^n \times \mathrm{SO}_n} K \Big( \frac{|x-y|}{\tilde R} \Big) \, f(y, B, t) \, B\, dy \, dB, 
\label{eq:sca_J}
\end{eqnarray}

We introduce a small parameter $\varepsilon \ll 1$ and make the scaling assumption $ \frac{1}{\tilde D} = \tilde R = \varepsilon$, while $\kappa$ is kept of order $1$. By Taylor's formula, we have $\tilde J_f = J_f + {\mathcal O}(\varepsilon^2)$ where $J_f(x,t) = \int_{\mathrm{SO}_n} A \, f(x, A, t) \, dA$. Since the map ${\mathcal P}$ is smooth on $\mathrm{M}_n^+$, we get ${\mathcal P}(\tilde J_f) = {\mathcal P}(J_f) + {\mathcal O}(\varepsilon^2)$. Inserting these scaling assumptions and neglecting the ${\mathcal O}(\varepsilon^2)$ terms in the above expansions (because they would have no influence on the result), we get the following perturbation problem:  
\begin{eqnarray}
&&\hspace{-1cm}
\partial_t f^\varepsilon +   A {\mathbf e}_1 \cdot \nabla_x f^\varepsilon = \frac{1}{\varepsilon} \big[ - \kappa \nabla_A \cdot \big( P_{T_A} ({\mathcal P} (J_{f^\varepsilon})) f^\varepsilon \big) + \Delta_A f^\varepsilon \big], \label{eq:sca_kin_2} \\
&&\hspace{-1cm}
J_f(x,t) = \int_{\mathrm{SO}_n} f(x, A, t) \, A\, dA. 
\label{eq:sca_J_2}
\end{eqnarray}
The goal of this paper is to provide the formal limit $\varepsilon \to 0$ of this problem. This problem is referred to as the hydrodynamic limit of the Fokker-Planck equation \eqref{eq:sca_kin_2}.

\setcounter{equation}{0}
\section{Hydrodynamic limit (I): main results and first steps of the proof}
\label{sec_hydro}

\subsection{Statement of the results}
\label{subsec:statement}

We will need the following 

\begin{definition}[von Mises distribution]
Let $\Gamma \in \mathrm{SO}_n$ and $\kappa >0$. The function $M_\Gamma$: $\mathrm{SO}_n \to [0,\infty)$ such that 
\begin{equation} 
M_\Gamma(A) = \frac{1}{Z} \exp(\kappa \Gamma \cdot A), \quad Z = \int_{\mathrm{SO}_n} \exp(\kappa \mathrm{Tr} (A)/2) \, dA, 
\label{eq:equi_VM}
\end{equation}
is called the von Mises distribution of orientation $\Gamma$ and concentration parameter $\kappa$. It is the density of a probability measure on $\mathrm{SO}_n$. 
\label{def:equi_VM}
\end{definition}

We note that $\int M_\Gamma(A) \, dA= \int \exp(\kappa \mathrm{Tr} (A)/2) \, dA$ does not depend on $\Gamma$ thanks to the translation invariance of the Haar measure.

The first main result of this paper is about the limit of the scaled kinetic System \eqref{eq:sca_kin_2}, \eqref{eq:sca_J_2} when $\varepsilon \to 0$. We will need the following notations: for two vector fields $X=(X_i)_{i=1}^n$ and $Y=(Y_i)_{i=1}^n$, we define the antisymmetric matrices $X \wedge Y = (X \wedge Y)_{ij}$ and $\nabla_x \wedge X = (\nabla_x \wedge X)_{ij}$ by 
$$ (X \wedge Y)_{ij} = X_i Y_j - X_j Y_i, \quad (\nabla_x \wedge X)_{ij} = \partial_{x_i} X_j - \partial_{x_j} X_i, \quad \forall i, \, j \in \{1, \ldots n\}. $$
Then, we have

\begin{theorem}
We suppose that there is a smooth solution $f^\varepsilon$ to System \eqref{eq:sca_kin_2}, \eqref{eq:sca_J_2}. We also suppose that $f^\varepsilon \to f^0$  as $\varepsilon \to 0$ as smoothly as needed. Then, there exist two functions $\rho$: ${\mathbb R}^n \times [0,\infty) \to [0,\infty)$ and $\Gamma$: ${\mathbb R}^n \times [0,\infty) \to \mathrm{SO}_n$ such that 
\begin{equation} 
f^0(x,A,t) = \rho(x,t) M_{\Gamma(x,t)}(A). 
\label{eq:equi_f0express}
\end{equation}
Furthermore, for appropriate real constants $c_1, \ldots, c_4$, $\rho$ and $\Gamma$ satisfy the following system of equations: 
\begin{eqnarray}
&&\hspace{-1cm}
\partial_t \rho + \nabla_x (\rho c_1 \Omega_1) = 0, \label{eq:cont_eq} \\
&&\hspace{-1cm}
\rho \big( \partial_t \Gamma + c_2 (\Omega_1 \cdot \nabla_x) \Gamma \big) = {\mathbb W} \Gamma,
\label{eq:Gamma_Gameq_second}
\end{eqnarray}
where 
\begin{equation}
\Omega_k(x,t) = \Gamma(x,t) {\mathbf e}_k, \quad k \in \{1, \ldots, n\},  
\label{eq:cont_Omdef}
\end{equation}
and where 
\begin{equation}
{\mathbb W} = - c_3 \nabla_x \rho \wedge \Omega_1 - c_4 \rho \big[ \big( \Gamma (\nabla_x \cdot \Gamma) \big) \wedge \Omega_1 + \nabla_x \wedge \Omega_1 \big].  
\label{eq:Gamma:W_express}
\end{equation}
The notation $\nabla_x \cdot \Gamma$ stands for the divergence of the matrix $\Gamma$, i.e. $(\nabla_x \cdot \Gamma)_i = \sum_{j=1}^n \partial_{x_i} \Gamma_{ij}$ and $\Gamma (\nabla_x \cdot \Gamma)$ is the vector arising from multiplying the vector $\nabla_x \cdot \Gamma$ on the left by the matrix~$\Gamma$. 
\label{thm:statmnt_main}
\end{theorem}

System \eqref{eq:cont_eq}, \eqref{eq:Gamma_Gameq_second} has been referred to in previous works \cite{degond2021body, degond2022hyperbolicity} as the Self-Organized Hydrodynamic model for Body attitude coordination (SOHB). It consists of coupled first order partial differential equation for the particle density $\rho$ and the average body attitude~$\Gamma$ and has been shown to be hyperbolic in \cite{degond2022hyperbolicity}. It models the system of interacting rigid bodies introduced in Section \ref{subsec_particle} as a fluid of which \eqref{eq:cont_eq} is the continuity equation. The velocity of the fluid is $c_1 \Omega_1$. Eq. \eqref{eq:Gamma_Gameq_second} is an evolution equation for the averaged body orientation of the particles within a fluid element, described by $\Gamma$. The left-hand side of~\eqref{eq:Gamma_Gameq_second} describes pure transport at velocity~$c_2 \Omega_1$. In general, $c_2 \not =  c_1$, which means that such transport occurs at a velocity different from the fluid velocity. The right-hand side appears as the multiplication of $\Gamma$ itself on the left by the antisymmetric matrix~${\mathbb W}$, which is a classical feature of rigid-body dynamics. The first term of \eqref{eq:Gamma:W_express} is the action of the pressure gradient which contributes to rotate~$\Gamma$  so as to align $\Omega_1$ with $- \nabla_x \rho$. The second term has a similar effect with the pressure gradient replaced by the vector $\Gamma (\nabla_x \cdot \Gamma)$ which encodes gradients of the mean body attitude $\Gamma$. Finally, the last term encodes self-rotations of the averaged body frame about the self-propulsion velocity $\Omega_1$. The last two terms do not have counterparts in classical fluid hydrodynamics. We refer to \cite{degond2021body, degond2021bulk, degond2022hyperbolicity} for a more detailed interpretation.

\begin{remark}
We stress that the density $\rho$ and the differentiation operator $\varrho$ defined by~\eqref{eq:rot_def_rho} have no relation and are distinguished by the different typography. The notation~$\varrho$ for \eqref{eq:rot_def_rho} is classical (see e.g. \cite{faraud2008Analysis}). 
\label{rem:rho_vs_varrho}
\end{remark}

The second main result of this paper is to provide explicit formulas for the coefficients $c_1, \ldots, c_4$. For this, we need to present additional concepts. Let $p = \lfloor \frac{n}{2} \rfloor$ the integer part of $\frac{n}{2}$ (i.e. $n=2p$ or $n = 2p+1$) and 
\begin{equation}
\epsilon_n = \left\{ \begin{array}{ccc} 0 & \mathrm{ if } & n=2p \\ 1 & \mathrm{ if } & n=2p +1 \end{array} \right. . 
\label{eq:Gamma_eps_def}
\end{equation}
Let ${\mathcal T} = [-\pi, \pi)^p$. Let $\Theta = (\theta_1, \ldots, \theta_p) \in {\mathcal T}$. We introduce 
\begin{eqnarray}
\hspace{-0.5cm} u_{2p} (\Theta) &=& \prod_{1 \leq j<k \leq p} \big( \cos \theta_j - \cos \theta_k \big)^2, \quad \text{ for } p \geq 2, 
\label{eq:Pi2p} \\
\hspace{-0.5cm} u_{2p+1}  (\Theta) &=&  \prod_{1 \leq j<k \leq p} \big( \cos \theta_j - \cos \theta_k \big)^2 \, \prod_{j=1}^p \sin^2 \frac{\theta_j}{2}, \quad \text{ for } p \geq 1,  
\label{eq:Pi2p+1}
\end{eqnarray}
and 
\begin{equation}
m(\Theta) = \exp \Big( \frac{\kappa}{2} \big( 2 \sum_{k=1}^p \cos \theta_k + \epsilon_n \big) \Big) \, u_n(\Theta). \label{eq:GCIId_m_def}
\end{equation}
Let $\nabla_\Theta$ (resp. $\nabla_\Theta \cdot$) denote the gradient (resp. divergence) operators with respect to~$\Theta$ of scalar (resp. vector) fields on ${\mathcal T}$. 
Then, we define $\alpha$: ${\mathcal T} \to {\mathbb R}^p$, with $\alpha = (\alpha_i)_{i=1}^p$ as a periodic solution of the following system: 
\begin{eqnarray}
&&\hspace{-1cm}
- \nabla_\Theta \cdot \big( m \nabla_\Theta \alpha_\ell \big) + m \sum_{k \not = \ell} \Big( \frac{\alpha_\ell - \alpha_k}{1 - \cos(\theta_\ell - \theta_k)} +  \frac{\alpha_\ell + \alpha_k}{1 - \cos(\theta_\ell + \theta_k)} \Big) \nonumber \\
&&\hspace{4cm}
+ \epsilon_n m \frac{\alpha_\ell}{1 - \cos \theta_\ell} = m \, \sin \theta_\ell, \quad \forall \ell \in \{1, \ldots, p \}, 
\label{eq:GCIId_strongform_alpha}
\end{eqnarray}
A functional framework which guarantees that $\alpha$ exists, is unique and satisfies an extra invariance property (commutation with the Weyl group) will be provided at Section \ref{subsec_GCIId}.

\begin{remark}
In the case of $\mathrm{SO}_3$, we have $p=1$ and a single unknown $\alpha_1(\theta_1)$. From \eqref{eq:GCIId_strongform_alpha}, we get that $\alpha_1$ satisfies 
$$ - \frac{\partial }{\partial \theta_1} \Big( m \frac{\partial \alpha_1}{\partial \theta_1} \Big) + \frac{m \, \alpha_1}{1 - \cos \theta_1} = m \sin \theta_1. $$
We can compare this equation with \cite[Eq. (4.16)]{degond2017new} and see that $\alpha_1$ coincides with the function $- \sin \theta \, \tilde \psi_0$ of \cite{degond2017new}. 
\label{rem:GCIId_compar_SO3}
\end{remark}

Thanks to these definitions, we have the 

\begin{theorem}
The constants $c_1, \ldots, c_4$ involved in \eqref{eq:cont_eq}, \eqref{eq:Gamma_Gameq_second}, \eqref{eq:Gamma:W_express} are given by 
\begin{eqnarray}
&&\hspace{-1cm}
c_1 = \frac{1}{n} \frac{\displaystyle \int \Big( 2 \sum \cos \theta_k + \epsilon_n \Big) \, m(\Theta) \, d \Theta}{\displaystyle \int \, m(\Theta) \, d \Theta}, \label{eq:equi_order_param_express} \\
&&\hspace{-1cm}
c_2 = - \frac{1}{n^2-4} \times \nonumber \\
&&\hspace{-1cm}
\frac{\displaystyle \int \Big[ -n \big( \sum \alpha_k \sin \theta_k \big) \big( 2 \sum \cos \theta_k + \epsilon_n \big) + 4 \big( \sum \alpha_k \sin \theta_k \cos \theta_k \big) \Big] m(\Theta) d \Theta}
{\displaystyle \int \big( \sum \alpha_k \sin \theta_k \big) m(\Theta) d \Theta}, \label{eq:Gamma:c2_express_3} \\
&&\hspace{-1cm}
c_3 = \frac{1}{\kappa}, \label{eq:Gamma:c3_express} \\
&&\hspace{-1cm}
c_4 = - \frac{1}{n^2-4} \times \nonumber \\
&&\hspace{-1cm}
\frac{\displaystyle \int \Big[ - \big( \sum \alpha_k \sin \theta_k \big) \big( 2 \sum \cos \theta_k + \epsilon_n \big) + n \big( \sum \alpha_k \sin \theta_k \cos \theta_k \big) \Big] m(\Theta) d \Theta}
{\displaystyle \int \big( \sum \alpha_k \sin \theta_k \big) m(\Theta) d \Theta}, \label{eq:Gamma:c4_express_3}
\end{eqnarray}
where the integrals are over $\Theta = (\theta_1, \ldots, \theta_p) \in {\mathcal T}$ and the sums over $k \in \{1, \ldots, p \}$. 
\label{thm:statmnt_main_csts}
\end{theorem}

\begin{remark} (i) Letting $\alpha_k(\Theta) = - \sin \theta_k$, we recover the formulas of \cite[Theorem 3.1]{degond2021body} for the coefficients $c_i$ (see also Remark \ref{rem:GCIId:mu_case_BGK}). 

\smallskip
\noindent
(ii) Likewise, restricting ourselves to dimension $n=3$, setting $\alpha_1 (\theta) = - \sin \theta \tilde \psi_0$ where~$\tilde \psi_0$ is defined in \cite[Prop. 4.6]{degond2017new} (see Remark \ref{rem:GCIId_compar_SO3}) the above formulas recover the formulas of \cite[Theorem 4.1]{degond2017new} (noting that in~\cite{degond2017new}, what was called $c_2$ is actually our $c_2 - c_4$). 

\smallskip
\noindent
(iii) Hence, the results of Theorem \ref{thm:statmnt_main_csts} are consistent with and generalize previous results on either lower dimensions or simpler models. 
\label{rem:GCIId_compar_SOn}
\end{remark}

We note that these formulas make $c_1, \ldots, c_4$ explicitely computable, at the expense of the resolution of System \eqref{eq:GCIId_strongform_alpha} and the computations of the integrals involved in the formulas above. In particular, it may be possible to compute numerical approximations of them for not too large values of $p$. For large values of $p$, analytical approximations will be required. Approximations of $\alpha$ may be obtained by considering the variational formulation \eqref{eq:GCIId_varform_alpha} and restricting the unknown and test function spaces to appropriate (possibly finite-dimensional) subspaces. 

The main objective of this paper is to prove Theorems \ref{thm:statmnt_main} and \ref{thm:statmnt_main_csts}. While the remainder of Section \ref{sec_hydro}, as well as Section \ref{sec_GCI} rely on the same framework as \cite{degond2017new}, the subsequent sections require completely new methodologies.

\subsection{Equilibria}
\label{subsec_equi}

For $f$: $\mathrm{SO}_n \to {\mathbb R}$ smooth enough, we define the collision operator 
\begin{equation} Q(f) = - \kappa \nabla_A \cdot \big( P_{T_A} ({\mathcal P} (J_f)) f \big) + \Delta_A f \quad \mathrm{with} \quad J_f = \int_{\mathrm{SO}_n} f \, A \, dA, 
\label{eq:equi_Q}
\end{equation}
so that \eqref{eq:sca_kin_2} can be recast into 
\begin{equation} 
\partial_t f^\varepsilon +   A {\mathbf e}_1 \cdot \nabla_x f^\varepsilon = \frac{1}{\varepsilon} Q(f^\varepsilon). 
\label{eq:equi_kin}
\end{equation}
It is clear that if $f^\varepsilon \to f^0$ as $\varepsilon \to 0$ strongly as well as all its first-order derivatives with respect to $x$ and second-order derivatives with respect to $A$, then, we must have $Q(f^0) = 0$. Solutions of this equation are called equilibria.  

We have the following lemma whose proof can be found in \cite[Appendix 7]{degond2021body} and is not reproduced here: 

\begin{lemma}
We have 
$$ \int_{\mathrm{SO}_n} A \, M_\Gamma(A) \, dA = c_1 \Gamma, \quad c_1 = \Big\langle \frac{ \mathrm{Tr}(A)}{n} \Big\rangle_{\exp(\kappa  \mathrm{Tr} (A)/2)},  $$
where for two functions $f$, $g$: $\mathrm{SO}_n \to {\mathbb R}$, with $g \geq 0$ and $g \not \equiv 0$, we note 
$$ \langle f(A) \rangle_{g(A)} = \frac{\int_{\mathrm{SO}_n} f(A) \, g(A) \, dA}{\int_{\mathrm{SO}_n} g(A) \, dA}. $$
The function $c_1$: ${\mathbb R} \to {\mathbb R}$, $\kappa \mapsto c_1(\kappa)$ is a nonnegative, nondecreasing function which satisfies $c_1(0) = 0$, and $\lim_{\kappa \to \infty} c_1(\kappa) = 1$. It is given by \eqref{eq:equi_order_param_express}.
\label{lem:equi_order_param}
\end{lemma}

From now on, we will use $\Gamma_f = {\mathcal P}(J_f)$. We have different expressions of $Q$ expressed in the following 

\begin{lemma}
We have 
\begin{eqnarray} 
Q(f) &=& - \kappa \nabla_A \cdot \big( P_{T_A} \Gamma_f f \big) + \Delta_A f \label{eq:equi_Q1}\\
&=& \nabla_A \cdot \Big[ M_{\Gamma_f} \nabla_A \Big( \frac{f}{M_{\Gamma_f}} \Big) \Big] \label{eq:equi_Q2}\\
&=& \nabla_A \cdot \Big[ f \nabla_A \big( - \kappa \Gamma_f \cdot A + \log f \big) \Big]. \label{eq:equi_Q3}
\end{eqnarray}
\label{lem:equi_Q_express}
\end{lemma}

\noindent
\textbf{Proof.} Formula \eqref{eq:equi_Q1} is nothing but \eqref{eq:equi_Q} with $\Gamma_f$ in place of ${\mathcal P}(J_f)$. To get the other two expressions, we note that 
\begin{equation} 
\nabla_A (\Gamma \cdot A) = P_{T_A} \Gamma, \quad \forall A, \, \Gamma \in \mathrm{SO}_n.
\label{eq:equi_Q_express_prf1}
\end{equation}
Then, \eqref{eq:equi_Q3} follows immediately and for \eqref{eq:equi_Q2} we have 
\begin{eqnarray*} 
\nabla_A \cdot \Big[ M_{\Gamma_f} \nabla_A \Big( \frac{f}{M_{\Gamma_f}} \Big) \Big]  &=& \Delta_A f - \nabla_A \cdot \big[f \nabla_A \big( \log(M_{\Gamma_f}) \big) \big] \\
&=& \Delta_A f - \kappa \nabla_A \cdot \big[f P_{T_{A}} \Gamma_f \big] = Q(f), 
\end{eqnarray*}
which ends the proof. \endproof

The following gives the equilibria of $Q$: 

\begin{lemma}
Let $f$: $\mathrm{SO}_n \to {\mathbb R}$ be smooth enough such that $f \geq 0$, $\rho_f >0$ and $\mathrm{det} J_f >0$. Then, we have 
$$ Q(f) = 0 \Longleftrightarrow \exists (\rho,\Gamma) \in (0,\infty) \times \mathrm{SO}_n \text{ such that } f = \rho M_{\Gamma}. $$
\label{lem:equi}
\end{lemma}

\noindent
\textbf{Proof.} Suppose $f$ fulfils the assumptions of the lemma and is such that $Q(f)=0$. Using~\eqref{eq:equi_Q2} and Stokes' theorem, this implies: 
$$ 0 =  \int_{\mathrm{SO}_n} Q(f) \frac{f}{M_{\Gamma_f}} \, dA = - \int_{\mathrm{SO}_n} \Big| \nabla_A \Big( \frac{f}{M_{\Gamma_f}} \Big) \Big|^2 \, M_{\Gamma_f} \, dA. $$
Hence, $f/M_{\Gamma_f}$ is constant. So, there exists $\rho \in (0,\infty)$ such that $f = \rho M_{\Gamma_f}$ which shows that $f$ is of the form $\rho M_{\Gamma}$ for some $(\rho,\Gamma) \in (0,\infty) \times \mathrm{SO}_n$. 

\smallskip
\noindent
Conversely, let $f = \rho M_\Gamma$ for some $(\rho,\Gamma) \in (0,\infty) \times \mathrm{SO}_n$. If we show that $\Gamma = \Gamma_f$, then, by \eqref{eq:equi_Q2} we deduce that $Q(f) = 0$. By Lemma \ref{lem:equi_order_param}, we have $J_{\rho M_\Gamma} = \rho c_1 \Gamma$ with $\rho c_1 > 0$. Thus, $\Gamma_f = {\mathcal P}(\rho c_1 \Gamma) = \Gamma$, which ends the proof. Note that knowing that $c_1 > 0$ is crucial in that step of the proof. \endproof

\begin{corollary}
Assume that $f^\varepsilon \to f^0$ as $\varepsilon \to 0$ strongly as well as all its first-order derivatives with respect to $x$ and second-order derivatives with respect to $A$, such that $\int f^0(x,v,t)  dv >0$, $\forall (x,t) \in {\mathbb R}^n \times [0,\infty)$. Then, there exists two functions $\rho$: ${\mathbb R}^n \times [0,\infty) \to (0,\infty)$ and $\Gamma$: ${\mathbb R}^n \times [0,\infty) \to \mathrm{SO}_n$ such that \eqref{eq:equi_f0express} holds. 
\label{cor:equi}
\end{corollary}
\noindent
\textbf{Proof.} This is an obvious consequence of Lemma \ref{lem:equi} since, for any given $(x,t) \in {\mathbb R}^n \times [0,\infty)$, the function $f^0(x,\cdot,t)$ satisfies $Q(f^0(x,\cdot,t)) = 0$. \endproof

Now, we are looking for the equations satisfied by $\rho$ and $\Gamma$.

\subsection{The continuity equation}
\label{subsec_cont}

\begin{proposition}
The functions $\rho$ and $\Gamma$ involved in \eqref{eq:equi_f0express} satisfy the continuity equation~\eqref{eq:cont_eq}.  
\label{prop_cont}
\end{proposition}

\noindent
\textbf{Proof.} By Stokes's theorem, for all second-order differentiable function $f$: $\mathrm{SO}_n \to {\mathbb R}$, we have $ \int_{\mathrm{SO}_n} Q(f) \, dA = 0$. Therefore, integrating \eqref{eq:equi_kin} with respect to $A$, we obtain, 
\begin{equation} 
\int_{\mathrm{SO}_n} (\partial_t + A {\mathbf e}_1 \cdot \nabla_x) f^\varepsilon \, dA = 0. 
\label{eq:cont_prf1}
\end{equation}
For any distribution function $f$, we define
$$ \rho_f(x,t) = \int_{\mathrm{SO}_n} f(x,A,t) \, dA. $$
Thus, with \eqref{eq:sca_J_2}, Eq. \eqref{eq:cont_prf1} can be recast into
\begin{equation}
\partial_t \rho_{f^\varepsilon} + \nabla_x \cdot (J_{f^\varepsilon} {\mathbf e}_1) = 0. 
\label{eq:cont_prf2}
\end{equation}
Now, given that the convergence of $f^\varepsilon \to f^0$ as $\varepsilon \to 0$ is supposed strong enough, and thanks to Lemma \ref{lem:equi_order_param}, we have 
$$ \rho_{f^\varepsilon} \to \rho_{f^0} = \rho_{\rho M_\Gamma} = \rho, \qquad J_{f^\varepsilon} \to J_{f^0} = J_{\rho M_\Gamma} = \rho c_1 \Gamma. $$
Then, passing to the limit $\varepsilon \to 0$ in \eqref{eq:cont_prf2} leads to \eqref{eq:cont_eq}. \endproof


\setcounter{equation}{0}
\Section{Generalized collision invariants: definition and existence}
\label{sec_GCI}

\subsection{Definition and first characterization}
\label{subsec_GCI}

Now, we need an equation for $\Gamma$. We see that the proof of Prop. \ref{prop_cont} can be reproduced if we can find functions $\psi$: $\mathrm{SO}_n \to {\mathbb R}$ such that for all second-order differentiable function $f$: $\mathrm{SO}_n \to {\mathbb R}$, we have
\begin{equation} 
\int_{\mathrm{SO}_n} Q(f) \, \psi \, dA = 0. 
\label{eq:GCI_CI}
\end{equation}
Such a function is called a collision invariant. In the previous proof, the collision invariant $\psi = 1$ was used. Unfortunately, it can be verified that the only collision invariants of $Q$ are the constants. Thus, the previous proof cannot be reproduced to find an equation for~$\Gamma$. In order to find more equations, we have to relax the condition that \eqref{eq:GCI_CI} must be satisfied for all functions $f$. This leads to the concept of generalized collision invariant (GCI). We first introduce a few more definitions. 

Given $\Gamma \in \mathrm{SO}_n$, we define the following linear Fokker-Planck operator, defined for second order differentiable functions $f$: $\mathrm{SO}_n \to {\mathbb R}$:  
$$ {\mathcal Q}(f,\Gamma) = \nabla \cdot \Big[ M_\Gamma \nabla \Big( \frac{f}{M_\Gamma} \Big) \Big] . $$
For simplicity, in the remainder of the present section as well as in Sections \ref{subsec:rotmore} and \ref{subsec_GCIId}, we will drop the subscript $A$ to the $\nabla$, $\nabla \cdot$ and $\Delta$ operators as all derivatives will be understood with respect to $A$. 

We note that 
\begin{equation} 
Q(f) = {\mathcal Q}(f,\Gamma_f). 
\label{eq:GCI_Qf}
\end{equation}

\begin{definition}
Given $\Gamma \in \mathrm{SO}_n$, a GCI associated with $\Gamma$ is a function $\psi$: $\mathrm{SO}_n \to {\mathbb R}$ such that 
\begin{equation} 
\int_{\mathrm{SO}_n} {\mathcal Q}(f,\Gamma) \, \psi \, dA = 0 \quad \text{for all} \quad  f: \, \mathrm{SO}_n \to {\mathbb R} \quad \text{such that} \quad P_{T_\Gamma} J_f = 0. 
\label{eq:GCI_def}
\end{equation}
The set ${\mathcal G}_\Gamma$ of GCI associated with $\Gamma$ is a vector space. 
\label{def_GCI_def}
\end{definition}

From this definition, we have the following lemma which gives a justification why this concept is useful for the hydrodynamic limit. 

\begin{lemma}
We have 
\begin{equation} 
\psi \in {\mathcal G}_{\Gamma_f} \quad \Longrightarrow \quad \int_{\mathrm{SO}_n} Q(f) \, \psi \, dA = 0. 
\label{eq:GCI_Ggamf}
\end{equation}
\label{lem:GCI_CI}
\end{lemma}

\noindent
\textbf{Proof.}. By \eqref{eq:GCI_Qf} and \eqref{eq:GCI_def}, it is enough to show that $P_{T_{\Gamma_f}} J_f = 0$. But $\Gamma_f = {\mathcal P}(J_f)$, so there exists a symmetric positive-definite matrix $S$ such that $J_f =  \Gamma_f S$. So, 
$$ P_{T_{\Gamma_f}} J_f = \Gamma_f \frac{\Gamma_f^T J_f - J_f^T \Gamma_f}{2} = \Gamma_f \frac{S-S^T}{2} = 0. $$
\endproof

The following lemma provides the equation solved by the GCI. 

\begin{lemma}
The function $\psi$: $\mathrm{SO}_n \to {\mathbb R}$ belongs to ${\mathcal G}_\Gamma$ if and only if $\exists P \in T_\Gamma$ such that 
\begin{equation} 
\nabla \cdot \big( M_\Gamma \nabla \psi \big) = P \cdot A \, M_\Gamma. 
\label{eq:GCI_eq}
\end{equation}
\label{lem:GCI_eq}
\end{lemma}

\noindent
\textbf{Proof.} On the one hand, we can write
$$ \int_{\mathrm{SO}_n} {\mathcal Q}(f,\Gamma) \, \psi \, dA = \int_{\mathrm{SO}_n} f \, {\mathcal Q}^*(\psi,\Gamma) \, dA, $$
where ${\mathcal Q}^*(\cdot,\Gamma)$ is the formal $L^2$-adjoint to ${\mathcal Q}(\cdot,\Gamma)$ and is given by
$$ {\mathcal Q}^*(\psi,\Gamma) = M_\Gamma^{-1} \nabla \cdot \big( M_\Gamma \nabla \psi \big). $$
On the other hand, we have 
\begin{eqnarray*} P_{T_\Gamma} J_f =0 & \Longleftrightarrow & P_{T_\Gamma} \int_{\mathrm{SO}_n} f \, A \, dA = 0 \\
& \Longleftrightarrow & \int_{\mathrm{SO}_n} f \, A \cdot P \, dA = 0, \quad \forall P \in T_\Gamma \\
& \Longleftrightarrow & f \in \{ A \mapsto A \cdot P \, \, | \, \, P \in T_\Gamma \}^\bot
\end{eqnarray*}
where the orthogonality in the last statement is meant in the $L^2$ sense. So, by \eqref{eq:GCI_def}, $\psi \in {\mathcal G}_\Gamma$ if and only if 
$$ \{ A \mapsto A \cdot P \, \, | \, \, P \in T_\Gamma \}^\bot \subset \{ {\mathcal Q}^*(\psi,\Gamma) \}^\bot, $$
or by taking orthogonals again, if and only if 
\begin{equation}
\mathrm{Span} \{ {\mathcal Q}^*(\psi,\Gamma) \} \subset \{ A \mapsto A \cdot P \, \, | \, \, P \in T_\Gamma \}, 
\label{eq:GCI_eq_prf1}
\end{equation}
because both sets in \eqref{eq:GCI_eq_prf1} are finite-dimensional, hence closed. Statement \eqref{eq:GCI_eq_prf1} is equivalent to the statement that there exists $P \in T_\Gamma$ such that \eqref{eq:GCI_eq} holds true. \endproof

\subsection{Existence and invariance properties}
\label{subsec_GCIe}

We now state an existence result for the GCI. First, we introduce the following spaces: $L^2(\mathrm{SO}_n)$ stands for the space of square integrable functions $f$: $\mathrm{SO}_n \to {\mathbb R}$ endowed with the usual $L^2$-norm $\|f\|^2_{L^2} = \int_{\mathrm{SO}_n} |f(A)|^2 \, dA$. Then, we define $H^1(\mathrm{SO}_n) = \{ f \in L^2(\mathrm{SO}_n) \, \, | \, \, \nabla f \in L^2(\mathrm{SO}_n)\}$ (where $\nabla f$ is meant in the distributional sense), endowed with the usual $H^1$-norm $\|f\|^2_{H^1} = \|f\|^2_{L^2} + \|\nabla f\|^2_{L^2}$. Finally, $H^1_0(\mathrm{SO}_n)$ are the functions of $H^1(\mathrm{SO}_n)$ with zero mean, i.e. $f \in H^1_0(\mathrm{SO}_n) \Longleftrightarrow f \in H^1(\mathrm{SO}_n)$ and 
\begin{equation}
\int_{\mathrm{SO}_n} f \, dA = 0 . 
\label{eq:GCI_zero_mean}
\end{equation}
We will solve \eqref{eq:GCI_eq} in the variational sense. We note that for $P \in T_\Gamma$, we have
$$ \int_{\mathrm{SO}_n} A \cdot P \, M_\Gamma(A) \, dA = c_1 \Gamma \cdot P = 0,  $$
i.e. the right-hand side of \eqref{eq:GCI_eq} satisfies \eqref{eq:GCI_zero_mean}. Hence, if $\psi$ is a smooth solution of \eqref{eq:GCI_eq} satisfying \eqref{eq:GCI_zero_mean}, it satisfies 
\begin{equation} 
\int_{\mathrm{SO}_n} M_\Gamma \, \nabla \psi \, \nabla \chi \, dA = - \int_{\mathrm{SO}_n} M_\Gamma \, A \cdot P \, \chi \, dA,
\label{eq:GCI_var_form}
\end{equation}
for all functions $\chi$ satisfying \eqref{eq:GCI_zero_mean}. This suggests to look for solutions of \eqref{eq:GCI_var_form} in $H^1_0(\mathrm{SO}_n)$. Indeed, we have

\begin{proposition}
For a given $P \in T_\Gamma$, there exists a unique $\psi \in H^1_0(\mathrm{SO}_n)$ such that \eqref{eq:GCI_var_form} is satisfied for all $\chi \in H^1_0(\mathrm{SO}_n)$. 
\label{prop:GCI_exist_var_form}
\end{proposition}

\noindent
\textbf{Proof.} This is a classical application of Lax-Milgram's theorem. We only need to verify that the bilinear form 
$$ a(\psi,\chi) = \int_{\mathrm{SO}_n} M_\Gamma \, \nabla \psi \, \nabla \chi \, dA, $$
is coercive on $H^1_0(\mathrm{SO}_n)$. Since $\mathrm{SO}_n$ is compact, there exists $C>0$ such that $M_\Gamma \geq C$. So, 
$$ a(\psi,\psi) \geq C \int_{\mathrm{SO}_n} |\nabla \psi|^2\, dA. $$
This is the quadratic form associated with the Laplace operator $- \Delta$ on $\mathrm{SO}_n$. But the lowest eigenvalue of $- \Delta$ is $0$ and its associated eigenspace are the constant functions. Then, there is a spectral gap and the next eigenvalue $\lambda_2$ is positive. Hence, we have 
\begin{equation} 
\int_{\mathrm{SO}_n} |\nabla \psi|^2\, dA \geq \lambda_2 \int_{\mathrm{SO}_n} |\psi|^2\, dA, \quad \forall \psi \in H^1_0(\mathrm{SO}_n),
\label{eq:GCI_exist_var_form_prf1}
\end{equation}
(see e.g. \cite[Section 4.3]{degond2023radial} for more detail). This implies the coercivity of $a$ on $H^1_0(\mathrm{SO}_n)$ and ends the proof. \endproof

\begin{remark}
Since the functions $A \mapsto M_\Gamma(A)$ and  $A \mapsto M_\Gamma(A) \, A \cdot P$ are $C^\infty$, by elliptic regularity, the unique solution of Prop. \ref{prop:GCI_exist_var_form} actually belongs to $C^\infty(\mathrm{SO}_n)$. 
\label{rem:GCI_elliptic_regul}
\end{remark}

\smallskip
For any $P \in T_\Gamma$, there exists $X \in \mathfrak{so}_n$ such that $P = \Gamma X$. We  denote by $\psi_X^\Gamma$ the unique solution of \eqref{eq:GCI_var_form} in $H^1_0(\mathrm{SO}_n)$ associated with $P = \Gamma X$. Then, we have the 

\begin{corollary}
The space ${\mathcal G}_\Gamma$ is given by 
\begin{equation}
{\mathcal G}_\Gamma = \mathrm{Span} \big( \{1\} \cup \{\psi_X^\Gamma \, \, | \, \, X \in \mathfrak{so}_n \} \big), 
\label{eq:GCI_Ggamma_express}
\end{equation}
and we have 
\begin{equation}
\mathrm{dim} \, {\mathcal G}_\Gamma = \mathrm{dim} \, \mathfrak{so}_n + 1 = \frac{n(n-1)}{2}+1. 
\label{eq:GCI_Ggamma_dim}
\end{equation}
\label{cor:GCI_space_GThet}
\end{corollary}

\noindent
\textbf{Proof.} If $\psi \in {\mathcal G}_\Gamma$, then, $\psi - \bar \psi \in {\mathcal G}_\Gamma \cap H^1_0(\mathrm{SO}_n)$ where $\bar \psi = \int_{\mathrm{SO}_n} \psi \, dA$. Then, $\exists X \in \mathfrak{so}_n$ such that $\psi - \bar \psi = \psi_X^\Gamma$, which leads to \eqref{eq:GCI_Ggamma_express}. Now, the map $\mathfrak{so}_n \to H^1_0(\mathrm{SO}_n)$, $X \mapsto \psi_X^\Gamma$ is linear and injective. Indeed, suppose $\psi_X^\Gamma = 0$. Then, inserting it into \eqref{eq:GCI_var_form}, we get that 
$$ \int_{\mathrm{SO}_n} M_\Gamma(A) \, A \cdot (\Gamma X) \, \chi(A) \, dA = 0, \quad \forall \chi \in H^1_0(\mathrm{SO}_n), $$
and by density, this is still true for all $\chi \in L^2(\mathrm{SO}_n)$. This implies that 
$$ M_\Gamma(A) \, A \cdot (\Gamma X) = 0, \quad \forall A \in \mathrm{SO}_n, $$
and since $M_\Gamma >0$, that $A \cdot (\Gamma X) = (\Gamma^T A) \cdot X = 0$, for all $A \in \mathrm{SO}_n$. Now the multiplication by $\Gamma^T$ on the left is a bijection of $\mathrm{SO}_n$, so we get that $X$ satisfies 
$$ A \cdot X = 0, \quad \forall A \in \mathrm{SO}_n. $$
Then, taking $A = e^{tY}$ with $Y \in \mathfrak{so}_n$ and differentiating with respect to $t$, we obtain 
$$ Y  \cdot X = 0, \quad \forall Y \in \mathfrak{so}_n, $$
which shows that $X=0$. Hence, ${\mathcal G}_\Gamma$ is finite-dimensional and~\eqref{eq:GCI_Ggamma_dim} follows.
\endproof

From now on, we will repeatedly use the following lemma. 

\begin{lemma}
(i) Let $g \in \mathrm{SO}_n$ and let $\ell_g$, $r_g$ and $\xi_g$ be the left and right translations and conjugation maps of $\mathrm{SO}_n$ respectively: 
\begin{equation} 
\ell_g(A) = gA, \quad r_g(A) = Ag, \quad \xi_g = \ell_g \circ r_{g^{-1}} = \ell_g \circ r_{g^T}, \quad \forall A \in \mathrm{SO}_n. 
\label{eq:GCI_conjug_def}
\end{equation}
Let $f$: $\mathrm{SO}_n \to {\mathbb R}$ be smooth. Then, we have for any $X \in \mathfrak{so}_n$: 
\begin{eqnarray}
\nabla (f \circ \ell_g) (A) \cdot AX &=& \nabla f (gA) \cdot gAX, 
\label{eq:GCI_deriv_compos_transl} \\
\nabla (f \circ \xi_g) (A) \cdot AX &=& \nabla f (gAg^T) \cdot gAXg^T. 
\label{eq:GCI_deriv_compos_conjug}
\end{eqnarray}

\smallskip
\noindent
(ii) If $f$ and $\varphi$: $\mathrm{SO}_n \to {\mathbb R}$ are smooth, then, 
\begin{eqnarray}
\nabla (f \circ \ell_g) (A) \cdot \nabla (\varphi \circ \ell_g) (A) &=& \nabla f(gA) \cdot \nabla \varphi (gA), \label{eq:GCI_grad_compos_transl} \\
\nabla (f \circ \xi_g) (A) \cdot \nabla (\varphi \circ \xi_g) (A) &=& \nabla f(gAg^T) \cdot \nabla \varphi (gAg^T), \label{eq:GCI_grad_compos_conjug} 
\end{eqnarray}
\label{lem:GCI_deriv_compos_conjug}
\end{lemma}

\noindent
\textbf{Proof.} (i) We show \eqref{eq:GCI_deriv_compos_transl}. \eqref{eq:GCI_deriv_compos_conjug} is shown in a similar way. By \eqref{eq:rot_operator_rho}, we have 
$$ \nabla  (f \circ \ell_g) (A) \cdot AX = \frac{d}{dt} \big(  (f \circ \ell_g)(A e^{tX}) \big)|_{t=0} = \frac{d}{dt} \big( f(g A e^{tX}) \big)|_{t=0} = \nabla  f(g A) \cdot g A X. $$ 

\smallskip
\noindent
(ii) Again, we show \eqref{eq:GCI_grad_compos_transl}, the proof of \eqref{eq:GCI_grad_compos_conjug} being similar. Applying \eqref{eq:GCI_deriv_compos_transl} twice, we have 
\begin{eqnarray*}
\nabla f(gA) \cdot \nabla \varphi(gA) &=& \nabla (f \circ \ell_g) (A) \cdot \big(g^T \nabla \varphi(gA) \big) = \nabla \varphi(gA) \cdot \big( g  \nabla (f \circ \ell_g) (A) \big) \\
&=& \nabla (\varphi \circ \ell_g) (A) \cdot \nabla (f \circ \ell_g) (A). 
\end{eqnarray*}

\vspace{-1cm}
\endproof

\begin{proposition}[translation invariance]
We have 
\begin{equation}
\psi_X^{\mathrm{I}}(A) = \psi_X^\Gamma(\Gamma A), \quad \forall A, \, \Gamma \in \mathrm{SO}_n, \quad \forall X \in \mathfrak{so}_n.  
\label{eq:GCI_transl}
\end{equation}
\label{prop:GCI_transl}
\end{proposition}

\noindent
\textbf{Proof.} $\psi = \psi_X^\Gamma$ is the unique solution in $H^1_0(\mathrm{SO}_n)$ of the following variational formulation: 
\begin{eqnarray*} 
&&\hspace{-1cm} 
\int_{\mathrm{SO}_n} \exp \big(\frac{\kappa}{2} \mathrm{Tr} (\Gamma^T A) \big)  \, \nabla \psi(A) \cdot \nabla \chi(A) \, dA \\
&&\hspace{1cm} 
= - \frac{1}{2} \int_{\mathrm{SO}_n} \exp \big(\frac{\kappa}{2} \mathrm{Tr} (\Gamma^T A) \big)  \, \mathrm{Tr} (A^T \Gamma X) \, \chi(A) \, dA, \quad
\forall \chi \in H^1_0(\mathrm{SO}_n). 
\end{eqnarray*}
By the change of variables $A' = \Gamma^T A$, the translation invariance of the Haar measure and~\eqref{eq:GCI_grad_compos_transl}, we get, dropping the primes for simplicity: 
\begin{eqnarray} 
&&\hspace{-1cm} 
\int_{\mathrm{SO}_n} \exp (\frac{\kappa}{2} \mathrm{Tr} A )  \, \nabla (\psi \circ \ell_\Gamma) (A) \cdot \nabla (\chi \circ \ell_\Gamma) (A) \, dA  \nonumber \\
&&\hspace{1cm} 
= - \frac{1}{2} \int_{\mathrm{SO}_n} \exp (\frac{\kappa}{2} \mathrm{Tr}  A )  \, \mathrm{Tr} (A^T X) \, \chi \circ \ell_\Gamma (A) \, dA, \quad \forall \chi \in H^1_0(\mathrm{SO}_n),
\label{eq:GCI_transl_prf1}
\end{eqnarray}
We remark that the mapping $H^1_0(\mathrm{SO}_n) \to H^1_0(\mathrm{SO}_n)$, $\chi \to \chi \circ \ell_\Gamma$ is a linear isomorphism and an isometry (the proof is analogous to the proof of Prop. \ref{prop:GCIId_L2inv_H1inv_prop} below and is omitted). Thus, we can replace $\chi \circ \ell_\Gamma$ in \eqref{eq:GCI_transl_prf1} by any test function $\tilde \chi \in H^1_0(\mathrm{SO}_n)$, which leads to a variational formulation for $\psi \circ \ell_\Gamma$ which is identical with that of $\psi_X^{\mathrm{I}}$. By the uniqueness of the solution of the variational formulation, this leads to \eqref{eq:GCI_transl} and finishes the proof. \endproof

\begin{proposition}[Conjugation invariance]
We have 
\begin{equation}
\psi_X^{\mathrm{I}}(gAg^T) = \psi_{g^TXg}^{\mathrm{I}}(A), \quad \forall A, \, g \in \mathrm{SO}_n.  
\label{eq:GCIId_conjug_invar}
\end{equation}
\label{prop:GCIId_conjug_invar}
\end{proposition}

\noindent
\textbf{Proof.} The proof is identical to that of Prop. \ref{prop:GCI_transl}. We start from the variational formulation for $\psi_X^{\mathrm{I}}$ and make the change of variables $A = g A' g^T$ in the integrals. Thanks to \eqref{eq:GCI_grad_compos_transl}, it yields a variational formulation for $\psi_X^{\mathrm{I}} \circ \xi_g$, which is noticed to be identical with that of $\psi_{g^TXg}^{\mathrm{I}}$. By the uniqueness of the solution of the variational formulation, we get \eqref{eq:GCIId_conjug_invar}. \endproof

From this point onwards, the search for GCI differs significantly from \cite{degond2017new} where the assumption of dimension $n=3$ was crucial. We will need further concepts about the rotation groups which are summarized in the next section.

\setcounter{equation}{0}
\section{Maximal torus and Weyl group}
\label{subsec:rotmore}

If $g \in \mathrm{SO}_n$, the conjugation map $\xi_g$ given by \eqref{eq:GCI_conjug_def} is a group isomorphism. Let $A$ and $B \in \mathrm{SO}_n$. We say that $A$ and $B$ are conjugate, and we write $A \sim B$, if and only if $\exists g \in \mathrm{SO}_n$ such that $B = gAg^T$. It is an equivalence relation. Conjugation classes can be described as follows. The planar rotation $R_\theta$ for $\theta \in {\mathbb R}/(2 \pi {\mathbb Z})$ is defined by
\begin{equation} 
R_\theta = \left( \begin{array}{rr} \cos \theta & - \sin \theta \\ \sin  \theta &  \cos \theta \end{array} \right). 
\label{eq:def_Rtheta}
\end{equation}
The set ${\mathcal T}=[- \pi, \pi)^p$ will be identified with the torus $({\mathbb R}/(2 \pi {\mathbb Z}))^p$. For $\Theta =: (\theta_1, \ldots , \theta_p) \in {\mathcal T}$, we define the matrix $A_\Theta$ blockwise by: 
\begin{itemize}
\item[$\bullet$] in the case $n = 2p$, $p \geq 2$, 
\begin{equation} A_\Theta = \left( \begin{array}{cccc} 
\scalebox{1.2}{$R_{\theta_1}$} &  &  & \scalebox{2.}{$0$} \\
& \scalebox{1.2}{$R_{\theta_2}$} &  & \\
&  &  \ddots  &  \\
\scalebox{2.}{$0$} &  &  &  \scalebox{1.2}{$R_{\theta_p}$}
\end{array} \right) \in \mathrm{SO}_{2p}{\mathbb R}, 
\label{eq:R2p}
\end{equation}
\item[$\bullet$] in the case $n = 2p+1$, $p \geq 1$, 
\begin{equation}
A_\Theta = \left( \begin{array}{ccccc} 
\scalebox{1.2}{$R_{\theta_1}$} &  &  & \scalebox{2.}{$0$} & 0 \\
& \scalebox{1.2}{$R_{\theta_2}$} &  & & \vdots \\
&  &  \ddots  &  & \vdots \\
\scalebox{2.}{$0$} &  &  &  \scalebox{1.2}{$R_{\theta_p}$} & 0 \\
0  &  \ldots & \ldots & 0 & 1 
\end{array} \right) \in \mathrm{SO}_{2p+1}{\mathbb R}. 
\label{eq:R2p+1}
\end{equation}
\end{itemize}
By classical matrix reduction theory, any $A \in \mathrm{SO}_n$ is conjugate to $A_\Theta$ for some $\Theta \in {\mathcal T}$. We define the subset ${\mathbb T}$ of $\mathrm{SO}_n$ by
$$ {\mathbb T} = \{ A_\Theta \, \, | \, \, \Theta \in {\mathcal T} \}. $$
${\mathbb T}$ is an abelian subgroup of $\mathrm{SO}_n$ and the map ${\mathcal T} \to {\mathbb T}$ is a group isomorphism. It can be shown that ${\mathbb T}$ is a maximal abelian subgroup of $\mathrm{SO}_n$ and for that reason, ${\mathbb T}$ is called a maximal torus. ${\mathbb T}$ is a Lie sub-group of $\mathrm{SO}_n$ and we denote by $\mathfrak{h}$ its Lie algebra. $\mathfrak{h}$ is a Lie subalgebra of $\mathfrak{so}_n$ given by 
$$ \mathfrak{h} = \big\{ \sum_{i=1}^p \alpha_i F_{2i-1 \, 2i} \, \, | \, \, (\alpha_1, \ldots \alpha_p) \in {\mathbb R}^p \big\}, $$
where we recall that $F_{ij}$ is defined by \eqref{eq:rot_def_Fij}. $\mathfrak{h}$ is an abelian subalgebra (i.e. $[X,Y] = 0$, $\forall X, \, Y \in \mathfrak{h}$) and is actually maximal among abelian subalgebras. In Lie algebra language, $\mathfrak{h}$ is a Cartan subalgebra of $\mathfrak{so}_n$. 

Let us describe the elements $g \in \mathrm{SO}_n$ that conjugate an element of ${\mathbb T}$ into an element of~${\mathbb T}$. Such elements form a group called the normalizer of ${\mathbb T}$ and denoted by $N({\mathbb T})$. Since~${\mathbb T}$ is abelian, elements of ${\mathbb T}$ conjugate an element of ${\mathbb T}$ to itself so we are rather interested in those elements of $N({\mathbb T})$ that conjugate an element of ${\mathbb T}$ to a different one. In other words, we want to describe the quotient group $N({\mathbb T})/{\mathbb T}$ (clearly, ${\mathbb T}$ is normal in $N({\mathbb T})$) which is a finite group called the Weyl group and denoted by $\mathfrak{W}$. The Weyl group differs in the odd and even dimension cases. It is generated by the following elements $g$ of $N({\mathbb T})$ (or stricly speaking by the cosets $g {\mathbb T}$ where $g$ are such elements) \cite{fulton2013representation}:
\begin{itemize} 
\item[$\bullet$] Case $n=2p$ even: 
\begin{itemize}
\item[-] elements $g = C_{ij}$, $1 \leq i < j \leq p$ where $C_{ij}$ exchanges $e_{2i-1}$ and $e_{2j-1}$ on the one hand, $e_{2i}$ and $e_{2j}$ on the other hand, and fixes all the other basis elements, (where $(e_i)_{i=1}^{2p}$ is the canonical basis of ${\mathbb R}^{2p}$). Conjugation of $A_{\Theta}$ by~$C_{ij}$ exchanges the blocks $R_{\theta_i}$ and $R_{\theta_j}$ of \eqref{eq:R2p}, i.e. exchanges $\theta_i$ and $\theta_j$. It induces an isometry of ${\mathbb R}^p$ (or ${\mathcal T}$), still denoted by $C_{ij}$ by abuse of notation, such that $C_{ij} (\Theta) = ( \ldots, \theta_j, \ldots \theta_i, \ldots)$. This isometry is the reflection in the hyperplane $\{e_i - e_j\}^\bot$;
\item[-] elements $g = D_{ij}$, $1 \leq i < j \leq p$ where $D_{ij}$ exchanges $e_{2i-1}$ and $e_{2i}$ on the one hand, $e_{2j-1}$ and $e_{2j}$ on the other hand, and fixes all the other basis elements. Conjugation of $A_{\Theta}$ by $D_{ij}$ changes the sign of $\theta_i$ in  $R_{\theta_i}$ and that of $\theta_j$ in $R_{\theta_j}$, i.e. changes $(\theta_i,\theta_j)$ into $(-\theta_i, -\theta_j)$. It induces an isometry of ${\mathbb R}^p$ (or ${\mathcal T}$), still denoted by $D_{ij}$ such that $D_{ij}(\Theta) = ( \ldots, - \theta_i, \ldots - \theta_j, \ldots)$. The transformation $C_{ij} \circ D_{ij} = D_{ij} \circ C_{ij}$  is the reflection in the hyperplane $\{e_i + e_j\}^\bot$; 

\end{itemize}
\item[$\bullet$] Case $n=2p+1$ odd: 
\begin{itemize}
\item[-] elements $g = C_{ij}$, $1 \leq i < j \leq p$ identical to those of the case $n=2p$;  
\item[-] elements $g = D_i$, $1 \leq i \leq p$, where $D_i$ exchanges $e_{2i-1}$ and $e_{2i}$ on the one hand, maps $e_{2p+1}$ into $-e_{2p+1}$ on the other hand, and fixes all the other basis elements. Conjugation of $A_{\Theta}$ by $D_i$ changes the sign of $\theta_i$ in  $R_{\theta_i}$, i.e. changes~$\theta_i$ into $-\theta_i$. It induces an isometry of ${\mathbb R}^p$ (or ${\mathcal T}$), still denoted by $D_i$ such that $D_i(\Theta) = ( \ldots, - \theta_i, \ldots)$. It is the reflection in the hyperplane $\{e_i\}^\bot$. 
\end{itemize}
\end{itemize}
The group of isometries of ${\mathbb R}^p$ (or ${\mathcal T}$) generated by $\{C_{ij}\}_{1 \leq i < j \leq p} \cup \{D_{ij}\}_{1 \leq i < j \leq p}$ in the case $n=2p$ and $\{C_{ij}\}_{1 \leq i < j \leq p} \cup \{D_i\}_{1=1}^p$ in the case $n=2p+1$ is still the Weyl group~$\mathfrak{W}$. Note that in the case $n=2p$, an element of $\mathfrak{W}$ induces only an even number of sign changes of~$\Theta$, while in the case $n=2p+1$, an arbitrary number of sign changes are allowed. $\mathfrak{W}$ is also generated by the orthogonal symmetries in the hyperplanes $\{e_i \pm e_j\}^\bot$ for $1 \leq i < j \leq p$ in the case $n=2p$. The elements of the set $\{\pm e_i \pm e_j\}_{1 \leq i< j \leq p }$ are called the roots of $\mathrm{SO}_{2p}$, while the roots of $\mathrm{SO}_{2p+1}$ are the elements of the set $\{\pm e_i \pm e_j\}_{1 \leq i< j \leq p } \cup \{\pm e_i\}_{1=1}^p$.

We also need one more definition. A closed Weyl chamber is the closure of a connected component of the complement of the union of the hyperplanes orthogonal to the roots. The Weyl group acts simply transitively on the closed Weyl chambers \cite[Sect. 14.1]{fulton2013representation}, i.e. for two closed Weyl chambers ${\mathcal W}_1$ and ${\mathcal W}_2$, there exists a unique $W \in \mathfrak{W}$ such that $W({\mathcal W}_1) = {\mathcal W}_2$. 
A distinguished closed Weyl chamber (that associated with a positive ordering of the roots) is given by \cite[Sect. 18.1]{fulton2013representation}: 
\begin{itemize}
\item[$\bullet$] Case $n=2p$ even: 
$$ {\mathcal W} = \big\{ \Theta \in {\mathbb R}^p \, \, | \, \, \theta_1 \geq \theta_2 \geq \ldots \geq \theta_{p-1} \geq |\theta_p| \geq 0 \big\}, $$
\item[$\bullet$] Case $n=2p+1$ odd: 
$$ {\mathcal W} = \big\{ \Theta \in {\mathbb R}^p \, \, | \, \, \theta_1 \geq \theta_2 \geq \ldots \geq \theta_{p-1} \geq \theta_p \geq 0 \big\}, $$
\end{itemize}
all other closed Weyl chambers being of the form $W({\mathcal W})$ for some element $W \in \mathfrak{W}$. We have
\begin{equation} 
{\mathbb R}^p = \bigcup_{W \in \mathfrak{W}} W \big( {\mathcal W} \big), 
\label{eq:rotmore_Weylchamb_union}
\end{equation}
and for any $W_1$, $W_2 \in \mathfrak{W}$, 
\begin{equation} 
W_1 \not = W_2 \, \, \Longrightarrow \, \, \mathrm{meas} \Big( W_1 \big( {\mathcal W} \big) \cap W_2 \big( {\mathcal W} \big) \Big)= 0, 
\label{eq:rotmore_Weylchamb_intersect}
\end{equation}
(where $\mathrm{meas}$ stands for the Lebesgue measure), the latter relation reflecting that the intersection of two Weyl chambers is included in a hyperplane. Defining ${\mathcal W}_{\mathrm{per}} = {\mathcal W} \cap [- \pi, \pi]^p$, we have  \eqref{eq:rotmore_Weylchamb_union} and \eqref{eq:rotmore_Weylchamb_intersect} with ${\mathbb R}^p$ replaced by $[- \pi, \pi]^p$ and ${\mathcal W}$ replaced by ${\mathcal W}_{\mathrm{per}}$. 

Class functions are functions $f$: $\mathrm{SO}_n \to {\mathbb R}$ that are invariant by conjugation, i.e. such that $f(gAg^T)=f(A)$, $\forall \, A, \, g \in \mathrm{SO}_n$. By the preceding discussion, a class function can be uniquely associated with a function $\varphi_f$: $\tilde {\mathcal T} =: {\mathcal T}/\mathfrak{W} \to {\mathbb R}$ such that $\varphi_f (\Theta) = f(A_\Theta)$. By a function on ${\mathcal T}/\mathfrak{W}$, we mean a function on ${\mathcal T}$ which is invariant by any isometry belonging to the Weyl group $\mathfrak{W}$. The Laplace operator $\Delta$ maps class functions to class functions. Hence, it generates an operator $L$ on $C^\infty(\tilde {\mathcal T})$ such that for any class function $f$ in $C^\infty(\mathrm{SO}_n )$, we have:
\begin{equation} 
L \varphi_f = \varphi_{\Delta f}. 
\label{eq:radlap_def}
\end{equation}
Expressions of the operator $L$ (called the radial Laplacian) are derived in \cite{degond2023radial}. They are recalled in Appendix \ref{appsec_direct_strong_form}. Class functions are important because they are amenable to a specific integration formula called the Weyl integration formula which states that if $f$ is an integrable class function on $\mathrm{SO}_n$, then, 
\begin{equation}
\int_{\mathrm{SO}_n{\mathbb R}} f(A) \, dA = \gamma_n \frac{1}{(2 \pi)^p} \int_{{\mathcal T}} f(A_\Theta) \, u_n (\Theta) \, d \Theta,
\label{eq:WIF}
\end{equation}
with~$u_n$ defined in~\eqref{eq:Pi2p} and~\eqref{eq:Pi2p+1}, and

$$ \gamma_n = \left\{ \begin{array}{lll} 
\frac{2^{(p-1)^2}}{p!} & \mathrm{ if } & n=2p \\
\mbox{} & & \\
\frac{2^{p^2}}{p!} & \mathrm{ if } & n=2p+1 \\
\end{array} \right. . $$

We now introduce some additional definitions. The adjoint representation of $\mathrm{SO}_n$ denoted by ``$\mathrm{Ad}$'' maps $\mathrm{SO}_n$ into the group $\mathrm{Aut}(\mathfrak{so}_n)$ of linear automorphisms of $\mathfrak{so}_n$ as follows: 
$$ \mathrm{Ad}(A) (Y) = A Y A^{-1},  \quad  \forall A \in \mathrm{SO}_n, \quad \forall Y \in \mathfrak{so}_n. $$
$\mathrm{Ad}$ is a Lie-group representation of $\mathrm{SO}_n{\mathbb R}$, meaning that 
$$ \mathrm{Ad}(A) \mathrm{Ad}(B) = \mathrm{Ad}(AB), \quad \forall A, \, B \in \mathrm{SO}_n. $$
We have 
$$ \mathrm{Ad}(A) (X) \cdot \mathrm{Ad}(A) (Y) = X \cdot Y, \quad  \forall A \in \mathrm{SO}_n, \quad \forall X, \, Y \in \mathfrak{so}_n, $$
showing that the inner product on $\mathfrak{so}_n$ is invariant by $\mathrm{Ad}$. The following identity, shown in \cite[Section 8.2]{faraud2008Analysis} will be key to the forthcoming analysis of the GCI. Let $f$ be a function $\mathrm{SO}_n \to V$, where $V$ is a finite-dimensional vector space over ${\mathbb R}$. Then, we have, using the definition~\eqref{eq:rot_def_rho} of~$\varrho$ :
\begin{equation}
\Big(\varrho\big(\mathrm{Ad}(g^{-1}) T - T \big) f\Big)(g) = \frac{d}{ds} \big( f(e^{sT}ge^{-sT}) \big) \big|_{s=0}, \quad \forall g \in \mathrm{SO}_n, \quad \forall T \in \mathfrak{so}_n. 
\label{eq:radlap_deriv_conjug}
\end{equation}

We finish with the following identity which will be used repeatedly. 
\begin{equation}
[F_{ij},F_{k \ell}] = \big( \delta_{jk} F_{i \ell} + \delta_{i \ell} F_{jk} - \delta_{ik} F_{j \ell} - \delta_{j \ell} F_{ik} \big). 
\label{eq:commutF}
\end{equation}

\setcounter{equation}{0}
\section{Generalized collision invariants associated with the identity}
\label{subsec_GCIId}

\subsection{Introduction of $\alpha = (\alpha_i)_{i=1}^p$ and first properties}
\label{subsec_GCIId_introduction_alpha}

This section is devoted to the introduction of the function $\alpha = (\alpha_i)_{i=1}^p$, ${\mathcal T} \to {\mathbb R}^p$ which will be eventually shown to solve System \eqref{eq:GCIId_strongform_alpha}. 

For simplicity, we denote $\psi_X^{\mathrm{I}}$ simply by $\psi_X$ and $M_{\mathrm{I}}$ by $M$. Let $A, \, \Gamma \in \mathrm{SO}_n$ be fixed. The map $\mathfrak{so}_n \to {\mathbb R}$, $X \mapsto \psi_X^\Gamma(A)$ is a linear form. Hence, there exists a map $\mu^\Gamma$: $\mathrm{SO}_n \to \mathfrak{so}_n$ such that 
\begin{equation} 
\psi_X^\Gamma(A) = \mu^\Gamma(A) \cdot X, \quad \forall A \in \mathrm{SO}_n, \quad \forall X \in \mathfrak{so}_n.  
\label{eq:GCIId_muGam_def}
\end{equation}
We abbreviate $\mu^{\mathrm{I}}$ into $\mu$. 

We define $H^1_0(\mathrm{SO}_n, \mathfrak{so}_n)$ as the space of functions $\chi$: $\mathrm{SO}_n \to \mathfrak{so}_n$ such that each component of $\chi$ in an orthonormal basis of $\mathfrak{so}_n$ is a function of $H^1_0(\mathrm{SO}_n)$ (with similar notations for $L^2(\mathrm{SO}_n, \mathfrak{so}_n)$ and $H^1(\mathrm{SO}_n, \mathfrak{so}_n)$). Obviously, the definition does not depend on the choice of the orthonormal basis. Now, let $\chi \in H^1(\mathrm{SO}_n, \mathfrak{so}_n)$. Then $\nabla \chi(A)$ (which is defined almost everywhere) can be seen as an element of $\mathfrak{so}_n \otimes T_A$ by the relation 
$$ \nabla \chi (A) \cdot (X \otimes AY) = \nabla (\chi \cdot X) (A) \cdot (AY), \quad \forall X, \, Y \in \mathfrak{so}_n, $$
where we have used \eqref{eq:tangent_at_A} to express an element of $T_A$ as $AY$ for $Y \in \mathfrak{so}_n$ and where we define the inner product on $\mathfrak{so}_n \otimes T_A$ by 
$$ \big( X \otimes AY \cdot X' \otimes AY' \big) = (X \cdot X') \, (A Y \cdot AY') = (X \cdot X') \, (Y \cdot Y'),  $$
for all $X, \, Y \, X', \, Y' \, \in \mathfrak{so}_n$. With this identification, if $(\Phi_i)_{i=1}^{\mathcal N}$ and $(\Psi_i)_{i=1}^{\mathcal N}$ (with ${\mathcal N} = \frac{n(n-1)}{2}$) are two orthonormal bases of $\mathfrak{so}_n$, then $(\Phi_i \otimes A \Psi_j)_{i,j = 1}^{\mathcal N}$ is an orthonormal basis of $\mathfrak{so}_n \otimes T_A$ and we can write 
\begin{equation} 
\nabla \chi (A) = \sum_{i,j=1}^{\mathcal N} \big( \nabla (\chi \cdot \Phi_i)(A) \cdot (A \Psi_j) \big) \, \Phi_i \otimes A \Psi_j. 
\label{eq:nabla_chi_express}
\end{equation}
Consequently, if $\mu \in H^1(\mathrm{SO}_n, \mathfrak{so}_n)$ is another function, we have, thanks to Parseval's formula and \eqref{eq:rot_operator_rho}:
\begin{eqnarray} 
\nabla \mu(A) \cdot \nabla \chi(A) &=& \sum_{i,j=1}^{\mathcal N} \big( \nabla (\mu \cdot \Phi_i) (A) \cdot A \Psi_j \big) \, \big( \nabla (\chi \cdot \Phi_i) (A) \cdot A \Psi_j \big) \nonumber \\
&=& \sum_{i,j=1}^{\mathcal N} \Big( \big( \varrho(\Psi_j) (\mu \cdot \Phi_i) \big) (A) \Big) \, \Big( \big( \varrho(\Psi_j) (\chi \cdot \Phi_i) \big) (A) \Big). \label{eq:GCIId_nabmu_nabchi_def}
\end{eqnarray}
In general, we will use \eqref{eq:GCIId_nabmu_nabchi_def} with identical bases $(\Phi_i)_{i=1}^{\mathcal N} = (\Psi_i)_{i=1}^{\mathcal N}$, but this is not necessary. The construction itself shows that these formulae are independent of the choice of the orthonormal bases of $\mathfrak{so}_n$.

Now, we have the following properties: 

\begin{proposition}[Properties of $\mu$]~

\noindent
(i) The function $\mu$ is the unique variational solution in $H^1_0(\mathrm{SO}_n, \mathfrak{so}_n)$ of the equation 
\begin{equation} 
M^{-1} \nabla \cdot \big( M \, \nabla \mu \big) (A) = \frac{A - A^T}{2}, \quad \forall A \in \mathrm{SO}_n, 
\label{eq:GCIId_equa_mu}
\end{equation}
where the differential operator at the left-hand side is applied componentwise. The variational formulation of \eqref{eq:GCIId_equa_mu} is given by 
\begin{equation}
\left\{ \begin{array}{l}
\displaystyle \mu \in H^1_0(\mathrm{SO}_n, \mathfrak{so}_n),  \\
\displaystyle \int_{\mathrm{SO}_n} \nabla \mu \cdot \nabla \chi \, M \, dA = - \int_{\mathrm{SO}_n}  
\frac{\displaystyle A-A^T}{\displaystyle 2} \cdot \chi \, M \, dA, \quad  \forall \chi \in H^1_0(\mathrm{SO}_n, \mathfrak{so}_n), 
\end{array} \right. 
\label{eq:GCIId_varform_mu}
\end{equation}
with the interpretation \eqref{eq:GCIId_nabmu_nabchi_def} of the left-hand side of \eqref{eq:GCIId_varform_mu}. 

\smallskip
\noindent
(ii) We have (conjugation invariance):
\begin{equation} 
\mu(gAg^T) = g \mu(A) g^T, \quad \forall A, \, g \in \mathrm{SO}_n.  
\label{eq:GCIId_conjug_invar_mu_0}
\end{equation}

\smallskip
\noindent
(iii) We have (translation invariance):
\begin{equation}
\mu^\Gamma(A) = \mu(\Gamma^T A), \quad \forall \Gamma, \, A \in \mathrm{SO}_n. 
\label{eq:GCIId_link_mu_mugam}
\end{equation}
\label{prop:mu_properties}
\end{proposition}

\noindent
\textbf{Proof.} (i) Since $X$ is antisymmetric, we have $A \cdot X = \frac{A-A^T}{2} \cdot X$. Hence, \eqref{eq:GCI_eq} (with $\Gamma = \mathrm{I}$) can be written
$$ \Big( M^{-1} \nabla \cdot \big( M \nabla \mu \big) - \frac{A-A^T}{2} \Big) \cdot X = 0 , \quad \forall X \in \mathfrak{so}_n. $$
Since the matrix to the left of the inner product is antisymmetric and the identity is true for all antisymmetric matrices $X$, we find \eqref{eq:GCIId_equa_mu}. We can easily reproduce the same arguments on the variational formulation \eqref{eq:GCI_var_form} (with $\Gamma = \mathrm{I}$), which leads to the variational formulation \eqref{eq:GCIId_varform_mu} for $\mu$. 

\smallskip
\noindent
(ii) \eqref{eq:GCIId_conjug_invar} reads 
$$ \mu(gAg^T) \cdot X = \mu(A) \cdot (g^T X g) = \big( g \mu(A) g^T \big) \cdot X, $$
hence \eqref{eq:GCIId_conjug_invar_mu_0}. 

\smallskip
\noindent
(iii) \eqref{eq:GCI_transl} reads 
$$ \mu(A) \cdot X = \mu^\Gamma(\Gamma A) \cdot X, $$
hence \eqref{eq:GCIId_link_mu_mugam}. \endproof

The generic form of a function satisfying \eqref{eq:GCIId_conjug_invar_mu_0} is given in the next proposition.

\begin{proposition} (i) Let $\chi$: $\mathrm{SO}_n \to \mathfrak{so}_n$ be a smooth map satisfying 
\begin{equation} 
\chi(gAg^T) = g \chi(A) g^T, \quad \forall A, \, g \in \mathrm{SO}_n.  
\label{eq:GCIId_conjug_invar_mu}
\end{equation}
Define $p$ such that $n=2p$ or $n=2p+1$. 

\smallskip
\noindent
(i) There exists a $p$-tuple $\tau = (\tau_i)_{i=1}^p$ of periodic functions $\tau_i$: ${\mathcal T} \to {\mathbb R}$ such that 
\begin{equation} 
\chi(A_\Theta) = \sum_{k=1}^p \tau_k(\Theta) F_{2k-1 \, 2k}, \quad \forall \Theta \in {\mathcal T}. 
\label{eq:GCIId:mu_generic}
\end{equation}
Furthermore, $\tau$ commutes with the Weyl group, i.e. 
\begin{equation}
\tau \circ W = W \circ \tau, \quad \forall W \in \mathfrak{W}. 
\label{eq:GCIId_tau_com_Weyl}
\end{equation}

\smallskip
\noindent
(ii) $\chi$ has expression
\begin{equation} 
\chi(A) = \sum_{k=1}^p \tau_k(\Theta) \, gF_{2k-1 \, 2k}g^T, 
\label{eq:GCIId:mu_generic_gene}
\end{equation}
where $g \in \mathrm{SO}_n$ and $\Theta \in {\mathcal T}$ are such that $A = g A_\Theta g^T$. 
\label{prop:GCIId:mu_generic}
\end{proposition}

\noindent
\textbf{Proof.} (i) Let $A$, $g \in {\mathbb T}$. Then, since ${\mathbb T}$ is abelian, $gAg^T=A$ and \eqref{eq:GCIId_conjug_invar_mu} reduces to 
\begin{equation}
g \chi(A) g^T = \chi(A), \quad \forall A, \, g \in {\mathbb T}.  
\label{eq:GCIId:mu_generic_prf1}
\end{equation}
Fixing $A$, letting $g = e^{tX}$ with $X \in \mathfrak{h}$ and differentiating \eqref{eq:GCIId:mu_generic_prf1} with respect to $t$, we get
$$ [X, \chi(A)] = 0, \quad \forall X \in \mathfrak{h}. $$
This means that $\mathfrak{h} \oplus (\chi(A) \,  {\mathbb R})$ is an abelian subalgebra of $\mathfrak{so}_n$. But $\mathfrak{h}$ is a maximal subalgebra of $\mathfrak{so}_n$. So, it implies that $\chi(A) \in \mathfrak{h}$. Hence, there exist functions $\tau_i$: ${\mathcal T} \to {\mathbb R}$, $\Theta \mapsto \tau_i(\Theta)$ for $i=1, \ldots, p$ such that \eqref{eq:GCIId:mu_generic} holds. 

Now, we show \eqref{eq:GCIId_tau_com_Weyl} on a set of generating elements of $\mathfrak{W}$. For this, we use \eqref{eq:GCIId_conjug_invar_mu} with such generating elements $g$, reminding that $\mathfrak{W} \approx N({\mathbb T})/{\mathbb T}$ as described in Section~\ref{subsec:rotmore}. We distinguish the two parity cases of $n$. 

\smallskip
\noindent
\textbf{Case $n=2p$ even.} In this case, the Weyl group is generated by $(C_{ij})_{1 \leq i < j \leq p}$ and $(D_{ij})_{1 \leq i < j \leq p}$ (see Section~\ref{subsec:rotmore}). First, 
we take $g = C_{ij}$ as defined in Section~\ref{subsec:rotmore}. Then, conjugation by $C_{ij}$ exchanges $F_{2i-1 \, 2i}$ and $F_{2j-1 \, 2j}$ and leaves $F_{2k-1 \, 2k}$ for $k \not = i, \, j$ invariant. On the other hand conjugation by $C_{ij}$ changes $A_\Theta$ into $A_{C_{ij} \Theta}$ where we recall that, by abuse of notation, we also denote by $C_{ij}$ the transformation of $\Theta$ generated by conjugation by $C_{ij}$. Thus, from~\eqref{eq:GCIId_conjug_invar_mu} and~\eqref{eq:GCIId:mu_generic} we get 
$$ \chi(A_{C_{ij} \Theta}) = \sum_{k \not = i, \, j} \tau_k (\Theta)  F_{2k-1 \, 2k} + \tau_i(\Theta) F_{2j-1 \, 2j} +  \tau_j(\Theta) F_{2i-1 \, 2i}. $$
On the other hand, direct application of \eqref{eq:GCIId:mu_generic} leads to 
$$ \chi(A_{C_{ij} \Theta}) = \sum_{k=1}^p \tau_k (C_{ij} \Theta) F_{2k-1 \, 2k}. $$
Equating these two expressions leads to 
\begin{eqnarray*}
\tau_k \big( C_{ij} (\Theta)\big) &=& \tau_k (\Theta), \quad \forall k \not = i, \, j. \\
\tau_i \big( C_{ij} (\Theta)\big) &=& \tau_j (\Theta),  \qquad\tau_j \big( C_{ij} (\Theta)\big) = \tau_i (\Theta), 
\end{eqnarray*}
Hence, we get 
\begin{equation} 
\tau \big( C_{ij}(\Theta) \big) =  C_{ij} \big( \tau(\Theta) \big). 
\label{eq:GCIId:mu_generic_prf3-1}
\end{equation}
Next, we take $g = D_{ij}$. Conjugation by $D_{ij}$ changes $F_{2i-1 \, 2i}$ into $-F_{2i-1 \, 2i}$ and $F_{2j-1 \, 2j}$ into $-F_{2j-1 \, 2j}$ and leaves $F_{2k-1 \, 2k}$ for $k \not = i, \, j$ invariant. Besides, conjugation by $D_{ij}$ changes $A_\Theta$ into $A_{D_{ij} \Theta}$. Thus, using the same reasoning as previously, we get  
\begin{eqnarray*}
\tau_k \big( D_{ij} \Theta \big) &=& \tau_k ( \Theta ), \quad \forall k \not = i, \, j, \\
\tau_i \big( D_{ij} \Theta \big) &=& - \tau_i ( \Theta ),  \qquad \tau_j \big( D_{ij} \Theta \big) = - \tau_j (  \Theta ).
\end{eqnarray*}
Hence, we find
$$ \tau \big( D_{ij}(\Theta) \big) =  D_{ij} \big( \tau(\Theta) \big). $$

\smallskip
\noindent
\textbf{Case $n=2p+1$ odd.} Here, the Weyl group is generated by $(C_{ij})_{1 \leq i < j \leq p}$ and $(D_i)_{1 =1}^p$. Taking $g=C_{ij}$ as in the previous case, we get \eqref{eq:GCIId:mu_generic_prf3-1} again. Now, taking $g=D_i$, conjugation by $D_i$ changes $F_{2i-1 \, 2i}$ into $-F_{2i-1 \, 2i}$ and leaves $F_{2k-1 \, 2k}$ for $k \not = i$ invariant. Besides, conjugation by $g$ changes $A_\Theta$ into $A_{D_i \Theta}$. Thus, we get  
\begin{eqnarray*}
\tau_k \big( D_i \Theta \big) &=& \tau_k (\Theta), \quad \forall k \not = i, \\
\tau_i \big( D_i \Theta \big) &=& - \tau_i ( \Theta ).  
\end{eqnarray*}
Thus, we finally get 
$$ \tau \big( D_i(\Theta) \big) =  D_i \big( \tau(\Theta) \big), $$
which ends the proof. 

\smallskip
\noindent
(ii) The fact that $\tau$ commutes with the Weyl group guarantees that formula \eqref{eq:GCIId:mu_generic_gene} is well defined, i.e. if $(g,\Theta)$ and $(g',\Theta')$ are two pairs in $\mathrm{SO}_n \times {\mathcal T}$ such that $A = g A_\Theta g^T = g' A_{\Theta'} {g'}^T$, then, the two expressions \eqref{eq:GCIId:mu_generic_gene} deduced from each pair are the same. Applying \eqref{eq:GCIId_conjug_invar_mu} to \eqref{eq:GCIId:mu_generic} shows that \eqref{eq:GCIId:mu_generic_gene} is necessary. It can be directly verified that \eqref{eq:GCIId:mu_generic_gene} satisfies~\eqref{eq:GCIId_conjug_invar_mu} showing that it is also sufficient. \endproof

The following corollary is a direct consequence of the previous discussion: 

\begin{corollary} 
Let $\mu$ be the solution of the variational formulation \eqref{eq:GCIId_equa_mu}. Then, there exists $\alpha = (\alpha_i)_{i=1}^p$: ${\mathcal T} \to {\mathbb R}^p$ such that
\begin{equation} 
\mu(A_\Theta) = \sum_{k=1}^p \alpha_k(\Theta) F_{2k-1 \, 2k}, \quad \forall \Theta \in {\mathcal T}, 
\label{eq:GCIId:mu_formula}
\end{equation}
and $\alpha$ commutes with the Weyl group, 
$$ \alpha \circ W = W \circ \alpha, \quad \forall W \in \mathfrak{W}. $$
\label{cor:GCIId:mu_formula}
\end{corollary}

\begin{remark}
The context of \cite{degond2021body} corresponds to $\mu(A) = \frac{A-A^T}{2}$, i.e. $\alpha_k(\Theta) = - \sin \theta_k$, see Remark \ref{rem:GCIId_compar_SOn}. 
\label{rem:GCIId:mu_case_BGK}
\end{remark}

Now, we wish to derive a system of PDEs for $\alpha$. There are two ways to achieve this aim. 
\begin{itemize}
\item[-] The first one consists of deriving the system in strong form by directly computing the differential operators involved in \eqref{eq:GCIId_equa_mu} at a point $A_\Theta$ of the maximal torus ${\mathbb T}$. This method applies the strategy exposed in \cite[Section 8.3]{faraud2008Analysis} and used in \cite{degond2023radial} to derive expressions of the radial Laplacian on rotation groups. However, this method does not give information on the well-posedness of the resulting system. We develop this method in Appendix \ref{appsec_direct_strong_form} for the interested reader and as a cross validation of the following results. 
\item[-] The second method, which is developed below, consists of deriving the system in weak form, using the variational formulation \eqref{eq:GCIId_varform_mu}. We will show that we can restrict the space of test functions $\chi$ to those satisfying the invariance relation~\eqref{eq:GCIId_conjug_invar_mu}. This will allow us to derive a variational formulation for the system satisfied by $\alpha$ which will lead us to its well-posedness and eventually to the strong form of the equations. 
\end{itemize}


\subsection{Reduction to a conjugation-invariant variational formulation}
\label{subsec_GCIId_conjug_invariant_varform}

We first define the following spaces: 
\begin{eqnarray*}
L^2_{\mathrm{inv}}(\mathrm{SO}_n, \mathfrak{so}_n) &=& \big\{ \chi \in L^2(\mathrm{SO}_n, \mathfrak{so}_n) \text{ such that } \\
&& \hspace{2cm} \chi \text{ satisfies \eqref{eq:GCIId_conjug_invar_mu} a.e. } A \in \mathrm{SO}_n, \, \, \forall g \in \mathrm{SO}_n \big\},  \\
H^1_{\mathrm{inv}}(\mathrm{SO}_n, \mathfrak{so}_n) &=& \big\{ \chi \in H^1(\mathrm{SO}_n, \mathfrak{so}_n) \text{ such that } \\
&& \hspace{2cm} \chi \text{ satisfies \eqref{eq:GCIId_conjug_invar_mu} a.e. } A \in \mathrm{SO}_n \, \, \forall g \in \mathrm{SO}_n  \big\},  
\end{eqnarray*}
where $\mathrm{a.e.}$ stands for ``for almost every''. Concerning these spaces, we have the

\begin{proposition} (i) $L^2_{\mathrm{inv}}(\mathrm{SO}_n, \mathfrak{so}_n)$ and $H^1_{\mathrm{inv}}(\mathrm{SO}_n, \mathfrak{so}_n)$ are closed subspaces of $L^2(\mathrm{SO}_n, \mathfrak{so}_n)$ and $H^1(\mathrm{SO}_n, \mathfrak{so}_n)$ respectively and consequently, are Hilbert spaces.  

\smallskip
\noindent
(ii) We have $H^1_{\mathrm{inv}}(\mathrm{SO}_n, \mathfrak{so}_n) \subset H^1_0(\mathrm{SO}_n, \mathfrak{so}_n)$. 
\label{prop:GCIId_L2inv_H1inv_prop}
\end{proposition}

\noindent
\textbf{Proof.} 
(i) Let $g \in \mathrm{SO}_n$. We introduce the conjugation map $\Xi_g$ mapping any function $\chi$: $\mathrm{SO}_n \to \mathfrak{so}_n$ to another function $\Xi_g \chi$: $\mathrm{SO}_n \to \mathfrak{so}_n$ such that 
$$ \Xi_g \chi(A) = g^T \chi(gAg^T) g, \quad \forall A \in \mathrm{SO}_n.$$ 
We prove that $\Xi_g$ is an isometry of $L^2(\mathrm{SO}_n, \mathfrak{so}_n)$ and of $H^1(\mathrm{SO}_n, \mathfrak{so}_n)$ for any $g \in \mathrm{SO}_n$. The result follows as 
$$ L^2_{\mathrm{inv}}(\mathrm{SO}_n, \mathfrak{so}_n) = \bigcap_{g \in \mathrm{SO}_n} \mathrm{ker}_{L^2} (\Xi_g - \mathrm{I}), \quad H^1_{\mathrm{inv}}(\mathrm{SO}_n, \mathfrak{so}_n) = \bigcap_{g \in \mathrm{SO}_n} \mathrm{ker}_{H^1} (\Xi_g - \mathrm{I}).  $$

Thanks to the cyclicity of the trace and the translation invariance of the Haar measure, we have 
\begin{eqnarray*}
\int_{\mathrm{SO}_n} |\Xi_g \chi(A)|^2 \, dA &=& \int_{\mathrm{SO}_n} \big( g^T \chi(gAg^T) g
 \big) \cdot \big( g^T \chi(gAg^T) g \big) \, dA \\
&=& \int_{\mathrm{SO}_n} |\chi(gAg^T)|^2  \, dA = \int_{\mathrm{SO}_n} |\chi(A)|^2  \, dA,  
\end{eqnarray*}
where, for $X \in \mathfrak{so}_n$, we have denoted by $|X| = (X \cdot X)^{1/2}$ the euclidean norm on $\mathfrak{so}_n$. This shows that $\Xi_g$ is an isometry of $L^2(\mathrm{SO}_n, \mathfrak{so}_n)$. 

Now, with \eqref{eq:GCIId_nabmu_nabchi_def} and \eqref{eq:GCIId_conjug_invar_mu}, we have
\begin{eqnarray*}
|\nabla (\Xi_g \chi) (A)|^2 &=& \frac{1}{2} \sum_{i,j=1}^n \nabla (\Xi_g \chi)_{ij} \cdot \nabla (\Xi_g \chi)_{ij} \\
&=& \frac{1}{2} \sum_{i,j,k,\ell,k',\ell'=1}^n g_{ki} \, g_{\ell j} \, g_{k' i} \, g_{\ell' j} \, \nabla (\chi_{k \ell} \circ \xi_g)(A)  \cdot \nabla (\chi_{k' \ell'} \circ \xi_g)(A) \\
&=& \frac{1}{2} \sum_{k,\ell=1}^n |\nabla \chi_{k \ell} (gAg^T)|^2 = |\nabla \chi (gAg^T)|^2, 
\end{eqnarray*}
where we used that $\sum_{i=1}^n g_{ki} g_{k'i} = \delta_{k k'}$ and similarly for the sum over $j$, as well as \eqref{eq:GCI_grad_compos_conjug}. Now, using the translation invariance of the Haar measure, we get 
$$ \int_{\mathrm{SO}_n} |\nabla (\Xi_g \chi) (A)|^2 \, dA = \int_{\mathrm{SO}_n} |\nabla \chi (A)|^2 \, dA, $$
which shows that $\Xi_g$ is an isometry of $H^1(\mathrm{SO}_n, \mathfrak{so}_n)$. 

\smallskip
\noindent
(ii) We use the fact that, for any integrable function $f$: $\mathrm{SO}_n \to \mathfrak{so}_n$, we have
\begin{equation} 
\int_{\mathrm{SO}_n} f(A) \, dA = \int_{\mathrm{SO}_n} \Big( \int_{\mathrm{SO}_n} f(gAg^T) \, dg \Big) dA. 
\label{eq:GCIId_varform_invar_prf1}
\end{equation}
Let $\chi \in H^1_{\mathrm{inv}}(\mathrm{SO}_n, \mathfrak{so}_n)$. With \eqref{eq:GCIId_conjug_invar_mu}, we have 
$$ \int_{\mathrm{SO}_n} \chi(gAg^T) \, dg = \int_{\mathrm{SO}_n} g \chi(A)  g^T\, dg. $$
We define the linear map $T$: $\mathfrak{so}_n \to {\mathbb R}$, $ X \mapsto \int_{\mathrm{SO}_n} (g \chi(A) g^T) \, dg \cdot X $. By translation invariance of the Haar measure, we have $T(hXh^T) = T(X)$, for all $h \in \mathrm{SO}_n$. Thus, $T$ intertwines the representation $\mathfrak{so}_n$ of $\mathrm{SO}_n$ (i.e. the adjoint representation $\mathrm{Ad}$) and its trivial representation~${\mathbb R}$. $\mathrm{Ad}$ is irreducible except for dimension $n=4$ where it decomposes into two irreducible representations. Neither of these representations is isomorphic to the trivial representation. Then, by Schur's Lemma, $T=0$. This shows that $\int_{\mathrm{SO}_n} \chi(gAg^T) \, dg =0$ and by application of \eqref{eq:GCIId_varform_invar_prf1}, that $\int_{\mathrm{SO}_n} \chi(A) \, dA =0$.
\endproof

We now show that the space $H^1_{\mathrm{inv}}(\mathrm{SO}_n, \mathfrak{so}_n)$ may replace $H^1_0(\mathrm{SO}_n, \mathfrak{so}_n)$ in the variational formulation giving $\mu$. More specifically, we have the 

\begin{proposition} 
(i) $\mu$ is the unique solution of the variational formulation \eqref{eq:GCIId_varform_mu} if and only if it is the unique solution of the variational formulation
\begin{equation}
\left\{ \begin{array}{l}
\displaystyle \mu \in H^1_{\mathrm{inv}}(\mathrm{SO}_n, \mathfrak{so}_n),  \\
\displaystyle \int_{\mathrm{SO}_n} \nabla \mu \cdot \nabla \chi \, M \, dA = - \int_{\mathrm{SO}_n}  
\frac{A-A^T}{2} \cdot \chi \, M \, dA, \quad  \forall \chi \in  H^1_{\mathrm{inv}}(\mathrm{SO}_n, \mathfrak{so}_n). 
\end{array} \right. 
\label{eq:GCIId_varform_mu_invar}
\end{equation}

\smallskip
\noindent
(ii) The variational formulation \eqref{eq:GCIId_varform_mu_invar} can be equivalently written 
\begin{equation}
\left\{ \begin{array}{l}
\displaystyle 
\mu \in H^1_{\mathrm{inv}}(\mathrm{SO}_n, \mathfrak{so}_n),  \\
\displaystyle 
\int_{{\mathcal T}} (\nabla \mu \cdot \nabla \chi) (A_\Theta) \, M(A_\Theta) \, u_n(\Theta) \, d\Theta  \\
\displaystyle 
\hspace{1cm} = - \int_{{\mathcal T}} \frac{A_\Theta-A_\Theta^T}{2} \cdot \chi (A_\Theta) \, M(A_\Theta) \, u_n(\Theta) \, d\Theta, \quad \forall \chi \in  H^1_{\mathrm{inv}}(\mathrm{SO}_n, \mathfrak{so}_n). 
\end{array} \right. 
\label{eq:GCIId_varform_mu_invar_torus}
\end{equation}
\label{prop:GCIId_varform_invar}
\end{proposition}

\noindent
\textbf{Proof.} (i) We remark that \eqref{eq:GCIId_varform_mu_invar} has a unique solution. Indeed, \eqref{eq:GCI_exist_var_form_prf1} can be extended componentwise to all $\psi \in H^1_0(\mathrm{SO}_n, \mathfrak{so}_n)$, and in particular, to all $\psi \in H^1_{\mathrm{inv}}(\mathrm{SO}_n, \mathfrak{so}_n)$. Thus, the bilinear form at the left-hand side of \eqref{eq:GCIId_varform_mu_invar} is coercive on $H^1_{\mathrm{inv}}(\mathrm{SO}_n, \mathfrak{so}_n)$. Existence and uniqueness follow from Lax-Milgram's theorem. 

\smallskip
Let $\mu$ be the solution of \eqref{eq:GCIId_varform_mu}. Since $\mu$ satisfies \eqref{eq:GCIId_conjug_invar_mu_0}, it belongs to $H^1_{\mathrm{inv}}(\mathrm{SO}_n, \mathfrak{so}_n)$. Furthermore, restricting \eqref{eq:GCIId_varform_mu} to test functions in $H^1_{\mathrm{inv}}(\mathrm{SO}_n, \mathfrak{so}_n)$, it satisfies \eqref{eq:GCIId_varform_mu_invar}. 

\smallskip
Conversely, suppose that $\mu$ is the unique solution of \eqref{eq:GCIId_varform_mu_invar}. We will use \eqref{eq:GCIId_varform_invar_prf1}. Let $\chi \in H^1_0(\mathrm{SO}_n, \mathfrak{so}_n)$. Thanks to \eqref{eq:GCI_grad_compos_conjug} and to the fact that $M$ is a class function, we have
\begin{eqnarray*}
&&\hspace{-1cm}
I(A) =: \int_{\mathrm{SO}_n} M(gAg^T) \, \nabla \mu (gAg^T) \cdot \nabla \chi(gAg^T) \, dg \\
&&\hspace{0cm}
= \frac{1}{2} \sum_{i,j=1}^n \int_{\mathrm{SO}_n} M(A) \, \nabla \mu_{ij} (gAg^T) \cdot \nabla \chi_{ij}(gAg^T) \, dg \\ 
&&\hspace{0cm}
= \frac{1}{2} \sum_{i,j=1}^n \int_{\mathrm{SO}_n} M(A) \, \nabla (\mu_{ij} \circ \xi_g) (A)  \cdot \nabla (\chi_{ij} \circ \xi_g)(A) \, dg .  
\end{eqnarray*}
Now, by \eqref{eq:GCIId_conjug_invar_mu_0}, we have $ (\mu_{ij} \circ \xi_g) (A) = \sum_{k,\ell = 1}^n g_{ik} \, g_{j \ell} \, \mu_{k \ell}(A)$. We deduce that 
\begin{eqnarray*}
I(A) &=& \frac{1}{2} \sum_{k,\ell = 1}^n M(A) \, \nabla \mu_{k \ell}(A) \cdot \nabla \Big( \sum_{i,j=1}^n  \int_{\mathrm{SO}_n} g_{ik} \, g_{j \ell} \,   (\chi_{ij} \circ \xi_g) \, dg \Big)(A)\\
&=& \frac{1}{2} \sum_{k,\ell = 1}^n M(A) \, \nabla \mu_{k \ell}(A) \cdot \nabla \bar \chi_{k \ell}(A) = M(A) \nabla \mu(A) \cdot \nabla \bar \chi(A), 
\end{eqnarray*}
where $\bar \chi$ is defined by
\begin{equation} 
\bar \chi(A) = \int_{\mathrm{SO}_n} g^T \chi(gAg^T) g \, dg, \quad \forall A \in \mathrm{SO}_n. 
\label{eq:GCIId_barchi_def}
\end{equation}
Similarly, we have 
\begin{eqnarray*}
&&\hspace{-1cm}
J(A) =: \int_{\mathrm{SO}_n} M(gAg^T) \, \Big( g \frac{A-A^T}{2} g^T \Big) \cdot \chi (gAg^T) \,  dg \\
&&\hspace{0cm}
= M(A) \,  \frac{A-A^T}{2} \cdot \Big( \int_{\mathrm{SO}_n} g^T \chi (gAg^T) g \, dg \Big) =  M(A) \,  \frac{A-A^T}{2} \cdot \bar \chi (A). 
\end{eqnarray*}
Applying \eqref{eq:GCIId_varform_invar_prf1} we get that 
\begin{eqnarray}
&&\hspace{-1cm}
\int_{\mathrm{SO}_n} \Big( \nabla \mu \cdot \nabla \chi + \frac{A-A^T}{2} \cdot \chi \Big) M \, dA  = \int_{\mathrm{SO}_n} (I(A) + J(A)) \, dA \nonumber \\
&&\hspace{3cm}
= \int_{\mathrm{SO}_n}  \Big( \nabla \mu \cdot \nabla \bar \chi + \frac{A-A^T}{2} \cdot \bar \chi \Big) M \,dA. \label{eq:GCIId_varform_invar_prf3}
\end{eqnarray}
Now, we temporarily assume that 
\begin{equation}
\bar \chi \in H^1_{\mathrm{inv}}(\mathrm{SO}_n, \mathfrak{so}_n), 
\label{eq:GCIId_varform_invar_prf2}
\end{equation}
Then, because $\mu$ is the unique solution of \eqref{eq:GCIId_varform_mu_invar}, the right-hand side of \eqref{eq:GCIId_varform_invar_prf3} is equal to zero. This shows that $\mu$ is the unique solution of \eqref{eq:GCIId_varform_mu}. 

Now, we show \eqref{eq:GCIId_varform_invar_prf2}. That $\bar \chi$ satisfies the invariance relation \eqref{eq:GCIId_conjug_invar_mu} is obvious. We now show that $ \| \bar \chi \|_{H^1} \leq  \| \chi \|_{H^1}$. 

We first show that $ \| \bar \chi \|_{L^2} \leq  \| \chi \|_{L^2}$. By Cauchy-Schwarz inequality,  Fubini's theorem and the translation invariance of the Haar measure, we have 
\begin{eqnarray*}
\int_{\mathrm{SO}_n} |\bar \chi(A)|^2 \, dA &=& \int_{\mathrm{SO}_n} \Big| \int_{\mathrm{SO}_n} g^T \chi(gAg^T) g \, dg \Big|^2 \, dA \leq \int_{\mathrm{SO}_n} \Big( \int_{\mathrm{SO}_n} | g^T \chi(gAg^T) g |^2 \, dg \Big) dA \\
&=&  \int_{\mathrm{SO}_n} \Big( \int_{\mathrm{SO}_n} | \chi(gAg^T) |^2 \, dA \Big) dg =   \int_{\mathrm{SO}_n} | \chi(A) |^2 \, dA . 
\end{eqnarray*}

We now show that $ \| \nabla \bar \chi \|_{L^2} \leq  \| \nabla \chi \|_{L^2}$. Differentiating \eqref{eq:GCIId_barchi_def} with respect to $A$ and using~\eqref{eq:GCI_deriv_compos_conjug}, we get 
\begin{eqnarray*}
&&\hspace{-1cm}
|\nabla \bar \chi(A)|^2  = \frac{1}{2} \sum_{k, \ell = 1}^n |\nabla \bar \chi_{k \ell}(A)|^2
= \frac{1}{2} \sum_{k,\ell=1}^n  \Big| \int_{\mathrm{SO}_n} \Big(\sum_{i,j=1}^n g_{ik} \, g_{j \ell} \, g^T \nabla  \chi_{ij}(gAg^T)  g \Big) \, dg \Big|^2. 
\end{eqnarray*}
Applying Cauchy-Schwarz formula, this leads to 
\begin{eqnarray*}
&&\hspace{-1cm}
|\nabla \bar \chi(A)|^2  \leq \frac{1}{2} \sum_{k,\ell=1}^n  \int_{\mathrm{SO}_n} \Big| \sum_{i,j=1}^n g_{ik} \, g_{j \ell} \, g^T \nabla  \chi_{ij}(gAg^T)  g \Big|^2 \, dg \\
&&\hspace{-0.5cm}
= \frac{1}{2} \sum_{i,j,i',j',k,\ell=1}^n \int_{\mathrm{SO}_n}  g_{ik} \, g_{j \ell} \, g_{i'k} \, g_{j' \ell} \, \big( g^T \nabla  \chi_{i' j'}(gAg^T) g \big) \cdot \big( g^T \nabla \chi_{ij} (gAg^T) g \big) \, dg  \\
&&\hspace{-0.5cm}
= \frac{1}{2} \sum_{i,j=1}^n \int_{\mathrm{SO}_n} \nabla  \chi_{i j}(gAg^T) \cdot \nabla \chi_{ij} (gAg^T) \, dg, 
\end{eqnarray*}
where we have applied that $\sum_{k=1}^n g_{ik} g_{i'k} = \delta_{i i'}$ and similarly for the sum over $\ell$. Thus, 
\begin{eqnarray*}
&&\hspace{-1cm}
 \int_{\mathrm{SO}_n} |\nabla \bar \chi(A)|^2 \, dA  \leq \frac{1}{2} \sum_{i,j=1}^n \int_{\mathrm{SO}_n} \Big( \int_{\mathrm{SO}_n} \nabla  \chi_{i j}(gAg^T) \cdot \nabla \chi_{ij} (gAg^T) \, dA \Big) \, dg \\
&&\hspace{-0.5cm}
=  \frac{1}{2} \sum_{i,j=1}^n \int_{\mathrm{SO}_n} \nabla  \chi_{i j}(A) \cdot \nabla \chi_{ij} (A) \, dA  =  \int_{\mathrm{SO}_n} | \nabla \chi (A)|^2 \, dA, 
\end{eqnarray*}
which shows the result and ends the proof of (i)

\smallskip
\noindent
(ii) Let $\chi \in  H^1_{\mathrm{inv}}(\mathrm{SO}_n, \mathfrak{so}_n)$. Then, the functions $A \mapsto M(A) \nabla \mu(A) \cdot \nabla \chi(A)$ and $A \mapsto M \, \frac{A-A^T}{2} \cdot \chi(A)$ are class functions (the proof relies on similar computations as those just made above and is omitted). Then, \eqref{eq:GCIId_varform_mu_invar_torus} is simply a consequence of Weyl's integration formula \eqref{eq:WIF}. This ends the proof. \endproof

\subsection{Derivation and well-posedness of System \eqref{eq:GCIId_strongform_alpha}
 for $\alpha$}
\label{subsec_GCIId_deriv_wellposed_alpha}

Now, we investigate how the condition $\chi \in H^1_{\mathrm{inv}}(\mathrm{SO}_n, \mathfrak{so}_n)$ translates onto $\tau$ and define the following spaces: 
\begin{itemize}
\item[$\bullet$] $C_{\mathrm{per}}^{\infty,\mathfrak{W}}({\mathcal T}, {\mathbb R}^p)$ is the set of periodic, $C^\infty$ functions ${\mathcal T} \to {\mathbb R}^p$ which commute with the Weyl group, i.e. such that \eqref{eq:GCIId_tau_com_Weyl} is satisfied in ${\mathcal T}$, for all $W \in \mathfrak{W}$, 
\item[$\bullet$] ${\mathcal H}$ is the closure of $C_{\mathrm{per}}^{\infty,\mathfrak{W}}({\mathcal T}, {\mathbb R}^p)$ for the norm 
$$ \| \tau \|_{\mathcal H}^2 =: \sum_{i=1}^p \int_{{\mathcal T}} |\tau_i(\Theta)|^2 \, u_n(\Theta) \, d \Theta, $$
\item[$\bullet$] ${\mathcal V}$ is the closure of $C_{\mathrm{per}}^{\infty,\mathfrak{W}}({\mathcal T}, {\mathbb R}^p)$ for the norm 
\begin{eqnarray*}
&&\hspace{-1cm}
 \| \tau \|_{\mathcal V}^2 =: \| \tau \|_{\mathcal H}^2 + \int_{{\mathcal T}} \Big\{ \sum_{i,j=1}^p \Big| \frac{\partial \tau_i}{\partial \theta_j}(\Theta) \Big|^2 + \sum_{1 \leq i < j \leq p} \Big( \frac{|(\tau_i - \tau_j)(\Theta)|^2}{1 - \cos(\theta_i - \theta_j)} \\
&&\hspace{3cm}
+ \frac{|(\tau_i + \tau_j)(\Theta)|^2}{1 - \cos(\theta_i + \theta_j)} \Big) + \epsilon_n \sum_{i=1}^p \frac{|\tau_i(\Theta)|^2}{1 - \cos \theta_i} \Big\} \, u_n(\Theta) \, d \Theta. 
\end{eqnarray*}
\end{itemize}

\begin{remark}
We note that for $\tau \in C_{\mathrm{per}}^{\infty,\mathfrak{W}}({\mathcal T}, {\mathbb R}^p)$, we have $\| \tau \|_{\mathcal V} < \infty$, which shows that the definition of ${\mathcal V}$ makes sense. Indeed Since $\tau$ commutes with $\mathfrak{W}$, we have 
\begin{eqnarray*}
&&\hspace{-1cm} 
(\tau_i - \tau_j) (\ldots , \theta_i , \ldots, \theta_j \ldots) = \tau_i (\ldots , \theta_i , \ldots, \theta_j \ldots) - \tau_i (\ldots , \theta_j , \ldots, \theta_i  \ldots) \\
&&\hspace{-1cm} 
= (\theta_j - \theta_i) \Big( \big( \frac{\partial }{\partial \theta_j} - \frac{\partial }{\partial \theta_i} \big) \tau_i \Big) (\ldots , \theta_i , \ldots, \theta_i \ldots) + {\mathcal O} \big((\theta_j - \theta_i)^2\big) =  {\mathcal O} \big(|\theta_j - \theta_i| \big), 
\end{eqnarray*}
as $\theta_j - \theta_i \to 0$, 
while 
$$ 1 - \cos (\theta_i - \theta_j) = 2 \sin^2 \Big( \frac{\theta_i - \theta_j}{2} \Big) = 
\frac{1}{2} |\theta_j - \theta_i|^2 + {\mathcal O} \big(|\theta_j - \theta_i|^4 \big). $$
Thus, 
$$ \frac{|(\tau_i - \tau_j)(\Theta)|^2}{ 1 - \cos (\theta_i - \theta_j)} < \infty. $$
The same computation holds when $\theta_j + \theta_i \to 0$ and, in the odd-dimensional case, when $\theta_i \to 0$. 
\label{rem:GCIId_normV_finite}
\end{remark}

\begin{proposition}
$\chi \in  H^1_{\mathrm{inv}}(\mathrm{SO}_n, \mathfrak{so}_n)$ if and only if $\chi$ is given by \eqref{eq:GCIId:mu_generic_gene} with $\tau \in {\mathcal V}$. 
\label{prop:GCIId_regularity_tau}
\end{proposition}

\noindent
\textbf{Proof.} Let $\chi \in C^{\infty}_\mathrm{inv}(\mathrm{SO}_n, \mathfrak{so}_n)$ be an element of the space of smooth functions satisfying the invariance relation \eqref{eq:GCIId_conjug_invar_mu}. Then, it is clear that $\tau$ associated with $\chi$ through \eqref{eq:GCIId:mu_generic} belongs to $C_{\mathrm{per}}^{\infty,\mathfrak{W}}({\mathcal T}, {\mathbb R}^p)$. 
For such $\chi$ and $\tau$, we temporarily assume that 
\begin{equation}
\| \chi \|_{H^1}^2 = \frac{\gamma_n}{(2 \pi)^p} \| \tau \|^2_{\mathcal V}. 
\label{eq:GCIId_regularity_tau_prf0}
\end{equation}
We also assume that
\begin{equation}
C^{\infty}_\mathrm{inv}(\mathrm{SO}_n, \mathfrak{so}_n) \, \,  \text{ is dense in } \, \, H^1_{\mathrm{inv}}(\mathrm{SO}_n, \mathfrak{so}_n). 
\label{eq:GCIId_regularity_tau_prf-1}
\end{equation}

Let now $\chi$ be an element of $H^1_{\mathrm{inv}}(\mathrm{SO}_n, \mathfrak{so}_n)$ and $(\chi^q)_{q \in {\mathbb N}}$ be a sequence of elements of $C^{\infty}_\mathrm{inv}(\mathrm{SO}_n, \mathfrak{so}_n)$ which converges to $\chi$ in $H^1_{\mathrm{inv}}(\mathrm{SO}_n, \mathfrak{so}_n)$. Then, the associated sequence $(\tau^q)_{q \in {\mathbb N}}$ is such that $\tau^q \in C_{\mathrm{per}}^{\infty,\mathfrak{W}}({\mathcal T}, {\mathbb R}^p)$ and by \eqref{eq:GCIId_regularity_tau_prf0} $(\tau^q)_{q \in {\mathbb N}}$ is a Cauchy sequence in ${\mathcal V}$. Thus, there is an element $\tau \in {\mathcal V}$ such that $\tau^q \to \tau$ in ${\mathcal V}$. Now, up to the extraction of subsequences, we have $\chi^q \to \chi$, a.e. in $\mathrm{SO}_n$ and $\tau^q \to \tau$, a.e. in ${\mathcal T}$. Hence 
$$ \chi^q (A_\Theta) = \sum_{i=1}^p \tau^q_i (\Theta) F_{2i-1 \, 2i} \to \sum_{i=1}^p \tau_i(\Theta) F_{2i-1 \, 2i} = \chi(A_\Theta) \quad \mathrm{a.e.} \quad \Theta \in {\mathcal T}, $$ 
and, since $\chi$ satisfies \eqref{eq:GCIId_conjug_invar_mu}, $\chi(A)$ is given by \eqref{eq:GCIId:mu_generic_gene}. Therefore, if $\chi$ belongs to $H^1_{\mathrm{inv}}(\mathrm{SO}_n, \mathfrak{so}_n)$, there exists $\tau \in {\mathcal V}$ such that $\chi$ is given by \eqref{eq:GCIId:mu_generic_gene}. 

Therefore, Prop. \ref{prop:GCIId_regularity_tau} will be proved if we prove the converse property, namely, 
\begin{equation}
\mathrm{For \, any \,\,} \tau \in {\mathcal V}, \mathrm{\, \, then \,\, } \chi \mathrm{\, \, given \, by \, \eqref{eq:GCIId:mu_generic} \, belongs \, to \,\, H^1_{\mathrm{inv}}(\mathrm{SO}_n, \mathfrak{so}_n)},  
\label{eq:chi_tau_conv}
\end{equation}
as well as Properties \eqref{eq:GCIId_regularity_tau_prf0} and \eqref{eq:GCIId_regularity_tau_prf-1}. 

\bigskip
\noindent
\textbf{Proof of \eqref{eq:GCIId_regularity_tau_prf-1}.} Let $\chi \in H^1_{\mathrm{inv}}(\mathrm{SO}_n, \mathfrak{so}_n)$. Since $C^\infty(\mathrm{SO}_n, \mathfrak{so}_n)$ is dense in $H^1(\mathrm{SO}_n, \mathfrak{so}_n)$, there exists a sequence $(\chi^q)_{q \in {\mathbb N}}$ in $C^\infty(\mathrm{SO}_n, \mathfrak{so}_n)$ such that $\chi^q \to \chi$. Now, $\bar \chi^q$ obtained from $\chi^q$ through \eqref{eq:GCIId_barchi_def} belongs to $C^{\infty}_\mathrm{inv}(\mathrm{SO}_n, \mathfrak{so}_n)$. And because the map $\chi \mapsto \bar \chi$ is continuous on $H^1(\mathrm{SO}_n, \mathfrak{so}_n)$ (see proof of Prop. \ref{prop:GCIId_varform_invar}), we get $\bar \chi^q \to \bar \chi$ as $q \to \infty$. But since $\chi$ satisfies \eqref{eq:GCIId:mu_generic}, we have $\bar \chi = \chi$, which shows the requested density result. 

\bigskip
\noindent
\textbf{Proof of \eqref{eq:GCIId_regularity_tau_prf0}.} Let $\chi \in C^{\infty}_\mathrm{inv}(\mathrm{SO}_n, \mathfrak{so}_n)$. Since the function $A \mapsto |\chi(A)|^2$ is a class function, we have, thanks to Weyl's integration formula \eqref{eq:WIF}:
\begin{equation}
\| \chi \|_{L^2}^2 = \frac{\gamma_n}{(2 \pi)^p} \int_{{\mathcal T}} |\chi(A_\Theta)|^2 \, u_n(\Theta) \, d \Theta = \frac{\gamma_n}{(2 \pi)^p} \int_{{\mathcal T}} \sum_{i=1}^p |\tau_i(\Theta)|^2 \, u_n(\Theta) \, d \Theta, 
\label{eq:GCIId_regularity_tau_prf1}
\end{equation}
which shows that $\| \chi \|^2_{L^2} = \frac{\gamma_n}{(2 \pi)^p} \| \tau \|^2_{\mathcal H}$. 

Now, we show that 
\begin{eqnarray}
\| \nabla \chi \|_{L^2}^2 &=& \frac{\gamma_n}{(2 \pi)^p} \int_{{\mathcal T}} \Big\{ \sum_{i,j=1}^p \Big|\frac{d \tau_i}{d \theta_j} (\Theta) \Big|^2 + \sum_{1 \leq i < j \leq p} \Big( \frac{|(\tau_i - \tau_j)(\Theta)|^2}{1 - \cos (\theta_i - \theta_j)} + \frac{|(\tau_i + \tau_j)(\Theta)|^2}{1 - \cos (\theta_i + \theta_j)} \Big) \nonumber \\
&& \hspace{6cm}+  \epsilon_n \sum_{i=1}^p \frac{\tau_i^2(\Theta)}{1 - \cos \theta_i} \, \Big\} u_n(\Theta) \, d \Theta. 
\label{eq:GCIId_regularity_tau_prf2}
\end{eqnarray}
The function $A \mapsto |\nabla \chi(A)|^2$ is a class function (the proof relies on similar computations to those made in the proof of Prop. \ref{prop:GCIId_varform_invar} and is omitted) and we can use Weyl's integration formula to evaluate $\| \nabla \chi \|_{L^2}^2$. For this, we need to compute $|\nabla \chi(A_\Theta)|^2$. This will be done by means of  \eqref{eq:GCIId_nabmu_nabchi_def}, which requires to find a convenient basis of $\mathfrak{so}_n$ and to compute the action of the derivation operator $\varrho$ on each term. The latter will be achieved thanks to~\eqref{eq:radlap_deriv_conjug}. Given that $\chi$ satisfies \eqref{eq:GCIId_conjug_invar_mu}, we get, for all $A \in \mathrm{SO}_n$ and $X \in \mathfrak{so}_n$:  
$$ \frac{d}{dt} \big( \chi ( e^{tX} A e^{-tX}) \big) \big|_{t=0} = \frac{d}{dt} \big( e^{tX} \chi (A) e^{-tX} \big) \big|_{t=0} = [X, \chi(A)]. $$
When inserted into \eqref{eq:radlap_deriv_conjug}, this leads to 
\begin{equation}
\Big(\varrho\big(\mathrm{Ad}(A^{-1}) X - X \big) \chi \Big)(A) = [X, \chi(A)], \quad \forall A \in \mathrm{SO}_n, \quad \forall X \in \mathfrak{so}_n. 
\label{eq:radlap_deriv_conjug_chi}
\end{equation}
We will also use the formula
\begin{eqnarray}
&&\hspace{-1cm}
\big( \varrho(F_{2k-1 \, 2k}) (\chi) \big)(A_\Theta) = \frac{d}{dt} \chi(A_\Theta e^{t F_{2k-1 \, 2k}}) \big|_{t=0} = \frac{d}{dt} \chi(A_{(\theta_1, \ldots, \theta_k-t, \ldots, \theta_p)}) \big|_{t=0} \nonumber \\
&&\hspace{-0.5cm}
= \sum_{\ell=1}^p \frac{d}{dt} \tau_\ell(\theta_1, \ldots, \theta_k-t, \ldots, \theta_p) \big|_{t=0} \, F_{2\ell-1 \, 2 \ell} = - \sum_{\ell=1}^p \frac{\partial \tau_\ell}{\partial \theta_k}(\Theta) \, F_{2\ell-1 \, 2 \ell}, \label{eq:GCIId_tildmuk_equation_prf4}
\end{eqnarray}
which is a consequence of \eqref{eq:GCIId:mu_generic}.

It will be convenient to first treat the special examples of dimension $n=3$ and $n=4$, before generalizing them to dimensions $n=2p$ and $n=2p+1$. 

\medskip
\paragraph{Case of $\mathrm{SO}_3$:} This is an even-dimensional case $n=2p+1$ with $p=1$. We have $\Theta = \theta_1$. Then, $A_\Theta = R_{\theta_1}$ (where $R_\theta$ is given by \eqref{eq:def_Rtheta}) and we have $\chi(A_\Theta) = \tau_1(\theta_1) F_{12}$. The triple $(F_{12}, G^+, G^-)$ with $G^\pm = \frac{1}{\sqrt{2}} (F_{13} \pm F_{23})$ is an orthonormal basis of $\mathfrak{so}_3$ which we will use to compute \eqref{eq:GCIId_nabmu_nabchi_def}. Thanks to \eqref{eq:GCIId_tildmuk_equation_prf4}, we first have 
$$ \big( \varrho(F_{12}) \chi \big)(A_\Theta) = - \frac{d \tau_1}{d \theta_1} (\theta_1) F_{12}. $$
Now, we apply \eqref{eq:radlap_deriv_conjug_chi} with $A=A_\Theta$ and $X=G^+$ or $G^-$. Since $A_\Theta^{-1} = A_{- \Theta}$, easy computations (see also \cite{degond2023radial}) lead to 
\begin{eqnarray*}
\mathrm{Ad}(A_{-\Theta}) G^+ - G^+ &=& (\cos \theta_1 - 1) G^+ + \sin \theta_1 G^-, \\
\mathrm{Ad}(A_{-\Theta}) G^- - G^- &=&  - \sin \theta_1 G^+ + (\cos \theta_1 - 1) G^-, 
\end{eqnarray*}
and, using \eqref{eq:commutF}, 
$$ [G^+, \chi(A_\Theta)] = - \tau_1(\theta_1) G^-, \qquad [G^-, \chi(A_\Theta)] = \tau_1(\theta_1) G^+.
$$ 
Thus
\begin{eqnarray*}
\Big( \big( (\cos \theta_1 - 1) \varrho(G^+) + \sin \theta_1 \varrho(G^-) \big) \chi \Big) (A_\Theta) &=& - \tau_1(\theta_1) G^-, \\
\Big( \big( - \sin \theta_1 \varrho(G^+) + (\cos \theta_1 - 1) \varrho(G^-) \big) \chi \Big) (A_\Theta) &=& \tau_1(\theta_1) G^+,
\end{eqnarray*}
which leads to 
\begin{eqnarray}
\big( \varrho(G^+) \chi \big) (A_\Theta) &=& - \frac{\tau_1(\theta_1)}{2 (1 - \cos \theta_1)} \big( \sin \theta_1 G^+ + (\cos \theta_1 - 1) G^- \big), \label{eq:varrhoG+}\\
\big( \varrho(G^-) \chi \big) (A_\Theta) &=& \frac{\tau_1(\theta_1)}{2 (1 - \cos \theta_1)} \big( (\cos \theta_1 - 1) G^+ - \sin \theta_1 G^- \big).  \label{eq:varrhoG-}
\end{eqnarray}
From \eqref{eq:GCIId_nabmu_nabchi_def}, it follows that 
$$ |\nabla \chi(A_\Theta) |^2 = \Big|\frac{d \tau_1}{d \theta_1} (\theta_1) \Big|^2 + \frac{\tau_1^2(\theta_1)}{1 - \cos \theta_1}, $$
which leads to \eqref{eq:GCIId_regularity_tau_prf2} for $n=3$.

\medskip
\paragraph{Case of $\mathrm{SO}_4$:} this is an odd-dimensional case $n=2p$ with $p=2$. We have $\Theta = (\theta_1, \theta_2)$ and 
$$ \chi(A_\Theta) = \tau_1(\theta_1,\theta_2) F_{12} + \tau_2(\theta_1,\theta_2) F_{34}.$$ 
The system $(F_{12}, F_{34}, H^+, H^-, K^+, K^-)$ with $H^\pm = \frac{1}{\sqrt{2}} (F_{13} \pm F_{24})$ and $K^\pm = \frac{1}{\sqrt{2}} (F_{14} \pm F_{23})$ is an orthonormal basis of $\mathfrak{so}_4$, which will be used to express \eqref{eq:GCIId_nabmu_nabchi_def}. Then, we have 
\begin{eqnarray}
\big( \varrho(F_{12}) \chi \big)(A_\Theta) &=& - \frac{d \tau_1}{d \theta_1} (\Theta) F_{12}
- \frac{d \tau_2}{d \theta_1} (\Theta) F_{34}, \label{eq:GCIId_regularity_tau_prf3}\\
\big( \varrho(F_{34}) \chi \big)(A_\Theta) &=& - \frac{d \tau_1}{d \theta_2} (\Theta) F_{12}
- \frac{d \tau_2}{d \theta_2} (\Theta) F_{34}. \label{eq:GCIId_regularity_tau_prf4}
\end{eqnarray}
Now, we compute:
\begin{eqnarray*}
\mathrm{Ad}(A_{-\Theta}) H^+ - H^+ &=& (c_1 c_2 + s_1 s_2 - 1) H^+ - (c_1 s_2 - s_1 c_2) K^-, \\
\mathrm{Ad}(A_{-\Theta}) H^- - H^- &=& - (c_1 s_2 + s_1 c_2) K^+ + (c_1 c_2 - s_1 s_2 - 1)H^-, \\
\mathrm{Ad}(A_{-\Theta}) K^+ - K^+ &=& (c_1 c_2 - s_1 s_2 - 1) K^+ + (c_1 s_2 + s_1 c_2) H^-, \\
\mathrm{Ad}(A_{-\Theta}) K^- - K^- &=& (c_1 s_2 - s_1 c_2) H^+ + (c_1 c_2 + s_1 s_2 - 1)K^-, 
\end{eqnarray*}
with $c_i = \cos \theta_i$ and $s_i = \sin \theta_i$, $i=1, \, 2$. We also compute, using \eqref{eq:commutF}:
\begin{eqnarray*}
[H^+, \chi(A_\Theta)] &=& (- \tau_1 + \tau_2) K^-, \quad [H^-, \chi(A_\Theta)] = (\tau_1 + \tau_2) K^+, \\
\mbox{} 
[K^+, \chi(A_\Theta)] &=& - (\tau_1 + \tau_2) H^-, \quad [K^-, \chi(A_\Theta)] = (\tau_1 - \tau_2) H^+. 
\end{eqnarray*}
where we omit the dependence of $\tau_i$ on $\Theta$ for simplicity. Applying \eqref{eq:radlap_deriv_conjug_chi}, we get two independent linear systems of equations for $((\varrho(H^+)\chi)(A_\Theta), (\varrho(K^-)\chi)(A_\Theta))$ on one hand, and $((\varrho(K^+)\chi)(A_\Theta), (\varrho(H^-)\chi)(A_\Theta))$ on the other hand, which can both easily be resolved into
\begin{eqnarray}
\big( \varrho(H^+)\chi \big)(A_\Theta) &=&  \frac{1}{2} (\tau_1-\tau_2) \Big( K^- - \frac{\sin(\theta_1-\theta_2)}{1-\cos(\theta_1-\theta_2)} H^+ \Big), \label{eq:GCIId_regularity_tau_prf5}\\
\big( \varrho(K^-)\chi \big)(A_\Theta) &=& \frac{1}{2} (\tau_1-\tau_2) \Big( - \frac{\sin(\theta_1-\theta_2)}{1-\cos(\theta_1-\theta_2)} K^- - H^+ \Big) , \label{eq:GCIId_regularity_tau_prf6}\\
\big( \varrho(H^-)\chi \big)(A_\Theta) &=&  \frac{1}{2} (\tau_1+\tau_2) \Big( - K^+ - \frac{\sin(\theta_1+\theta_2)}{1-\cos(\theta_1+\theta_2)} H^- \Big), \label{eq:GCIId_regularity_tau_prf7}\\
\big( \varrho(K^+)\chi \big)(A_\Theta) &=& \frac{1}{2} (\tau_1+\tau_2) \Big( - \frac{\sin(\theta_1+\theta_2)}{1-\cos(\theta_1+\theta_2)} K^+ + H^- \Big) . \label{eq:GCIId_regularity_tau_prf8}
\end{eqnarray}
Taking the squared norms in $\mathfrak{so}_4$ of \eqref{eq:GCIId_regularity_tau_prf3} to \eqref{eq:GCIId_regularity_tau_prf8}, we get 
$$|\nabla \chi(A_\Theta) |^2 = \sum_{i,j=1}^2 \Big|\frac{d \tau_i}{d \theta_j} \Big|^2 + \frac{|\tau_1 - \tau_2|^2}{1 - \cos (\theta_1 - \theta_2)} + \frac{|\tau_1 + \tau_2|^2}{1 - \cos (\theta_1 + \theta_2)} , $$
which leads to \eqref{eq:GCIId_regularity_tau_prf2} for $n=4$.

\medskip
\paragraph{Case of $\mathrm{SO}_{2p}$:} Define $H_{jk}^\pm = \frac{1}{\sqrt{2}} (F_{2j-1 \, 2k-1} \pm F_{2j \, 2k})$ and $K_{jk}^\pm = \frac{1}{\sqrt{2}} (F_{2j-1 \, 2k} \pm F_{2j \, 2k-1})$ for $1 \leq j < k \leq p$. Then, the system $\big( (F_{2j-1 \, 2j})_{j=1, \ldots, p}, (H_{jk}^+, H_{jk}^-, K_{jk}^+, K_{jk}^-)_{1 \leq j < k \leq p} \big)$ is the orthonormal basis of $\mathfrak{so}_{2p} {\mathbb R}$ which will be used to evaluate \eqref{eq:GCIId_nabmu_nabchi_def}. Then, we remark that the computations of $\mathrm{Ad}(A_{-\Theta}) H_{jk}^\pm$ and $\mathrm{Ad}(A_{- \Theta}) K_{jk}^\pm$ only involve the $4 \times 4$ matrix subblock corresponding to line and column indices belonging to $\{ 2j-1, 2j \} \cup \{ 2k-1, 2k \}$. Thus, restricted to these $4 \times 4$ matrices, the computations are identical to those done in the case of $\mathrm{SO}_{4}{\mathbb R}$. This directly leads to \eqref{eq:GCIId_regularity_tau_prf2} for $n=2p$. 

\medskip
\paragraph{Case of $\mathrm{SO}_{2p+1}$:} this case is similar, adding to the previous basis the elements $G_j^{\pm} = \frac{1}{\sqrt{2}} (F_{2j-1 \, 2p+1} \pm F_{2j \, 2p+1})$. These additional terms contribute to terms like in the $\mathrm{SO}_3$ case, which leads to \eqref{eq:GCIId_regularity_tau_prf2} for $n=2p+1$. 

\medskip
This finishes the proof of \eqref{eq:GCIId_regularity_tau_prf2}.

\bigskip
\noindent
\textbf{Proof of \eqref{eq:chi_tau_conv}.} Consider $\tau \in {\mathcal V}$ and a sequence $(\tau^q)_{q \in {\mathbb N}}$ of elements of $C_{\mathrm{per}}^{\infty,\mathfrak{W}}({\mathcal T}, {\mathbb R}^p)$ which converges to $\tau$ in ${\mathcal V}$. Let $\chi^q$ be associated with $\tau^q$ through \eqref{eq:GCIId:mu_generic_gene}. From \eqref{eq:GCIId:mu_generic}, we see that the function $\Theta \to \chi^q(A_\Theta)$ belongs to $C^{\infty}({\mathcal T}, \mathfrak{so}_n)$. However, we cannot deduce directly from it that $\chi^q$ belongs to $C^{\infty}_\mathrm{inv}(\mathrm{SO}_n, \mathfrak{so}_n)$. Indeed, for a given $A \in \mathrm{SO}_n$, the pair $(g,\Theta) \in \mathrm{SO}_n \times {\mathcal T}$ such that $A = g A_\Theta g^T$ is not unique, and consequently, the map $\mathrm{SO}_n \times {\mathcal T} \to \mathrm{SO}_n$, $(g, \Theta) \mapsto g A_\Theta g^T$ is not invertible. Therefore, we cannot deduce that $\chi^q$ is smooth from the smoothness of the map $\mathrm{SO}_n \times {\mathcal T} \to \mathfrak{so}_n$, $(g, \Theta) \mapsto g \chi^q(A_\Theta) g^T$. The result is actually true but the proof requires a bit of topology (see Remark \ref{rem:topological_proof} below). Here, we give a proof of \eqref{eq:chi_tau_conv} that avoids proving that $\chi^q$ is smooth and which is thus simpler. 

Temporarily dropping the superscript $q$ from $\chi^q$, we first show that $\chi$ is a differentiable function $\mathrm{SO}_n \to \mathfrak{so}_n$ at any point $A \in {\mathbb T}$ (which means that the derivatives in all directions of the tangent space $T_A = A \mathfrak{so}_n$ of $\mathrm{SO}_n$ at $A \in {\mathbb T}$ are defined, not only those derivatives in the direction of an element of the tangent space $T_A{\mathbb T} = A \mathfrak{h}$ of ${\mathbb T}$ at $A$). Indeed, this is a consequence of the proof of \eqref{eq:GCIId_regularity_tau_prf0} above. This proof shows that for an orthonormal basis $(\Phi_i)_{i=1}^{\mathcal N}$ (with ${\mathcal N} = \mathrm{dim} \, \mathfrak{so}_n$) whose precise definition depends on $n$ (see proof of  \eqref{eq:GCIId_regularity_tau_prf0}), then $(\varrho(\Phi_i) (\chi))(A)$ exists for all $A \in {\mathbb T}$ and all $i \in \{1, \ldots {\mathcal N}\}$, which is exactly saying that the derivatives of $\chi$ in all the direction of $T_A$ exist for all $A \in {\mathbb T}$. If we refer to the $n=3$ case for instance, $(\varrho(F_{12}) (\chi))(A_\Theta)$ exists because $(d \tau_1/d \theta)(\theta_1)$ exists since $\tau$ is $C^\infty$. Then,~\eqref{eq:varrhoG+} and \eqref{eq:varrhoG-} tell us that $(\varrho(G^+) (\chi))(A_\Theta)$ and $(\varrho(G^-) (\chi))(A_\Theta)$ exist, except may be when $\cos \theta_1 = 1$. But, because $\tau$ commutes with the Weyl group, Eqs.~\eqref{eq:varrhoG+} and \eqref{eq:varrhoG-} lead to finite values of $(\varrho(G^\pm) (\chi))(A_\Theta)$ when $\cos \theta_1 = 1$ thanks to a Taylor expansion similar to that made in Remark \ref{rem:GCIId_normV_finite}. It is straightforward to see that dimensions $n \geq 4$ can be treated in a similar fashion. 

From this, we deduce that $\nabla \chi(A)$ exists for all $A \in \mathrm{SO}_n$. Indeed, by construction, $\chi$ satisfies \eqref{eq:GCIId_conjug_invar_mu}. So, we deduce that
\begin{equation} 
\nabla (\chi \circ \xi_g) (A) = \nabla (g \chi g^T) (A), \quad \forall A, \, g \in \mathrm{SO}_n, 
\label{eq:na_chi_circ_xi_equals}
\end{equation}
where $\xi_g$ is defined by \eqref{eq:GCI_conjug_def}. Applying \eqref{eq:nabla_chi_express} with the basis $(\Phi_i)_{i=1}^{\mathcal N} = (\Psi_i)_{i=1}^{\mathcal N} = (F_{ij})_{1 \leq i < j \leq n}$, we get, thanks to \eqref{eq:GCI_deriv_compos_conjug}, 
\begin{equation} 
\nabla (\chi \circ \xi_g) (A) = \sum_{k < \ell} \sum_{k' < \ell'} \big( \nabla \chi_{k \ell}  (g A g^T) \cdot g A F_{k' \ell'} g^T \big) \, F_{k \ell} \otimes A F_{k' \ell'}, 
\label{eq:na_chi_circ_xi}
\end{equation}
while 
\begin{equation} 
\nabla (g \chi g^T)(A) = \sum_{k < \ell} \sum_{k' < \ell'} \big( \nabla \chi_{k \ell}(A) \cdot A F_{k' \ell'} \big) \,  (g F_{k \ell} g^T) \otimes A F_{k' \ell'}, 
\label{eq:na_g_chi_gT}
\end{equation}
where $\sum_{k < \ell}$ means the sum over all pairs $(k,\ell)$ such that $1 \leq k < \ell \leq n$ and similarly for $\sum_{k' < \ell'}$. Thus, using the fact that the basis $(F_{k \ell} \otimes A F_{k' \ell'})_{k<\ell, k'<\ell'}$ is an orthonormal basis of $\mathfrak{so}_n \otimes T_A$ and that the two expressions \eqref{eq:na_chi_circ_xi} and \eqref{eq:na_g_chi_gT} are equal thanks to \eqref{eq:na_chi_circ_xi_equals}, we get
\begin{equation} 
\nabla \chi_{k \ell}(gAg^T) \cdot g A F_{k' \ell'} g^T = \sum_{k_1 < \ell_1} \big( \nabla \chi_{k_1 \ell_1} (A) \cdot A F_{k' \ell'} \big) \big( g F_{k_1 \ell_1} g^T \cdot F_{k \ell} \big). 
\label{eq:na_chi_kl_gAgT}
\end{equation}
We apply this formula with $A = A_\Theta$. Because we know that $\nabla \chi (A_\Theta)$ exists for all $\Theta \in {\mathcal T}$, the right-hand side of \eqref{eq:na_chi_kl_gAgT} is well-defined. Thus, the left-hand side of \eqref{eq:na_chi_kl_gAgT} tells us that $\nabla \chi(A)$ is defined (by its components on the basis $(F_{k \ell} \otimes A F_{k' \ell'})_{k<\ell, k'<\ell'}$) for all $A$ such that there exists $(g,\Theta) \in \mathrm{SO}_n \times {\mathcal T}$ with $A = g A_\Theta g^T$. But of course, such $A$ range over the whole group $\mathrm{SO}_n$, which shows that $\nabla \chi$ exists everywhere. 

Now, we note that formula \eqref{eq:GCIId_regularity_tau_prf0} applies to $\chi$. Indeed, because $\chi$ satisfies \eqref{eq:GCIId_conjug_invar_mu}, $|\chi|^2$ and $|\nabla \chi|^2$ are class functions, so they only depend on their values on ${\mathbb T}$, which are given by the same formulas in terms of $\tau$ as those found in the proof of \eqref{eq:GCIId_regularity_tau_prf0}. Hence, the value of $\| \chi \|_{H^1}^2$ is still given by \eqref{eq:GCIId_regularity_tau_prf0}. In particular, it is finite and so, we have that $\chi \in H^1_{\mathrm{inv}}(\mathrm{SO}_n, \mathfrak{so}_n)$. 

Now, putting the superscipt $q$ back, we have $\chi^q \in H^1_{\mathrm{inv}}(\mathrm{SO}_n, \mathfrak{so}_n)$ and up to an unimportant constant, $\| \chi^q \|_{H^1} = \| \tau^q \|_{\mathcal V}$. Since $\tau^q \to \tau$ in ${\mathcal V}$, $\chi^q$ is a Cauchy sequence in $H^1_{\mathrm{inv}}(\mathrm{SO}_n, \mathfrak{so}_n)$. Thus, it converges to an element $\chi \in H^1_{\mathrm{inv}}(\mathrm{SO}_n, \mathfrak{so}_n)$ and by the same reasoning as when we proved the direct implication, we deduce that $\chi$ and $\tau$ are related with each other by \eqref{eq:GCIId:mu_generic_gene}, which ends the proof. \endproof 

\begin{remark}
We sketch a direct proof that, for any $\tau \in C_{\mathrm{per}}^{\infty,\mathfrak{W}}({\mathcal T}, {\mathbb R}^p)$, the function $\chi$ given by \eqref{eq:GCIId:mu_generic_gene} belongs to $C^{\infty}_\mathrm{inv}(\mathrm{SO}_n, \mathfrak{so}_n)$. We consider the mapping 
$\Pi$: $\mathrm{SO}_n \times {\mathcal T} \to \mathrm{SO}_n$, $(g, \Theta) \mapsto g A_\Theta g^T$. Suppose $g' \in g {\mathbb T}$, i.e. $\exists \Upsilon \in {\mathcal T}$ such that $g' = g A_\Upsilon$. Then $g' A_\Theta (g')^T = g A_\Theta g^T$ because ${\mathbb T}$ is abelian. Thus, $\Pi$ defines a map $\tilde \Pi$: $\mathrm{SO}_n/{\mathbb T} \times {\mathcal T} \to \mathrm{SO}_n$, $(g {\mathbb T}, \Theta) \mapsto g A_\Theta g^T$, where $\mathrm{SO}_n/{\mathbb T}$ is the quotient set, whose elements are the cosets $g {\mathbb T}$ for $g \in \mathrm{SO}_n$. From the discussion of the Weyl group in Section \ref{subsec:rotmore}, we know that generically (i.e. for all $\Theta$ except those belonging to the boundary of one of the Weyl chambers), the map $\tilde \Pi$ is a $\mathrm{Card} (\mathfrak{W})$-sheeted covering of $\mathrm{SO}_n$ (see also \cite[p. 443]{fulton2013representation}). Thus, it is locally invertible and its inverse is smooth. Let us generically denote by $(\tilde \Pi)^{-1}$ one of these inverses. Now, we introduce $\bar \chi$: $\mathrm{SO}_n \times {\mathcal T} \to \mathfrak{so}_n$, $(g, \Theta) \mapsto \sum_{k=1}^p \tau(\Theta) g F_{2k-1 \, 2k} g^T$. This map is smooth. Then, it is easy to see that $\bar \chi$ defines a map $\tilde \chi$: $\mathrm{SO}_n/{\mathbb T} \times {\mathcal T} \to \mathfrak{so}_n$ in the same way as $\Pi$ did and likewise, $\tilde \chi$ is smooth. Then, by construction, we can write locally $\chi = \tilde \chi \circ (\tilde \Pi)^{-1}$ in the neighborhood of any $A$ such that the associated $\Theta$'s do not belong to the boundary of one of the Weyl chambers. Since the maps $\tilde \chi$ and $(\tilde \Pi)^{-1}$ are smooth we deduce that $\chi$ is smooth. We need now to apply a special treatment when $\Theta$ belongs to the boundary of one of the Weyl chambers, because some of the sheets of the covering $\tilde \Pi$ intersect there. We will skip these technicalities in this remark. The case $\cos \theta_1 = 1$ in dimension $n=3$ that we encountered above and that required a Taylor expansion to show that $\nabla \chi (A_\Theta)$ existed is a perfect illustration of the kind of degeneracy which appears at the boundary of the Weyl chambers. 
\label{rem:topological_proof}
\end{remark}

We can now write the variational formulation obeyed by $\alpha$ in the following:

\begin{proposition}
Let $\mu$ be the unique solution of the variational formulation \eqref{eq:GCIId_varform_mu_invar_torus}. Then, $\mu$ is given by \eqref{eq:GCIId:mu_formula} where $\alpha = (\alpha_i)_{i=1}^p$ is the unique solution of the following variational formulation 
\begin{equation}
\left\{ \begin{array}{l}
\displaystyle \alpha \in {\mathcal V},  \\
\displaystyle {\mathcal A}(\alpha,\tau) = {\mathcal L}(\tau), \quad \forall \tau \in {\mathcal V}, 
\end{array} \right. 
\label{eq:GCIId_varform_alpha}
\end{equation}
and with
\begin{eqnarray}
{\mathcal A}(\alpha,\tau) &=& \int_{{\mathcal T}} \Big\{ \sum_{i,j=1}^p \frac{\partial \alpha_i}{\partial \theta_j} \frac{\partial \tau_i}{\partial \theta_j} + \sum_{1 \leq i < j \leq p} \Big( \frac{(\alpha_i - \alpha_j)(\tau_i - \tau_j)}{1 - \cos(\theta_i - \theta_j)} \nonumber \\
&&\hspace{1.5cm}
+  \frac{(\alpha_i + \alpha_j)(\tau_i + \tau_j)}{1 - \cos(\theta_i + \theta_j)} \Big) + \epsilon_n \sum_{i=1}^p  \frac{\alpha_i \tau_i}{1 - \cos \theta_i} \Big\} m(\Theta) \, d \Theta,  \label{eq:GCIId_bilinform_A} \\
{\mathcal L}(\tau) &=& \int_{{\mathcal T}}  \Big( \sum_{i=1}^p  \sin \theta_i \,  \tau_i \Big) m(\Theta) \, d \Theta \label{eq:GCIId_linform_L} . 
\end{eqnarray}
We recall that $m(\Theta)$ is given by \eqref{eq:GCIId_m_def}.
\label{prop:GCIId_varform_tau}
\end{proposition}

\noindent
\textbf{Proof.} ${\mathcal A}$ and ${\mathcal L}$ are just the expression of the left-hand and right-hand sides of \eqref{eq:GCIId_varform_mu_invar_torus} when $\mu$ and $\chi$ are given the expressions \eqref{eq:GCIId:mu_formula} and \eqref{eq:GCIId:mu_generic} respectively. The computation of \eqref{eq:GCIId_bilinform_A} follows closely the computations made in the proof of Prop. \ref{prop:GCIId_regularity_tau} and is omitted. That of \eqref{eq:GCIId_linform_L}  follows from 
\begin{equation}
\frac{A_\Theta-A_\Theta^T}{2} = - \sum_{\ell=1}^p \sin \theta_\ell \, F_{2\ell-1 \, 2\ell}. 
\label{eq:GCIId_tildmuk_equation_prf2}
\end{equation}
which is a consequence of \eqref{eq:R2p}, \eqref{eq:R2p+1}.

To show that the variational formulation \eqref{eq:GCIId_varform_alpha} is well-posed we apply Lax-Milgram's theorem. ${\mathcal A}$ and ${\mathcal L}$ are clearly continuous bilinear and linear forms on ${\mathcal V}$ respectively. To show that ${\mathcal A}$ is coercive, it is enough to show a Poincar\'e inequality
$$ {\mathcal A} (\tau, \tau) \geq C \| \tau \|_{\mathcal H}^2, \quad \forall \tau \in {\mathcal V}, $$
for some $C >0$. Suppose that $p \geq 2$. Since $1 - \cos (\theta_i \pm \theta_j) \leq 2$, we have 
\begin{eqnarray*} 
{\mathcal A} (\tau, \tau) &\geq& \int_{{\mathcal T}} \sum_{1 \leq i < j \leq p} \Big( \frac{|\tau_i - \tau_j|^2}{1 - \cos(\theta_i - \theta_j)} 
+  \frac{|\tau_i + \tau_j|^2}{1 - \cos(\theta_i + \theta_j)} \Big) m(\Theta) \, d \Theta  \\
& \geq & \frac{1}{2} \sum_{1 \leq i < j \leq p} \int_{{\mathcal T}} \big( | \tau_i-\tau_j |^2 + | \tau_i+\tau_j |^2 \big) m(\Theta) \, d \Theta \\
&=&  \sum_{1 \leq i < j \leq p} \int_{{\mathcal T}}  \big( | \tau_i |^2 + | \tau_j |^2 \big) m(\Theta) \, d \Theta = (p-1) \| \tau \|_{\mathcal H}^2, 
\end{eqnarray*}
which shows the coercivity in the case $p \geq 2$. Suppose now that $p=1$, i.e., $n=3$. Then, by the same idea, 
$$ {\mathcal A}(\tau,\tau) \geq \int_{{\mathcal T}} \frac{|\tau_1|^2}{1 - \cos \theta_1} \,  m(\Theta) \, d \Theta \geq \frac{1}{2} \int_{{\mathcal T}} |\tau_1|^2 \,  m(\Theta) \, d \Theta = \frac{1}{2}  \| \tau \|_{\mathcal H}^2,  $$
which shows the coercivity in the case $p=1$ and ends the proof. \endproof

The variational formulation \eqref{eq:GCIId_varform_alpha} gives the equations satisfied by $\alpha$ in weak form. It is desirable to express this system in strong form and show that the latter is given by System \eqref{eq:GCIId_strongform_alpha}. This is done in the following proposition. 

\begin{proposition}
Let $\alpha$ be the unique solution of the variational formulation \eqref{eq:GCIId_varform_alpha}. Then, $\alpha$ is a distributional solution of System \eqref{eq:GCIId_strongform_alpha} on the open set 
$$ {\mathcal O} = \big\{ \Theta \in {\mathcal T} \, \, | \, \, \theta_\ell \not = \pm \theta_k, \, \, \forall k \not = \ell \text{ and,  in the case } n \text{ even, } \theta_\ell \not = 0, \, \, \forall \ell \in \{1, \ldots, p \} \big\}. $$
\label{prop:GCIId_strongform_alpha}
\end{proposition}

\noindent
\textbf{Proof.} A function $f$: ${\mathcal T} \to {\mathbb R}$ is invariant by the Weyl group if and only if $f \circ W = f$, $\forall W \in \mathfrak{W}$. It is easily checked that the functions inside the integrals defining ${\mathcal A}$ and ${\mathcal L}$ in~\eqref{eq:GCIId_bilinform_A} and \eqref{eq:GCIId_linform_L} are invariant by the Weyl group. 
If $f \in L^1({\mathcal T})$ is invariant by the Weyl group, we can write, using \eqref{eq:rotmore_Weylchamb_union} and \eqref{eq:rotmore_Weylchamb_intersect} (with ${\mathbb R}^p$ replaced by ${\mathcal T}$ and ${\mathcal W}$ by ${\mathcal W}_{\mathrm{per}}$):
\begin{equation} 
\int_{{\mathcal T}} f(\Theta) \, d \Theta = 
\sum_{W \in \mathfrak{W}} \int_{{\mathcal W}_{\mathrm{per}}} f \circ W (\Theta) \, d \Theta= \mathrm{Card} (\mathfrak{W})  \int_{{\mathcal W}_{\mathrm{per}}} f (\Theta) \, d \Theta, 
\label{eq:GCIId_strongform_alpha_prf0}
\end{equation}
where $\mathrm{Card} (\mathfrak{W})$ stands for the order of $\mathfrak{W}$. Thus, up to a constant which will factor out from \eqref{eq:GCIId_varform_alpha}, the integrals over ${\mathcal T}$ which are involved in the definitions of ${\mathcal A}$ and ${\mathcal L}$ in \eqref{eq:GCIId_bilinform_A} and \eqref{eq:GCIId_linform_L} can be replaced by integrals over ${\mathcal W}_{\mathrm{per}}$.

Now, given $\ell \in \{1, \ldots p\}$, we wish to recover the $\ell$-th equation of System \eqref{eq:GCIId_strongform_alpha} by testing the variational formulation \eqref{eq:GCIId_varform_alpha} with a convenient $p$-tuple of test functions $\tau = (\tau_1, \ldots, \tau_p)$. A natural choice is by taking $\tau_\ell = \varphi (\Theta)$ for a  given $\varphi \in C^\infty_c (\mathrm{Int}({\mathcal W}_{\mathrm{per}}))$ (where $C^\infty_c$ stands for the space of infinitely differentiable functions with compact support and $\mathrm{Int}$ for the interior of a set), while $\tau_i = 0$ for $i \not = \ell$. However, this only defines $\tau$ on ${\mathcal W}_{\mathrm{per}}$ and we have to extend it to the whole domain ${\mathcal T}$ and show that it defines a valid test function $\tau \in {\mathcal V}$.

More precisely, we claim that we can construct a unique $\tau \in {\mathcal V}$ such that
\begin{equation}
\tau_i (\Theta) = \varphi (\Theta)  \delta_{i \ell}, \quad \forall \Theta \in \mathrm{Int}({\mathcal W}_{\mathrm{per}}), \quad \forall i \in \{1, \ldots p\}. 
\label{eq:GCIId_strongform_alpha_prf1}
\end{equation}
Indeed, we can check that  
$$ {\mathcal O} = \bigcup_{W \in \mathfrak{W}} \mathrm{Int} \big(W({\mathcal W}_{\mathrm{per}}) \big). $$ 
Thus, if $\Theta \in {\mathcal O}$, there exists a unique pair $(W,\Theta_0)$ with $W \in \mathfrak{W}$ and $\Theta_0 \in \mathrm{Int}({\mathcal W}_{\mathrm{per}})$ such that $\Theta = W(\Theta_0)$. Then, by the fact that $\tau$ must commute with the elements of the Weyl group, we necessarily have 
\begin{equation}
\tau(\Theta) = \tau \circ W (\Theta_0) = W \circ \tau (\Theta_0). 
\label{eq:GCIId_strongform_alpha_prf2}
\end{equation}
Since $\tau (\Theta_0)$ is determined by \eqref{eq:GCIId_strongform_alpha_prf1}, then $\tau(\Theta)$ is determined by \eqref{eq:GCIId_strongform_alpha_prf2} in a unique way. It remains to determine $\tau$ on ${\mathcal T} \setminus {\mathcal O}$. But since $\varphi$ is compactly supported in $\mathrm{Int}({\mathcal W}_{\mathrm{per}})$ its value on the boundary of $\partial {\mathcal W}_{\mathrm{per}}$ of ${\mathcal W}_{\mathrm{per}}$ is equal to $0$. By the action of the Weyl group on $\partial {\mathcal W}_{\mathrm{per}}$, we deduce that $\tau$ must be identically $0$ on ${\mathcal T} \setminus {\mathcal O}$. In particular, $\tau$ is periodic on ${\mathcal T}$. So, the so-constructed $\tau$ is in $C_{\mathrm{per}}^{\infty,\mathfrak{W}}({\mathcal T}, {\mathbb R}^p)$ and thus, in ${\mathcal V}$. The so-constructed $\tau$ is unique because we proceeded by necessary conditions throughout all this reasoning. 

We use the so-constructed $\tau$ as a test function in \eqref{eq:GCIId_varform_alpha} with ${\mathcal T}$ replaced by ${\mathcal W}_{\mathrm{per}}$ in~\eqref{eq:GCIId_bilinform_A} and \eqref{eq:GCIId_linform_L}. This leads to 
\begin{eqnarray*}
&&\hspace{-1cm}
\int_{{\mathcal W}_{\mathrm{per}}} \Big\{ \sum_{k=1}^p \frac{\partial \alpha_\ell}{\partial \theta_k} \frac{\partial \varphi}{\partial \theta_k} + \Big[ \sum_{k \not = \ell} \Big( \frac{(\alpha_\ell - \alpha_k)}{1 - \cos(\theta_\ell - \theta_k)} +  \frac{(\alpha_\ell + \alpha_k)}{1 - \cos(\theta_\ell + \theta_k)} \Big) \\
&&\hspace{3cm}
 + \epsilon_n \frac{\alpha_\ell }{1 - \cos \theta_\ell} \Big] \varphi \Big\} m(\Theta) \, d \Theta = \int_{{\mathcal W}_{\mathrm{per}}}  \sin \theta_\ell \,  \varphi \, m(\Theta) \, d \Theta,   
\end{eqnarray*}
for all $\ell \in \{1, \ldots, p\}$, for all $\varphi \in C^\infty_c (\mathrm{Int}({\mathcal W}_{\mathrm{per}}) )$, which is equivalent to saying that $\alpha$ is a distributional solution of \eqref{eq:GCIId_strongform_alpha} on $\mathrm{Int}({\mathcal W}_{\mathrm{per}})$. Now, in \eqref{eq:GCIId_strongform_alpha_prf0}, ${\mathcal W}_{\mathrm{per}}$ can be replaced by $W({\mathcal W}_{\mathrm{per}})$ for any $W \in \mathfrak{W}$. This implies that $\alpha$ is a distributional solution of \eqref{eq:GCIId_strongform_alpha} on the whole open set ${\mathcal O}$, which ends the proof. \endproof

\begin{remark}
Because of the singularity of System \eqref{eq:GCIId_strongform_alpha} it is delicate to give sense of it on ${\mathcal T} \setminus {\mathcal O}$. Observe however that since $\mu \in C^\infty(\mathrm{SO}_n, \mathfrak{so}_n)$ (by elliptic regularity), then $\tau_i$: $\Theta \mapsto \mu(A_\Theta) \cdot F_{2i-1 \, 2i}$ belongs to $C^\infty({\mathcal T})$. 
\label{rem:GCIId_tildmuk_equation}
\end{remark}

\setcounter{equation}{0}
\section{Hydrodynamic limit II: final steps of the proof}
\label{subsec:Gamma}

\subsection{Use of the generalized collision invariant}
\label{subsec:Gamma_use_GCI}

Here, we note a slight confusion we have made so far between two different concepts. The reference frame $({\mathbf e}_1, \ldots, {\mathbf e}_n)$ is used to define a rotation (temporarily noted as $\gamma$) which maps this frame to the average  body frame $(\Omega_1, \ldots, \Omega_n)$ (i.e. $\Omega_j = \gamma({\mathbf e}_j)$, see~\eqref{eq:cont_Omdef}). Now to identify the rotation $\gamma$ with a rotation matrix $\Gamma$, we can use a coordinate basis $({\mathbf f}_1, \ldots, {\mathbf f}_n)$ which is different from the reference frame $({\mathbf e}_1, \ldots, {\mathbf e}_n)$. The rotation matrix~$\Gamma$ is defined by $\gamma({\mathbf f}_j) = \sum_{i=1}^n \Gamma_{ij} {\mathbf f}_i$. So far, we have identified the rotation $\gamma$ and the matrix~$\Gamma$, but of course, this requires to specify the coordinate basis $({\mathbf f}_1, \ldots, {\mathbf f}_n)$. Note that $\Gamma$ can be recovered from $({\mathbf e}_1, \ldots, {\mathbf e}_n)$ and $(\Omega_1, \ldots, \Omega_n)$ by $\Gamma = T S^T$ where $S$ and $T$ are the transition matrices from $({\mathbf f}_1, \ldots, {\mathbf f}_n)$ to $({\mathbf e}_1, \ldots, {\mathbf e}_n)$ and $(\Omega_1, \ldots, \Omega_n)$ respectively.

We denote by $e_{1 \, m}$ the $m$-th coordinate of ${\mathbf e}_1$ in the coordinate basis $({\mathbf f}_1, \ldots, {\mathbf f}_n)$ and we define a matrix ${\mathbb P}$ and a four-rank tensor ${\mathbb S}$ as follows: 
\begin{eqnarray}
{\mathbb P} &=& \Gamma^T (\nabla_x \rho \otimes {\mathbf e}_1) - (\nabla_x \rho \otimes {\mathbf e}_1)^T \Gamma + \kappa \rho \Gamma^T \partial_t \Gamma, \label{eq:Gamma_Pdef}\\
{\mathbb S}_{i j m q} &=& \frac{1}{2} \sum_{k, \ell = 1}^n \Gamma_{ki} \frac{\partial \Gamma_{kj}}{\partial x_\ell} \big(\Gamma_{\ell m}  e_{1 \, q} +  \Gamma_{\ell q}  e_{1 \, m} \big) 
, \label{eq:Gamma_Sdef}
\end{eqnarray}
where in \eqref{eq:Gamma_Sdef}, ${\mathbb S}$ is read in the coordinate basis $({\mathbf f}_1, \ldots, {\mathbf f}_n)$. The symbol $\otimes$ denotes the tensor product of two vectors. Hence, the matrix $\nabla_x \rho \otimes {\mathbf e}_1$ has entries $(\nabla_x \rho \otimes {\mathbf e}_1)_{ij} = (\nabla_x \rho)_i e_{1 \, j}$. 
We note that ${\mathbb P}$ is an antisymmetric matrix while ${\mathbb S}$ is antisymmetric with respect to $(i,j)$ and symmetric with respect to $(m,q)$. We denote by ${\mathcal S}_n$ the space of $n \times n$ symmetric matrices. Now, we define two maps $L$ and $B$ as follows: 
\begin{itemize}
\item[$\bullet$] $L$ is a linear map $\mathfrak{so}_n \to \mathfrak{so}_n$ such that 
\begin{equation}
L(P) = \int_{\mathrm{SO}_n} (A \cdot P) \, \mu(A) \, M(A) \, dA, \quad \forall P \in  \mathfrak{so}_n. 
\label{eq:Gamm_Ldef}
\end{equation}
\item[$\bullet$] $B$: $\mathfrak{so}_n \times \mathfrak{so}_n \to {\mathcal S}_n$ is bilinear and defined by 
\begin{equation}
B(P,Q) = \int_{\mathrm{SO}_n} (A \cdot P) \, (\mu(A) \cdot Q) \, \frac{A + A^T}{2} \, M(A) \, dA, \quad \forall P, \, Q \in \mathfrak{so}_n. 
\label{eq:Gamma_Bdef}
\end{equation}
\end{itemize}
We can now state a first result about the equation satisfied by $\Gamma$. 

\begin{proposition}
The functions $\rho$ and $\Gamma$ involved in \eqref{eq:equi_f0express} satisfy the following equations: 
\begin{equation}
L({\mathbb P})_{rs} + \kappa \rho \sum_{m,q = 1}^n B_{m q}({\mathbb S}_{\cdot \cdot m q}, F_{rs}) = 0, \quad \forall r, s \in \{1, \ldots, n \}, 
\label{eq:Gamma_Gameq_first}
\end{equation}
where ${\mathbb S}_{\cdot \cdot m q}$ stands for the antisymmetric matrix $({\mathbb S}_{\cdot \cdot m q})_{ij} = {\mathbb S}_{i j m q}$, $F_{rs}$ is given by \eqref{eq:rot_def_Fij} and $B_{m q}(P,Q)$ is the $(m,q)$-th entry of the symmetric matrix $B(P,Q)$ (for any $P, \, Q \in \mathfrak{so}_n$). 
\label{prop:Gamma_Gameq_first}
\end{proposition}

\noindent
\textbf{Proof.} Recalling the definition \eqref{eq:GCIId_muGam_def} of $\mu^\Gamma$, we have, thanks to \eqref{eq:GCI_Ggamf}
$$ \int_{\mathrm{SO}_n} Q(f^\varepsilon) \, \mu^{\Gamma_{f^\varepsilon}} \, dA = 0. $$
It follows from \eqref{eq:equi_kin} that
$$ \int_{\mathrm{SO}_n} \Big[ \partial_t f^\varepsilon +   (A {\mathbf e}_1) \cdot \nabla_x f^\varepsilon \Big] \, \mu^{\Gamma_{f^\varepsilon}} \, dA = 0. 
$$
Letting $\varepsilon \to 0$ and noting that $\mu^\Gamma$ is a smooth function of $\Gamma$ and that $\Gamma_{f^\varepsilon} \to \Gamma$, we get 
\begin{equation} 
\int_{\mathrm{SO}_n} \Big[ \partial_t f^0 +   (A {\mathbf e}_1) \cdot \nabla_x f^0 \Big] \, \mu^{\Gamma} \, dA = 0. 
\label{eq:Gamma_Gameq_first_prf1}
\end{equation}
With \eqref{eq:equi_Q_express_prf1}, we compute
$$ \partial_t f^0 +   (A {\mathbf e}_1) \cdot \nabla_x f^0 = M_\Gamma \Big\{ \partial_t \rho + (A {\mathbf e}_1) \cdot \nabla_x \rho  + \kappa \rho P_{T_\Gamma} A \cdot \big[ \partial_t \Gamma + \big((A {\mathbf e}_1) \cdot \nabla_x \big) \Gamma \big] \Big\}, $$
where $(A {\mathbf e}_1) \cdot \nabla_x$ stands for the operator $\sum_{i=1}^n (A {\mathbf e}_1)_i \partial_{x_i}$. We insert this expression into~\eqref{eq:Gamma_Gameq_first_prf1} and we make the change of variables $A' = \Gamma^T A$ with $dA = dA'$ owing to the translation invariance of the Haar measure. With \eqref{eq:GCIId_link_mu_mugam} and the fact that $M_\Gamma(A) = M(\Gamma^T A)$ and $P_{T_\Gamma} (\Gamma A) = \Gamma \frac{A-A^T}{2}$, we get (dropping the primes for clarity): 
$$ \int_{\mathrm{SO}_n}  \Big\{ \partial_t \rho + (\Gamma A {\mathbf e}_1) \cdot \nabla_x \rho + \kappa \rho \frac{A-A^T}{2} \cdot \Big( \Gamma^T \big[ \partial_t \Gamma + \big((\Gamma A {\mathbf e}_1) \cdot \nabla_x \big) \Gamma \big] \Big) \Big\} \, M(A) \, \mu(A) \, dA = 0. $$

Now, changing $A$ to $A^T$ in \eqref{eq:GCIId_equa_mu}, we notice that $\mu(A^T)$ and $- \mu(A)$ satisfy the same variational formulation. By uniqueness of its solution, we deduce that $\mu(A^T) = - \mu(A)$.

Then, making the change of variables $A' = A^T= A^{-1}$ and remarking that the Haar measure on $\mathrm{SO}_n$ is invariant by group inversion, we have, thanks to the fact that $M(A) = M(A^T)$, 
$$ \int_{\mathrm{SO}_n}  \Big\{ \partial_t \rho + (\Gamma A^T {\mathbf e}_1) \cdot \nabla_x \rho - \kappa \rho \frac{A-A^T}{2} \cdot \Big( \Gamma^T \big[ \partial_t \Gamma + \big((\Gamma A^T {\mathbf e}_1) \cdot \nabla_x \big) \Gamma \big] \Big) \Big\} \, M(A) \, \mu(A) \, dA = 0. $$
Subtracting the last two equations and halfing the result, we get 
\begin{eqnarray}
&&\hspace{-1cm} 
0= \int_{\mathrm{SO}_n}  \Big\{ (\Gamma \frac{A-A^T}{2} {\mathbf e}_1) \cdot \nabla_x \rho + \kappa \rho \frac{A-A^T}{2} \cdot (\Gamma^T \partial_t \Gamma)  \nonumber \\
&&\hspace{-1cm} 
+ \kappa \rho \frac{A-A^T}{2} \cdot  \Big( \Gamma^T \big[ \big( (\Gamma \frac{A+A^T}{2} {\mathbf e}_1) \cdot \nabla_x \big) \Gamma \big] \Big) \Big\} \, M(A) \, \mu(A) \, dA =: \textrm{\textcircled{a}} + \textrm{\textcircled{b}} + \textrm{\textcircled{c}} . \label{eq:Gamma_Gameq_first_prf2}
\end{eqnarray}
We have 
\begin{eqnarray*} 
(\Gamma \frac{A-A^T}{2} {\mathbf e}_1) \cdot \nabla_x \rho &=& (\frac{A-A^T}{2} {\mathbf e}_1) \cdot (\Gamma^T \nabla_x \rho) = 2 \frac{A-A^T}{2} \cdot \big( \Gamma^T (\nabla_x \rho \otimes {\mathbf e}_1)  \big) \\
&=& \frac{A-A^T}{2} \cdot \big( \Gamma^T (\nabla_x \rho \otimes {\mathbf e}_1) -  (\nabla_x \rho \otimes {\mathbf e}_1)^T \Gamma \big), 
\end{eqnarray*}
where the first two dots are vector inner products and the last two ones are matrix inner products. The factor $2$ in the last expression of the first line arises because of the definition~\eqref{eq:rot_inner_product} of the matrix inner product. Finally, the second line is just a consequence of the fact that $A - A^T$ is an antisymmetric matrix. Thus, the first two terms in \eqref{eq:Gamma_Gameq_first_prf2} can be written as 
\begin{eqnarray}
\textrm{ \textcircled{a}} + \textrm{\textcircled{b}} &=& \int_{\mathrm{SO}_n} \frac{A-A^T}{2} \cdot \big[  \Gamma^T (\nabla_x \rho \otimes {\mathbf e}_1) -  (\nabla_x \rho \otimes {\mathbf e}_1)^T \Gamma + \kappa \rho \Gamma^T \partial_t \Gamma \Big] \, M(A) \, \mu(A) \, dA \nonumber \\
&=& \int_{\mathrm{SO}_n} (A \cdot {\mathbb P}) \, M(A) \, \mu(A) \, dA = L({\mathbb P}). \label{eq:Gamma_Gameq_first_prf3}
\end{eqnarray}

To compute \textcircled{c}, we first state the following lemma: 

\begin{lemma}
For a vector $w \in {\mathbb R}^n$, we have  
\begin{equation}
\big( \Gamma^T (w \cdot \nabla_x) \Gamma \big)_{ij} = \sum_{k, \ell = 1}^n \Gamma_{ki} w_\ell  \frac{ \partial \Gamma_{kj}}{\partial x_\ell}. 
\label{eq:Gamma_DGamw}
\end{equation}
\label{lem:Gamma_DGamw}
\end{lemma}

\noindent
This lemma is not totally obvious as the differentiation is that of a matrix in $\mathrm{SO}_n$. It is proved in \cite{degond2023radial} in its adjoint form (with $\Gamma^T$ to the right of the directional derivative). The proof of the present result is similar and is skipped.

Then, applying this lemma, we have
\begin{eqnarray*}
&&\hspace{-1cm}
\Big( \Gamma^T \big[ \big( (\Gamma \frac{A+A^T}{2} {\mathbf e}_1) \cdot \nabla_x \big) \Gamma \big] \Big)_{ij} = \sum_{k=1}^n \Gamma_{ki} \sum_{\ell, m, q = 1}^n \Gamma_{\ell m} \frac{A_{mq}+ A_{qm}}{2} e_{1 \, q} \frac{\partial \Gamma_{kj}}{\partial x_\ell} \\
&&\hspace{2cm}
= \sum_{m, q = 1}^n \frac{A_{mq}+ A_{qm}}{2} \, \tilde {\mathbb S}_{ijmq} = \sum_{m, q = 1}^n \frac{A_{mq}+ A_{qm}}{2} \, {\mathbb S}_{ijmq}, 
\end{eqnarray*}
with 
$$ \tilde {\mathbb S}_{ijmq} = \sum_{k, \ell = 1}^n \Gamma_{ki} \Gamma_{\ell m} e_{1 \, q} \frac{\partial \Gamma_{kj}}{\partial x_\ell} \quad \mathrm{ and } \quad {\mathbb S}_{ijmq} = \frac{\tilde {\mathbb S}_{ijmq} + \tilde {\mathbb S}_{ijqm}}{2}. $$
Thus, 
$$ 
\textrm{\textcircled{c}} = \kappa \rho \sum_{i,j,m,q = 1}^n \int_{\mathrm{SO}_n} \Big( \frac{A - A^T}{2} \Big)_{ij}  \Big( \frac{A + A^T}{2} \Big)_{mq} \, M(A) \, \mu(A) \, dA \, \, {\mathbb S}_{ijmq}. $$
Now, because $\mu(A)_{rs} = (\mu(A) \cdot F_{rs})$ and ${\mathbb S}_{ijmq}$ is antisymmetric with respect to $(i,j)$, the $(r,s)$ entry of the matrix \textcircled{c} is given by 
\begin{eqnarray} 
\textrm{\textcircled{c}}_{rs} &=& \kappa \rho \sum_{m,q = 1}^n \int_{\mathrm{SO}_n} (A \cdot {\mathbb S}_{\cdot \cdot mq}) \, (\mu(A) \cdot F_{rs}) \, \Big( \frac{A + A^T}{2} \Big)_{mq} \, M(A)  \, dA \nonumber \\
&=& \kappa \rho \sum_{m,q = 1}^n B_{mq}({\mathbb S}_{\cdot \cdot mq}, F_{rs}). \label{eq:Gamma_Gameq_first_prf4}
\end{eqnarray}
Now, combining \eqref{eq:Gamma_Gameq_first_prf3} and \eqref{eq:Gamma_Gameq_first_prf4}, we get \eqref{eq:Gamma_Gameq_first}. \endproof

\subsection{Expressions of the linear map $L$ and bilinear map $B$}
\label{subsec:Gamma_express_LB}

Now, we give expressions of $L$ and $B$ defined in~\eqref{eq:Gamm_Ldef} and~\eqref{eq:Gamma_Bdef}.

\begin{proposition} (i) We have 
\begin{equation}
L(P) = C_2 \, P, \quad \forall P \in \mathfrak{so}_n , 
\label{eq:Gamma_L(P)_express}
\end{equation}
with 
\begin{eqnarray}
&&\hspace{-1.2cm}
C_2 = \frac{2}{n(n-1)} \int_{\mathrm{SO}_n} (\mu(A) \cdot A) \, M(A) \, dA \label{eq:Gamma_C2_express1} \\
&&\hspace{-1.2cm}
= - \frac{2}{n(n-1)}  \frac{ \displaystyle \int_{{\mathcal T}} \Big( \sum_{k=1}^p \alpha_k(\Theta) \sin \theta_k \Big) \, m(\Theta) \, d \Theta}
{ \displaystyle \int_{{\mathcal T}} m(\Theta) \, d \Theta}, \label{eq:Gamma_C2_express2}
\end{eqnarray}
where $m$ is given by \eqref{eq:GCIId_m_def}. 

\smallskip
\noindent
(ii) We have 
\begin{equation}
B(P,Q) = C_3 \mathrm{Tr}(PQ) \mathrm{I} + C_4 \Big( \frac{PQ+QP}{2} - \frac{1}{n} \mathrm{Tr}(PQ) \mathrm{I} \Big) \,\quad \forall P, \, Q \in \mathfrak{so}_n , 
\label{eq:Gamma_B(P,Q)_express}
\end{equation}
with 
\begin{eqnarray}
&&\hspace{-1cm}
C_3 = - \frac{1}{n^2(n-1)} \int_{\mathrm{SO}_n} (\mu(A) \cdot A) \, \mathrm{Tr} (A) \, M(A) \, dA \label{eq:Gamma_C3_express1} \\
&&\hspace{-1cm}
=\frac{1}{n^2(n-1)} \frac{ \displaystyle \int_{{\mathcal T}} \Big( \sum_{k=1}^p \alpha_k(\Theta) \sin \theta_k \Big) \, \Big( 2 \sum_{k=1}^n \cos \theta_k + \epsilon_n \Big) \, m(\Theta) \, d \Theta}
{ \displaystyle \int_{{\mathcal T}} m(\Theta) \, d \Theta}, \nonumber \\
&&\hspace{-1cm}
\label{eq:Gamma_C3_express2}
\end{eqnarray}
and
\begin{equation}
C_4 = \frac{2n}{n^2-4} \big( - 2 C_3 + C'_4 \big), 
\label{eq:Gamma_C4_express0}
\end{equation}
where
\begin{eqnarray}
&&\hspace{-1cm}
C'_4 = \frac{1}{n(n-1)} \int_{\mathrm{SO}_n} \mathrm{Tr} \Big\{ \mu(A) \Big( \frac{A+A^T}{2} \Big) \Big(\frac{A-A^T}{2} \Big) \Big\} \, M(A) \, dA \label{eq:Gamma_C4_express1} \\
&&\hspace{-1cm}
=\frac{2}{n(n-1)} \, \frac{ \displaystyle \int_{{\mathcal T}} \Big( \sum_{k=1}^p \alpha_k(\Theta) \sin \theta_k \cos \theta_k \Big) \, m(\Theta) \, d \Theta}
{ \displaystyle \int_{{\mathcal T}} m(\Theta) \, d \Theta}, \label{eq:Gamma_C4_express2}
\end{eqnarray}
and where $\epsilon_n$ is given by \eqref{eq:Gamma_eps_def}. 
\label{prop:Gamma_L(P)_B(P,Q)_express}
\end{proposition}

\noindent
\textbf
{Proof.} The proof of \eqref{eq:Gamma_L(P)_express} and \eqref{eq:Gamma_B(P,Q)_express} relies on Lie group representations and Schur's Lemma. We refer to \cite[Sect. 6]{degond2021body} for a list of group representation concepts which will be useful for what follows. The proof will follow closely \cite[Sect. 8 \& 9]{degond2021body} but with some differences which will be highlighted when relevant.

\smallskip
\noindent
(i) Proof of \eqref{eq:Gamma_L(P)_express}. Using \eqref{eq:GCIId_conjug_invar_mu_0} and the translation invariance (on both left and right) of the Haar measure, we easily find that 
\begin{equation} 
L(gPg^T) = g L(P) g^T, \quad \forall P \in \mathfrak{so}_n,  \quad \forall g \in \mathrm{SO}_n. 
\label{eq:Gamma_L(P)_B(P,Q)_express_prf1}
\end{equation}
Denote $V = {\mathbb C}^n$ (the standard complex representation of $\mathrm{SO}_n$) and by $\Lambda^2(V) = V \wedge V$ its exterior square. We recall that $\Lambda^2(V) = \mathrm{Span}_{\mathbb C} \{ v \wedge w \, | \, (v,w) \in V^2 \}$ where $v \wedge w = v \otimes w - w \otimes v$ is the antisymmetrized tensor product of $v$ and $w$. Clearly, $\Lambda^2(V) = \mathfrak{so}_n{\mathbb C}$ where $\mathfrak{so}_n{\mathbb C}$ is the complexification of $\mathfrak{so}_n$. 
So, by linearity, we can extend~$L$ into a linear map $\Lambda^2(V) \to \Lambda^2(V)$, which still satisfies \eqref{eq:Gamma_L(P)_B(P,Q)_express_prf1} (with now $P \in \Lambda^2(V)$). Thus, $L$ intertwines the representation $\Lambda^2(V)$. 
\begin{itemize}
\item[$\bullet$] For $n \geq 3$ and $n \not = 4$, $\Lambda^2(V)$ is an irreducible representation of $\mathrm{SO}_n$. Thus, by Schur's Lemma, we have \eqref{eq:Gamma_L(P)_express} (details can be found in \cite[Sect. 8]{degond2023radial}). 
\item[$\bullet$] For $n=4$, $\Lambda^2(V)$ is not an irreducible representation of $\mathrm{SO}_n$. Still, \eqref{eq:Gamma_L(P)_express} remains true but its proof requires additional arguments developed in Section~\ref{appsubsec_proof_of_L}. 
\end{itemize}

\medskip
\noindent
Now, we show \eqref{eq:Gamma_C2_express1} and \eqref{eq:Gamma_C2_express2}. Taking the matrix inner product of \eqref{eq:Gamma_L(P)_express} with $P$, we get 
$$ \int_{\mathrm{SO}_n} (A \cdot P) \, (\mu(A) \cdot P) \, M(A) \, dA = C_2 (P \cdot P). $$
For $P=F_{ij}$ with $i \not = j$, this gives
$$ C_2 = \int_{\mathrm{SO}_n} \frac{A_{ij} - A_{ji}}{2} \, \mu(A)_{ij} \, M(A) \, dA. $$
Averaging this formula over all pairs $(i,j)$ with $i \not = j$ leads to~\eqref{eq:Gamma_C2_express1}. Now, thanks to~\eqref{eq:GCIId_conjug_invar_mu_0}, the function $A \mapsto (A \cdot \mu(A)) \, M(A)$ is a class function. So, we can apply Weyl's integration formula \eqref{eq:WIF}. For $A = A_\Theta$, we have,
\begin{equation} 
\mathrm{Tr} (A_\Theta) = 2 \sum_{k=1}^n \cos \theta_k + \epsilon_n, 
\label{eq:Gamma_L(P)_B(P,Q)_express_prf19}
\end{equation}
so that 
\begin{equation} 
M(A_\Theta) = \frac{1}{Z} \exp \Big( \frac{\kappa}{2} \big( 2 \sum_{k=1}^n \cos \theta_k + \epsilon_n \big) \Big). 
\label{eq:Gamma_L(P)_B(P,Q)_express_prf20}
\end{equation}
Besides, dotting \eqref{eq:GCIId:mu_formula} with \eqref{eq:GCIId_tildmuk_equation_prf2}, 
 we get 
\begin{equation}
\big( A_\Theta \cdot \mu(A_\Theta) \big) = - \sum_{k=1}^p \alpha_k(\Theta) \, \sin \theta_k. 
\label{eq:Gamma_L(P)_B(P,Q)_express_prf21}
\end{equation}
Expressing the integral involved in $Z$ (see \eqref{eq:equi_VM}) using Weyl's integration formula \eqref{eq:WIF} as well and collecting \eqref{eq:Gamma_L(P)_B(P,Q)_express_prf20} and \eqref{eq:Gamma_L(P)_B(P,Q)_express_prf21} into \eqref{eq:Gamma_C2_express1}, we get \eqref{eq:Gamma_C2_express2}.

\smallskip
\noindent (ii) Proof of \eqref{eq:Gamma_B(P,Q)_express}. By contrast to \cite[Sect. 9]{degond2023radial}, the bilinear form $B$ is not symmetric. Thus, we decompose 
\begin{eqnarray}
&&\hspace{-1cm} 
B(P,Q) = B_s(P,Q) + B_a(P,Q), \nonumber \\
&&\hspace{-1cm} 
B_s(P,Q) = \frac{1}{2} \big( B(P,Q) + B(Q,P) \big) , \quad B_a(P,Q) = \frac{1}{2} \big( B(P,Q) - B(Q,P) \big). \label{eq:Gamma_L(P)_B(P,Q)_express_prf9}
\end{eqnarray}
Again, using \eqref{eq:GCIId_conjug_invar_mu_0}, we get
\begin{equation} 
B(gPg^T, gQg^T) = g B(P,Q) g^T, \quad \forall P, \, Q \in \mathfrak{so}_n,  \quad \forall g \in \mathrm{SO}_n,
\label{eq:Gamma_L(P)_B(P,Q)_express_prf10}
\end{equation}
and similar for $B_s$ and $B_a$. Both $B_s$ and $B_a$ can be extended by bilinearity to the complexifications of $\mathfrak{so}_n$ and ${\mathcal S}_n$ which are $\Lambda^2(V)$ and $\vee^2(V)$ respectively (where $\vee^2(V) = V \vee V$ is the symmetric tensor square of $V$, spanned by elements of the type $v \vee w = v \otimes w + w \otimes v$, for $v$, $w$ in $V$). By the universal property of the symmetric and exterior products, $B_s$ and~$B_a$ generate linear maps $\tilde B_s$: $\vee^2(\Lambda^2(V)) \to \vee^2(V)$ and $\tilde B_a$: $\Lambda^2(\Lambda^2(V)) \to \vee^2(V)$ which intertwine the $\mathrm{SO}_n$ representations. By decomposing $\vee^2(\Lambda^2(V))$, $\Lambda^2(\Lambda^2(V))$ and $\vee^2(V)$ into irreducible $\mathrm{SO}_n$ representations and applying Schur's lemma, we are able to provide generic expressions of $\tilde B_s$ and $\tilde B_a$. 

For $\tilde B_s$, this decomposition was done in \cite[Sect. 9]{degond2021body} and we get the following result. 
\begin{itemize}
\item[$\bullet$] For $n \geq 3$ and $n \not = 4$, there exists real constants $C_3$ and $C_4$ such that 
\begin{equation}
B_s(P,Q) = C_3 \mathrm{Tr}(PQ) \mathrm{I} + C_4 \Big( \frac{PQ+QP}{2} - \frac{1}{n} \mathrm{Tr}(PQ) \mathrm{I} \Big) \,\quad \forall P, \, Q \in \mathfrak{so}_n . 
\label{eq:Gamma_L(P)_B(P,Q)_express_prf11}
\end{equation}

\item[$\bullet$] For $n=4$, we still have \eqref{eq:Gamma_L(P)_B(P,Q)_express_prf11} but this requires additional arguments developed in Section \ref{appsubsec_proof_of_Bs}. 
\end{itemize}

For $\tilde B_a$, thanks to Pieri's formula \cite[Exercise 6.16]{fulton2013representation} we have $\Lambda^2(\Lambda^2(V)) = {\mathbb S}_{(2,1,1)}$ where ${\mathbb S}_{(2,1,1)}$ denotes the Schur functor (or Weyl module) associated to partition $(2,1,1)$ of $4$. As a Schur functor, ${\mathbb S}_{(2,1,1)}$ is irreducible over $\mathfrak{sl}_n {\mathbb C}$, the Lie algebra of the group of unimodular matrices $\mathrm{SL}_n{\mathbb C}$. We apply Weyl's contraction method \cite[Sect. 19.5]{fulton2013representation} to decompose it into irreducible representations over $\mathfrak{so}_n {\mathbb C}$.

It can be checked that all contractions with respect to any pair of indices of $\Lambda^2(\Lambda^2(V))$ are either $0$ or coincide (up to a sign), with the single contraction ${\mathcal K}$ defined as follows: 
\begin{eqnarray*} 
{\mathcal K}: \, \, \Lambda^2(\Lambda^2(V)) &\to& \Lambda^2(V) \\ 
(v_1 \wedge v_2) \wedge (v_3 \wedge v_4) & \mapsto & (v_1 \cdot v_3) v_2 \wedge v_4 +  (v_2 \cdot v_4) v_1 \wedge v_3 \\
&& \hspace{0cm} -  (v_1 \cdot v_4) v_2 \wedge v_3 - (v_2 \cdot v_3) v_1 \wedge v_4, \quad \forall (v_1, \ldots, v_4) \in V^4. 
\end{eqnarray*}  
${\mathcal K}$ is surjective as soon as $n \geq 3$. Indeed, Let $(e_i)_{i=1}^n$ be the canonical basis of $V$. Then, 
$$ {\mathcal K} \big( (e_i \wedge e_j) \wedge (e_i \wedge e_k) \big) = e_j \wedge e_k, \quad \forall i,j,k \quad \text{all distinct}. $$
We have $\ker {\mathcal K} = {\mathbb S}_{[2,1,1]}$ where ${\mathbb S}_{[2,1,1]}$ denotes the intersection of ${\mathbb S}_{(2,1,1)}$ with all the kernels of contractions with respect to pairs of indices (see \cite[Sect. 19.5]{fulton2013representation}) and consequently 
$$ \Lambda^2(\Lambda^2(V)) \cong {\mathbb S}_{[2,1,1]} \oplus \Lambda^2(V), $$
is a decomposition of $\Lambda^2(\Lambda^2(V))$ in subrepresentations. We must discuss the irreducibility of ${\mathbb S}_{[2,1,1]}$ and $\Lambda^2(V)$ according to the dimension. 
\begin{itemize}
\item[$\bullet$] If $n \geq 7$,  by \cite[Theorem 19.22]{fulton2013representation}, ${\mathbb S}_{[2,1,1]}$ and $\Lambda^2(V)$ are respectively the irreducible representations of $\mathfrak{so}_n {\mathbb C}$ of highest weights $2 {\mathcal L}_1 + {\mathcal L}_2 + {\mathcal L}_3$ and ${\mathcal L}_1 + {\mathcal L}_2$ (where $({\mathcal L}_i)_{i=1}^p$ is the basis of the weight space, i.e. is the dual basis (in $\mathfrak{h}^*_{\mathbb C}$) of the basis $(F_{2i-1,2i})_{i=1}^p$ of $\mathfrak{h}_{\mathbb C}$, the complexification of $\mathfrak{h}$). On the other hand, $\vee^2(V) = {\mathbb S}_{[2]} \oplus {\mathbb C} \, \mathrm{I}$ (where ${\mathbb S}_{[2]}$ is the space of symmetric trace-free matrices) is the decomposition of $\vee^2(V)$ into irreducible representations over $\mathfrak{so}_n {\mathbb C}$ and corresponds to decomposing a symmetric matrix into a trace-free matrix and a scalar one. We note that ${\mathbb S}_{[2]}$ has highest weight $2 {\mathcal L}_1$ while ${\mathbb C} \, \mathrm{I}$ has highest weight $0$. By \cite[Prop. 26.6 \& 27.7]{fulton2013representation} the corresponding representations of $\mathfrak{so}_n$ are irreducible and real. Hence, they are irreducible representations of $\mathrm{SO}_n$. Since 
the weights of ${\mathbb S}_{[2,1,1]}$ and $\Lambda^2(V)$ and those of ${\mathbb S}_{[2]}$ and ${\mathbb C} \, \mathrm{I}$ are different, no irreducible subrepresentation of $\Lambda^2(\Lambda^2(V))$ can be isomorphic to an irreducible subrepresentation of $\vee^2(V)$. Consequently, by Schur's Lemma we have 
\begin{equation}
B_a(P,Q) = 0 \quad \forall P, \, Q \in \mathfrak{so}_n , 
\label{eq:Gamma_L(P)_B(P,Q)_express_prf15}
\end{equation}
\item[$\bullet$] Case $n \in \{3, \ldots, 6 \}$. 
\begin{itemize}
\item[-] Case $n=3$. We have $\Lambda^2(V) \cong V$ by the isomorphism $\eta$: $\Lambda^2(V) \to V$ such that $(\eta(v \wedge w) \cdot z) = \mathrm{det}(v,w,z)$, $\forall (v,w,z) \in V^3$. Consequently $\Lambda^2(\Lambda^2(V)) \cong V$ as well. But $V$ is an irreducible representation of $\mathrm{so}_3{\mathbb C}$ with highest weight~${\mathcal L}_1$. Hence, it 
can be isomorphic to neither ${\mathbb S}_{[2]}$ nor ${\mathbb C} \, \mathrm{I}$. Therefore, by Schur's lemma, \eqref{eq:Gamma_L(P)_B(P,Q)_express_prf15} is true again. 
\item[-] Case $n=4$. By  \cite[p. 297]{fulton2013representation}, the partitions $(2,1,1)$ and $(2)$ are associated in the sense of Weyl. Hence ${\mathbb S}_{[2,1,1]} \cong {\mathbb S}_{[2]}$ as $\mathfrak{so}_4{\mathbb C}$ representations. Since $\vee^2(V)$ decomposes into irreducible representations according to $\vee^2(V) = {\mathbb S}_{[2]} \oplus {\mathbb C} \mathrm{I}$, we see that Schur's lemma allows the possibility of a non-zero $\tilde B_a$ mapping the component ${\mathbb S}_{[2,1,1]}$ of $\Lambda^2(\Lambda^2(V))$ into the component ${\mathbb S}_{[2]}$ of $\vee^2(V)$. Likewise, $\Lambda^2(V)$ is reducible. To show that \eqref{eq:Gamma_L(P)_B(P,Q)_express_prf15} is actually true requires additional arguments which are developed in Section \ref{appsubsec_proof_of_Ba}. 
\item[-] Case $n=5$. Like in the case $n=4$, we find that the partitions $(2,1,1)$ and $(2,1)$ are associated and thus, ${\mathbb S}_{[2,1,1]} \cong {\mathbb S}_{[2,1]}$. The latter is an irreducible representation of $\mathfrak{so}_5{\mathbb C}$ of highest weight $2 {\mathcal L}_1 + {\mathcal L}_2$. Likewise, $\Lambda^2(V)$ is an irreducible representation of highest weight ${\mathcal L}_1 + {\mathcal L}_2$. By \cite[Prop. 26.6]{fulton2013representation} these are real irreducible representation of $\mathfrak{so}_5$. Thus, they are irreducible representations of $\mathrm{SO}_5$. Having different highest weights from those of ${\mathbb S}_{[2]}$ or ${\mathbb C} {\mathrm I}$, by the same argument as in the case $n \geq 7$, \eqref{eq:Gamma_L(P)_B(P,Q)_express_prf15} holds true. 
\item[-] Case $n=6$. In this case, the partition $(2,1,1)$ is self-associated. By \cite[Theorem 19.22 (iii)]{fulton2013representation}, ${\mathbb S}_{[2,1,1]}$ decomposes into the direct sum of two non-isomorphic representations of $\mathfrak{so}_6{\mathbb C}$ of highest weights $2 {\mathcal L}_1 + {\mathcal L}_2 + {\mathcal L}_3$ and $2 {\mathcal L}_1 + {\mathcal L}_2 - {\mathcal L}_3$. On the other hand, $\Lambda^2(V)$ is an irreducible representation of highest weight ${\mathcal L}_1 + {\mathcal L}_2$. By \cite[27.7]{fulton2013representation} the corresponding representations of $\mathfrak{so}_6$ are irreducible and real and thus generate irreducible representations of $\mathrm{SO}_6$. Having different weights from those of ${\mathbb S}_{[2]}$ and ${\mathbb C} {\mathrm I}$, the same reasoning applies again and \eqref{eq:Gamma_L(P)_B(P,Q)_express_prf15} holds true. 
\end{itemize}
\end{itemize}
Finally by adding \eqref{eq:Gamma_L(P)_B(P,Q)_express_prf10} and \eqref{eq:Gamma_L(P)_B(P,Q)_express_prf15}, we get \eqref{eq:Gamma_B(P,Q)_express}, which finishes the proof of Point (ii). 

\medskip
\noindent
We now show \eqref{eq:Gamma_C3_express1} and \eqref{eq:Gamma_C3_express2}. Taking the trace of \eqref{eq:Gamma_B(P,Q)_express}, we get
$$ C_3 = \frac{ \mathrm{Tr} \big( B(P,Q) \big) }{n \, \mathrm{Tr} (PQ) }, \quad \forall P, \, Q \in \mathfrak{so}_n. $$
Taking $P = Q = F_{ij}$ for $i \not = j$ and owing to the fact that $\mathrm{Tr} (F_{ij}^2) = -2 F_{ij} \cdot F_{ij} = -2$, we find 
$$ C_3 = - \frac{1}{2n} \int_{\mathrm{SO}_n} \frac{A_{ij} - A_{ji}}{2} \, \mu(A)_{ij} \, \mathrm{Tr}(A) \, M(A) \, dA. $$
Now, averaging this formula over all pairs $(i,j)$ with $i \not = j$ leads to~\eqref{eq:Gamma_C3_express1}. Again, the function $A \mapsto (A \cdot \mu(A)) \, \mathrm{Tr}(A) \, M(A)$ is a class function and Weyl's integration formula~\eqref{eq:WIF} can be applied. Inserting  \eqref{eq:Gamma_L(P)_B(P,Q)_express_prf19}, \eqref{eq:Gamma_L(P)_B(P,Q)_express_prf20}, \eqref{eq:Gamma_L(P)_B(P,Q)_express_prf21} into \eqref{eq:Gamma_C3_express1} leads to \eqref{eq:Gamma_C3_express2}.

\medskip
\noindent
We finish with showing \eqref{eq:Gamma_C4_express0}-\eqref{eq:Gamma_C4_express2}. We now insert $P=F_{ij}$ and $Q=F_{i \ell}$ with $i \not = j$, $i \not = \ell$ and $j \not = \ell$. Then, $F_{ij} \cdot F_{i \ell} = 0$, so that $\mathrm{Tr}(F_{ij} F_{i \ell}) = 0$. A small computation shows that 
$$ \frac{F_{ij} F_{i \ell} + F_{i \ell} F_{ij}}{2} = - \frac{E_{j \ell} + E_{\ell j}}{2}, $$
where $E_{j \ell}$ is the matrix with $(j ,\ell)$ entry equal to $1$ and the other entries equel to $0$. 
It follows that 
\begin{equation} 
- C_4 \frac{E_{j \ell} + E_{\ell j}}{2} = \int_{\mathrm{SO}_n} \Big( \frac{A-A^T}{2} \Big)_{ij} \, \mu(A)_{i \ell} \, \frac{A + A^T}{2} \, M(A) \, dA. 
\label{eq:Gamma_L(P)_B(P,Q)_express_prf22}
\end{equation}
Now, taking $P=Q=F_{ij}$ with $i \not = j$, and noting that $F_{ij}^2 = -(E_{ii}+E_{jj})$ and $\mathrm{Tr} (F_{ij}^2) = -2$, we get 
\begin{equation} 
- 2 C_3 \mathrm{I} + C_4 \Big( - (E_{ii} + E_{jj}) + \frac{2}{n} \mathrm{I} \Big) = \int_{\mathrm{SO}_n} \Big( \frac{A-A^T}{2} \Big)_{ij} \, \mu(A)_{ij} \, \frac{A + A^T}{2} \, M(A) \, dA. 
\label{eq:Gamma_L(P)_B(P,Q)_express_prf23}
\end{equation}
Take $i$, $j$ with $i \not = j$ fixed. For any $\ell \not = j$, taking the $(\ell,j)$-th entry of \eqref{eq:Gamma_L(P)_B(P,Q)_express_prf22}, we get
\begin{equation} 
\frac{C_4}{2} = \int_{\mathrm{SO}_n} \mu(A)_{i \ell} \Big(\frac{A+A^T}{2} \Big)_{\ell j} \Big(\frac{A-A^T}{2}\Big)_{ji} \, M(A) \, dA . \label{eq:Gamma_L(P)_B(P,Q)_express_prf24}
\end{equation}
Likewise, taking the $(j,j)$-th entry of \eqref{eq:Gamma_L(P)_B(P,Q)_express_prf23}, we get
\begin{equation} 
2 C_3 - C_4 \Big( - 1 + \frac{2}{n} \Big) = \int_{\mathrm{SO}_n} \mu(A)_{ij} \Big(\frac{A+A^T}{2} \Big)_{jj} \Big(\frac{A-A^T}{2}\Big)_{ji} \, M(A) \, dA. 
\label{eq:Gamma_L(P)_B(P,Q)_express_prf25}
\end{equation}
Now, summing \eqref{eq:Gamma_L(P)_B(P,Q)_express_prf24} over $\ell \not \in \{i, j\}$ and adding \eqref{eq:Gamma_L(P)_B(P,Q)_express_prf25}, we find
$$
2 C_3 + C_4 \frac{n^2 - 4}{2n} = \sum_{\ell=1}^n \int_{\mathrm{SO}_n} \mu(A)_{i \ell} \Big(\frac{A+A^T}{2} \Big)_{\ell j} \Big(\frac{A-A^T}{2}\Big)_{ji} \, M(A) \, dA. 
$$
Then, averaging this equation over all pairs $(i,j)$ such that $i \not = j$, we get \eqref{eq:Gamma_C4_express0} with $C'_4$ given by \eqref{eq:Gamma_C4_express1}. Again, one can check that the function $A \mapsto \mathrm{Tr} \{ \mu(A) ( \frac{A+A^T}{2} ) (\frac{A-A^T}{2}) \}  M(A)$ is a class function. Thanks to \eqref{eq:GCIId:mu_formula}, \eqref{eq:GCIId_tildmuk_equation_prf2} and to the fact that 
\begin{equation} 
\frac{A_\Theta + A_\Theta^T}{2} = \sum_{k=1}^p \cos \theta_k (E_{2k-1 \, 2k-1} + E_{2k \, 2k}) +\epsilon_n E_{n\, n}, 
\label{eq:Gamma_L(P)_B(P,Q)_express_prf26}
\end{equation}
we get 
$$ \mu(A_\Theta) \Big( \frac{A_\Theta+A_\Theta^T}{2} \Big) \Big( \frac{A_\Theta-A_\Theta^T}{2} \Big) = \sum_{k=1}^p \alpha_k(\Theta) \, \cos \theta_k  \, \sin \theta_k \,  (E_{2k-1 \, 2k-1} + E_{2k \, 2k}), $$
and thus, 
$$ \mathrm{Tr} \Big\{ \mu(A_\Theta) \Big( \frac{A_\Theta+A_\Theta^T}{2} \Big) \Big( \frac{A_\Theta-A_\Theta^T}{2} \Big) \Big\} = 2 \sum_{k=1}^p \alpha_k(\Theta) \, \cos \theta_k  \, \sin \theta_k, $$
which leads to \eqref{eq:Gamma_C4_express2}. \endproof

\subsection{Final step: establishment of \eqref{eq:Gamma_Gameq_second}}
\label{subsec:Gamma_final_step}

Now, we can finish with the following 

\begin{proposition}
The functions $\rho$ and $\Gamma$ involved in \eqref{eq:equi_f0express} satisfy \eqref{eq:Gamma_Gameq_second}. The constants $c_2$ and $c_4$ are given by 
\begin{eqnarray}
&&\hspace{-1cm}
c_2 = - \frac{2}{C_2} \Big(C_3 - \frac{C_4}{n} \Big) \label{eq:Gamma:c2_express} \\
&&\hspace{-1cm}
= \frac{1}{n^2-4} \frac{\displaystyle \int \Big[ n (\mu(A) \cdot A) \mathrm{Tr}(A) + 2 \mathrm{Tr} \Big(\mu(A) \frac{A+A^T}{2} \frac{A-A^T}{2} \Big) \Big] M(A) dA}
{\displaystyle \int (\mu(A) \cdot A) M(A) dA} \label{eq:Gamma:c2_express_2}\\
&&\hspace{-1cm}
c_4 = \frac{C_4}{2 C_2} \label{eq:Gamma:c4_express} \\
&&\hspace{-1cm}
= \frac{1}{2(n^2-4)} \frac{\displaystyle \int \Big[ 2 (\mu(A) \cdot A) \mathrm{Tr}(A) + n \mathrm{Tr} \Big(\mu(A) \frac{A+A^T}{2} \frac{A-A^T}{2} \Big) \Big] M(A) dA}
{\displaystyle \int (\mu(A) \cdot A) M(A) dA} \label{eq:Gamma:c4_express_2}\\
\end{eqnarray}
where $C_2$, $C_3$ and $C_4$ are given in Prop. \ref{prop:Gamma_L(P)_B(P,Q)_express}. Then, $c_2$, $c_3$ and $c_4$ are given by \eqref{eq:Gamma:c2_express_3},~\eqref{eq:Gamma:c3_express} and \eqref{eq:Gamma:c4_express_3} respectively. 
\label{prop:Gamma_Gameq_second}
\end{proposition}

\noindent
\textbf{Proof.} We simplify \eqref{eq:Gamma_Gameq_first} in light of \eqref{eq:Gamma_L(P)_express}
and \eqref{eq:Gamma_B(P,Q)_express}. From \eqref{eq:Gamma_L(P)_express}, we have $L({\mathbb P}) = C_2 {\mathbb P}$. With \eqref{eq:Gamma_Pdef} and recalling that $\Omega_1$ is defined by \eqref{eq:cont_Omdef}, this leads to
\begin{eqnarray} 
\rho \partial_t \Gamma &=& \frac{1}{\kappa}  \Big[ - \big( ( \nabla_x \rho \otimes {\mathbf e}_1 ) \Gamma^T - \Gamma ({\mathbf e}_1 \otimes \nabla_x \rho) \big) + \frac{1}{C_2} \, \Gamma L({\mathbb P}) \Gamma^T \Big] \, \Gamma \nonumber \\
&=& \Big[ - \frac{1}{\kappa} \nabla_x \rho \wedge \Omega_1 + \frac{1}{C_2 \kappa} \, \Gamma L({\mathbb P}) \Gamma^T \Big] \, \Gamma. \label{eq:Gamma_Gameq_second_prf-1}
\end{eqnarray}
Now, we compute $\Gamma L({\mathbb P}) \Gamma^T$ thanks to \eqref{eq:Gamma_Gameq_first} and the expressions \eqref{eq:Gamma_B(P,Q)_express} of $B$ and \eqref{eq:Gamma_Sdef} of~${\mathbb S}$. We have 
$$ B(P,Q) = C'_3 \mathrm{Tr}(PQ) \mathrm{I} + C_4 \frac{PQ+QP}{2}, \quad \forall P, \, Q \in \mathfrak{so}_n, $$
with $C'_3 = C_3 - \frac{C_4}{n}$. This leads to 
\begin{equation} 
\sum_{m,q=1}^n B_{mq} ({\mathbb S}_{\cdot \cdot m q},F_{rs}) = \sum_{m,q=1}^n \Big[ C'_3 \mathrm{Tr}({\mathbb S}_{\cdot \cdot m q}F_{rs}) \delta_{mq} + C_4 \frac{ \big({\mathbb S}_{\cdot \cdot m q} \, F_{rs} + F_{rs} \, {\mathbb S}_{\cdot \cdot m q} \big)_{mq}}{2} \Big]. 
\label{eq:Gamma_Gameq_second_prf0}
\end{equation}
We compute, for $m$, $q$, $r$, $s$, $u$, $v$ in $\{1, \ldots, n\}$: 
$$\big({\mathbb S}_{\cdot \cdot m q} \, F_{rs}\big)_{uv} = {\mathbb S}_{u r m q} \delta_{sv} - {\mathbb S}_{u s mq} \delta_{rv}, \quad 
\big(F_{rs} \, {\mathbb S}_{\cdot \cdot m q}\big)_{uv} = {\mathbb S}_{s v m q} \delta_{ru} - {\mathbb S}_{r v m q} \delta_{su}. 
$$
Hence, $ \mathrm{Tr}({\mathbb S}_{\cdot \cdot m q}F_{rs}) = - 2 {\mathbb S}_{r s m q}$ and 
\begin{eqnarray} 
\sum_{m,q=1}^n \mathrm{Tr}({\mathbb S}_{\cdot \cdot m q}F_{rs}) \delta_{mq} &=& -2 \sum_{m=1}^n {\mathbb S}_{r s m m} = - 2 \sum_{m,k,\ell=1}^n \Gamma_{kr} \frac{\partial \Gamma_{ks}}{\partial x_\ell} \Gamma_{\ell m} e_{1 \, m} \nonumber \\
&=& - 2 \sum_{k,\ell=1}^n \Gamma_{kr} \frac{\partial \Gamma_{ks}}{\partial x_\ell} \Omega_{1 \ell}
= - 2 \sum_{k=1}^n \Gamma_{kr} \, (\Omega_1 \cdot \nabla_x) \Gamma_{ks},
\label{eq:Gamma_Gameq_second_prf1}
\end{eqnarray}
where we have used Lemma \ref{lem:Gamma_DGamw}. On the other hand, we have 
\begin{eqnarray}
&&\hspace{-1cm}
\sum_{m,q=1}^n \frac{ \big({\mathbb S}_{\cdot \cdot m q} \, F_{rs} + F_{rs} \, {\mathbb S}_{\cdot \cdot m q} \big)_{mq}}{2} = \frac{1}{2} \sum_{m,q=1}^n  \big( {\mathbb S}_{m r m q} \delta_{sq} - {\mathbb S}_{m s m q} \delta_{rq} + {\mathbb S}_{s q m q} \delta_{rm} - {\mathbb S}_{r q m q} \delta_{sm} \big) \nonumber \\
&&\hspace{0cm}
= \frac{1}{2} \sum_{m=1}^n  \big( {\mathbb S}_{m r m s} - {\mathbb S}_{m s m r} + {\mathbb S}_{s m r m} - {\mathbb S}_{r m s m} \big) = \sum_{m=1}^n  \big( {\mathbb S}_{m r m s} - {\mathbb S}_{m s m r} \big) \nonumber \\
&&\hspace{0cm}
= \frac{1}{2} \sum_{k, \ell, m=1}^n \Big( \Gamma_{km} \frac{\partial \Gamma_{kr}}{\partial x_\ell} \big( \Gamma_{\ell m} e_{1 \, s} + \Gamma_{\ell s} e_{1 \, m} \big) - 
\Gamma_{km} \frac{\partial \Gamma_{ks}}{\partial x_\ell} \big( \Gamma_{\ell m} e_{1 \, r} + \Gamma_{\ell r} e_{1 \, m} \big) \Big)
. \label{eq:Gamma_Gameq_second_prf2}
\end{eqnarray}
Inserting \eqref{eq:Gamma_Gameq_second_prf1} and \eqref{eq:Gamma_Gameq_second_prf2} into \eqref{eq:Gamma_Gameq_second_prf0} and using \eqref{eq:Gamma_Gameq_first} and Lemma \ref{lem:Gamma_DGamw} leads to
\begin{eqnarray*}
&&\hspace{-1cm}
L({\mathbb P})_{rs} = - \kappa \rho \Big\{ -2 C'_3 \sum_{k=1}^n \Gamma_{kr} \, (\Omega_1 \cdot \nabla_x) \Gamma_{ks} \\
&&\hspace{-0.5cm}
+ \frac{C_4}{2} \sum_{k, \ell, m=1}^n \Big( \Gamma_{km} \frac{\partial \Gamma_{kr}}{\partial x_\ell} \big( \Gamma_{\ell m} e_{1 \, s} + \Gamma_{\ell s} e_{1 \, m} \big) - 
\Gamma_{km} \frac{\partial \Gamma_{ks}}{\partial x_\ell} \big( \Gamma_{\ell m} e_{1 \, r} + \Gamma_{\ell r} e_{1 \, m} \big) \Big) \Big\}. 
\end{eqnarray*}
Thus,
\begin{eqnarray*}
&&\hspace{-1cm}
\big( \Gamma L({\mathbb P}) \Gamma^T \big)_{ij} = \sum_{r, s = 1}^n \Gamma_{ir} L({\mathbb P})_{rs} \Gamma_{js} \\
&&\hspace{-0.5cm}
= - \kappa \rho \Big\{ - 2 C'_3 \sum_{s=1}^n (\Omega_1 \cdot \nabla_x) \Gamma_{is} \, \Gamma_{js} + \frac{C_4}{2} \Big( \sum_{k, r, s=1}^n \big( \frac{\partial \Gamma_{kr}}{\partial x_k} \Gamma_{ir} \Gamma_{j s} e_{1 \, s} - \frac{\partial \Gamma_{ks}}{\partial x_k} \Gamma_{ir} \Gamma_{j s} e_{1 \, r} \big)  \\
&&\hspace{5cm}
 + \sum_{k, m, r=1}^n \big( \Gamma_{km} \frac{\partial \Gamma_{kr}}{\partial x_j} \Gamma_{ir} e_{1 \, m} - \Gamma_{km} \frac{\partial \Gamma_{kr}}{\partial x_i} \Gamma_{jr} e_{1 \, m} \big) \Big) \Big\}, 
\end{eqnarray*}
where we have used Lemma \ref{lem:Gamma_DGamw} again as well as that $\sum_{r=1}^n \Gamma_{jr} \Gamma_{ir} = \delta_{ij}$ and similar identities. We note that 
$$ \sum_{k, r = 1}^n \frac{\partial \Gamma_{kr}}{\partial x_k} \Gamma_{ir} = \big( \Gamma (\nabla_x \cdot \Gamma) \big)_i, $$
with $\nabla_x \cdot \Gamma$ being the divergence of the tensor $\Gamma$ defined in the statement of Prop. \ref{prop:Gamma_Gameq_second}. On the other hand, since 
$$ \sum_{k=1}^n \Gamma_{km} \frac{\partial \Gamma_{kr}}{\partial x_j} = - \sum_{k=1}^n  \frac{\partial \Gamma_{km}}{\partial x_j} \Gamma_{kr}, $$
we have 
$$ \sum_{k, m, r=1}^n \Gamma_{km} \frac{\partial \Gamma_{kr}}{\partial x_j} \Gamma_{ir} e_{1 \, m} = - \sum_{m=1}^n \frac{\partial \Gamma_{im}}{\partial x_j} e_{1 \, m} = - \frac{\partial \Omega_{1 \, i}}{\partial x_j}, $$
because ${\mathbf e}_1$ does not depend on space nor time. Thus, we get 
\begin{equation}
\Gamma L({\mathbb P}) \Gamma^T =  - \kappa \rho \Big\{ - 2 C'_3 (\Omega_1 \cdot \nabla_x) \Gamma \, \Gamma^T+ \frac{C_4}{2}  \big( \Gamma (\nabla_x \cdot \Gamma) \big) \wedge \Omega_1   
 + \nabla_x \wedge \Omega_1 \Big) \Big\}.  
\label{eq:Gamma_Gameq_second_prf3}
\end{equation}
Inserting \eqref{eq:Gamma_Gameq_second_prf3} into \eqref{eq:Gamma_Gameq_second_prf-1} yields 
\eqref{eq:Gamma_Gameq_second} with \eqref{eq:Gamma:W_express} and formulas \eqref{eq:Gamma:c2_express}, \eqref{eq:Gamma:c3_express}, \eqref{eq:Gamma:c4_express} for the coefficients. Formulas \eqref{eq:Gamma:c2_express_2}, \eqref{eq:Gamma:c2_express_3}, \eqref{eq:Gamma:c4_express_2}, \eqref{eq:Gamma:c4_express_3} follow with a little bit of algebra from the formulas of Prop. \ref{prop:Gamma_L(P)_B(P,Q)_express}.\endproof

\subsection{Case of dimension $n=4$}
\label{appsec_dim=4}

In this subsection, we prove Prop. \ref{prop:Gamma_L(P)_B(P,Q)_express} for the special case $n=4$. 

\subsubsection{Proof of \eqref{eq:Gamma_L(P)_express}}
\label{appsubsec_proof_of_L}

If $n=4$, there exists an automorphism $\beta$ of $\Lambda^2(V)$ (with $V = {\mathbb C}^4$) characterized by the following relation (see \cite[Sect. 8]{degond2021body})
\begin{equation} 
\big(\beta(v_1 \wedge v_2)\big) \cdot (v_3 \wedge v_4) = \mathrm{det} (v_1, v_2, v_3, v_4), \quad \forall (v_1, \ldots, v_4) \in V^4, 
\label{eq:dim4_L_prf0}
\end{equation}
and where the dot product in $\Lambda^2(V)$ extends that of $\mathfrak{so}_4$, namely $(v_1 \wedge v_2) \cdot (v_3 \wedge v_4) = (v_1 \cdot v_3) (v_2 \cdot v_4)  -  (v_1 \cdot v_4) (v_2 \cdot v_3)$. In addition, $\beta$ intertwines $\Lambda^2(V)$ and itself as $\mathrm{SO}_4$ representations. It is also an involution (i.e. $\beta^{-1} = \beta$) with 
\begin{equation} 
\beta(F_{12}) = F_{34}, \quad \beta(F_{13}) = -F_{24}, \quad \beta (F_{14}) = F_{23}. 
\label{eq:dim4_L_prf1}
\end{equation}
The map $\beta$ has eigenvalues $\pm 1$ with associated eigenspaces $\Lambda_\pm$ such that
$$ \Lambda_\pm = \mathrm{Span} \{ F_{12} \pm F_{34}, \, F_{13} \mp F_{24}, \, F_{14} \pm F_{23} \}. $$
Thus, $\mathrm{dim} \Lambda_\pm = 3$, each $\Lambda_\pm$ is an irreducible representation of $\mathrm{SO}_4$ and $\beta$ is the orthogonal symmetry in $\Lambda_+$. Thanks to this, it was shown in \cite[Sect. 8]{degond2021body} that for a map $L$: $\mathfrak{so}_4 \to \mathfrak{so}_4$ satisfying \eqref{eq:Gamma_L(P)_B(P,Q)_express_prf1}, there exist two real constants $C_2$, $C'_2$ such that 
$$ L(P)= C_2 \, P + C'_2 \, \beta(P), \quad \forall P \in \mathfrak{so}_4. $$
In \cite[Sect. 8]{degond2021body} it was used that $L$ commutes, not only with conjugations with elements of $\mathrm{SO}_4$ (inner automorphisms) through \eqref{eq:Gamma_L(P)_B(P,Q)_express_prf1}, but also with those of $\mathrm{O}_4 \setminus \mathrm{SO}_4$ (outer automorphisms). This simple observation allowed us to conclude that $C'_2 = 0$. However, here, it is not true any more that $L$ commutes with outer automorphisms. Hence, we have to look for a different argument to infer that $C'_2 = 0$. This is what we develop now.

Taking the inner product of $L(P)$ with $\beta(P)$ and using \eqref{eq:Gamm_Ldef}, we get 
$$ \int_{\mathrm{SO}_4} (A \cdot P) \, (\mu(A) \cdot \beta(P)) \, M(A) \, dA = C_2 (P \cdot \beta(P)) + C'_2 (\beta(P), \beta(P)). $$
We apply this equality with $P=F_{ij}$ for $i \not = j$ and we note that $(\beta(F_{ij}) \cdot F_{ij})=0$, $\forall i, j$, and that $(\beta(P). \beta(P)) = (P.P)$ and $(\mu(A) \cdot \beta(P)) = (\beta \circ \mu(A) \cdot P)$ due to the fact that $\beta$ is an orthogonal self-adjoint transformation of $\mathfrak{so}_4$. This leads to 
$$ C'_2 = \int_{\mathrm{SO}_4} \frac{A_{ij} - A_{ji}}{2} \, (\beta \circ \mu (A))_{ij} \, M(A) \, dA. $$
Averaging over $(i,j)$ such that $i \not = j$, we get 
\begin{equation} 
C'_2 = \frac{1}{6} \int_{\mathrm{SO}_4}  \Big( \frac{A-A^T}{2} \cdot \big(\beta \circ \mu(A)\big) \Big) \, M(A) \, dA. 
\label{eq:dim4_L_prf1.5}
\end{equation}
Due to the fact that $\beta$ is an intertwining map, the function $A \mapsto ( A \cdot (\beta \circ \mu(A)) ) \, M(A)$ is a class function. Thus, we can apply Weyl's integration formula \eqref{eq:WIF}. Using \eqref{eq:GCIId_tildmuk_equation_prf2}, \eqref{eq:GCIId:mu_generic}, \eqref{eq:Gamma_L(P)_B(P,Q)_express_prf20}, \eqref{eq:dim4_L_prf1} and the fact that $(F_{ij})_{i<j}$ forms an orthonormal basis of $\mathfrak{so}_4$, we get 
\begin{equation} 
C'_2 = - \,  \frac{\displaystyle \int_{{\mathcal T}_2} \big( \sin \theta_1 \alpha_2 (\Theta) + \sin \theta_2 \alpha_1 (\Theta) \big) \, m(\Theta) \, d \Theta}
{\displaystyle 6 \int_{{\mathcal T}_2} m(\Theta) \, d \Theta}, 
\label{eq:dim4_L_prf2}
\end{equation}
with $\Theta = (\theta_1, \theta_2)$ and $ m(\Theta) = e^{\kappa (\cos \theta_1 + \cos \theta_2)} \, (\cos \theta_1 - \cos \theta_2)^2$. Now, we define 
\begin{equation} 
\tilde \alpha_1 (\theta_1, \theta_2) = - \alpha_1 (-\theta_1, \theta_2), \quad \tilde \alpha_2 (\theta_1, \theta_2) = \alpha_2 (-\theta_1, \theta_2). 
\label{eq:dim4_tildealpha_def}
\end{equation}
We can check that $\tilde \alpha = (\tilde \alpha_1, \tilde \alpha_2)$ belongs to ${\mathcal V}$. Furthermore, using $\tilde \tau$ as a test function where~$\tilde \tau$ is obtained from $\tau$ by the same formulas \eqref{eq:dim4_tildealpha_def}, we realize that $\tilde \alpha$ is another solution of the variational formulation \eqref{eq:GCIId_varform_alpha}. Since the solution of \eqref{eq:GCIId_varform_alpha} is unique and equal to $\alpha$, we deduce that $\tilde \alpha = \alpha$, hence, 
\begin{equation} 
\alpha_1 (- \theta_1, \theta_2) = - \alpha_1(\theta_1, \theta_2), \quad \alpha_2 (- \theta_1, \theta_2) = \alpha_2(\theta_1, \theta_2). 
\label{eq:sym_al1_al2}
\end{equation}
Thus, changing $\theta_1$ into $-\theta_1$ in the integrals, the numerator of \eqref{eq:dim4_L_prf2} is changed in its opposite, while the denominator is unchanged. It results that $C'_2=0$, ending the proof. \endproof

\subsubsection{Proof of \eqref{eq:Gamma_L(P)_B(P,Q)_express_prf11}}
\label{appsubsec_proof_of_Bs}

In \cite[Sect. 9]{degond2021body} a symmetric bilinear map $B_s$: $\mathfrak{so}_4 \times \mathfrak{so}_4 \to {\mathcal S}_4$ satisfying \eqref{eq:Gamma_L(P)_B(P,Q)_express_prf10} is shown to be of the form 
\begin{equation} 
B_s(P,Q) = C_3 \mathrm{Tr}(PQ) \mathrm{I} + C_4 \Big( \frac{PQ+QP}{2} - \frac{1}{4} \mathrm{Tr}(PQ) \mathrm{I} \Big) + C_5 (\beta(P) \cdot Q) \mathrm{I}, \quad \forall P, \, Q \in \mathfrak{so}_4, 
\label{eq:dim4_Bs_prf1}
\end{equation}
where $C_3$, $C_4$ and $C_5$ are real constants and $\beta$ is the map defined by \eqref{eq:dim4_L_prf0}. To show that $C_5=0$, we adopt the same method as in the previous section. 
Taking the trace of \eqref{eq:dim4_Bs_prf1} and applying the resulting formula with $P=F_{ij}$ and $Q = \beta(F_{ij})$, we get, with \eqref{eq:Gamma_L(P)_B(P,Q)_express_prf9} and~\eqref{eq:Gamma_Bdef}: 
\begin{eqnarray*}
&&\hspace{-1cm}
\frac{1}{2} \int_{\mathrm{SO}_4} \Big\{  \big( \frac{A-A^T}{2} \cdot F_{ij} \big) \big( \beta \circ \mu(A) \cdot F_{ij} \big) \\
&&\hspace{2cm}
+  \Big( \beta \big( \frac{A-A^T}{2} \big) \cdot F_{ij} \Big) \big( \mu(A) \cdot F_{ij} \big) \Big\} \mathrm{Tr}(A) \, M(A) \, dA = 4 C_5. 
\end{eqnarray*}
Averaging the result over $i$, $j$, such that $i \not = j$, we get 
\begin{eqnarray*}
8 C_5 &=& \frac{1}{6} \int_{\mathrm{SO}_4}  \Big\{ \Big( \frac{A-A^T}{2} \cdot \beta \circ \mu(A) \Big) + \Big( \beta \big( \frac{A-A^T}{2} \big) \cdot \mu(A) \Big) \Big\} \, \mathrm{Tr}(A) \, M(A) \, dA \\
&=& \frac{1}{3} \int_{\mathrm{SO}_4}  \Big( \frac{A-A^T}{2} \cdot \beta \circ \mu(A) \Big)  \, \mathrm{Tr}(A) \, M(A) \, dA. 
\end{eqnarray*}
This formula is similar to \eqref{eq:dim4_L_prf1.5} but for the additional factor $\mathrm{Tr}(A)$. This factor adds one more factor to the formula corresponding to \eqref{eq:dim4_L_prf2} and this additional factor is an even function of the $\theta_k$. Thus, the conclusion remains that the corresponding integral vanishes by antisymmetry, and this leads to $C_5=0$. \endproof

\subsubsection{Proof of \eqref{eq:Gamma_L(P)_B(P,Q)_express_prf15}}
\label{appsubsec_proof_of_Ba}

We have already shown that $\Lambda^2(\Lambda^2(V)) = W \oplus Z$, where $W \cong {\mathbb S}_{[2]}$ and $Z \cong \Lambda(V)$. If $n=4$, $\Lambda(V)$ is not irreducible. Instead, it decomposes into two irreducible representations of highest weights ${\mathcal L_1} + {\mathcal L_2}$ and ${\mathcal L_1} - {\mathcal L_2}$. On the other hand, ${\mathcal S}_2$ decomposes into the irreducible representations ${\mathbb S}_{[2]}$ and ${\mathbb C} \, \mathrm{I}$, which have highest weights $2 {\mathcal L}_1$ and $0$. Thus, by Schur's lemma, a non-zero intertwining map $\tilde B_a$: $\Lambda^2(\Lambda^2(V)) \to {\mathcal S}_2$ must be an isomorphism between $W$ and ${\mathbb S}_{[2]}$ and equal to zero on the complement $Z$ of $W$ in $\Lambda^2(\Lambda^2(V))$. We now identify such a map. 

If $n=4$, there exists an isomorphism $\zeta$: $\Lambda^3(V) \to V$ such that $\zeta(v_1 \wedge v_2 \wedge v_3) \cdot v_4 = \mathrm{det}(v_1, v_2, v_3, v_4)$, $\forall (v_1, \ldots, v_4) \in V^4$. Now, we define the map ${\mathcal R}$ as follows: 
\begin{eqnarray*} 
{\mathcal R}: \, \, \Lambda^2(\Lambda^2(V)) &\to& \vee^2(V) 
\\ 
(v_1 \wedge v_2) \wedge (v_3 \wedge v_4) & \mapsto & \zeta(v_1 \wedge v_2 \wedge v_3) \vee v_4 - \zeta(v_1 \wedge v_2 \wedge v_4) \vee v_3 
\\
&& \hspace{-2cm} -  \zeta(v_3 \wedge v_4 \wedge v_1) \vee v_2 + \zeta(v_3 \wedge v_4 \wedge v_2) \vee v_1,  \quad \forall (v_1, \ldots, v_4) \in V^4. 
\end{eqnarray*}  
${\mathcal R}$ intertwines the representations $\Lambda^2(\Lambda^2(V))$ and $\vee^2(V)$ of $\mathrm{so}_4{\mathbb C}$. Thus, its kernel and image are subrepresentations of $\Lambda^2(\Lambda^2(V))$ and $\vee^2(V)$ of $\mathrm{so}_4{\mathbb C}$ respectively. One has 
\begin{eqnarray*}
\mathrm{Tr} \big\{  {\mathcal R} \big( (v_1 \wedge v_2) \wedge (v_3 \wedge v_4) \big)  \big\} &=& 2 \{ \zeta(v_1 \wedge v_2 \wedge v_3) \cdot v_4 - \zeta(v_1 \wedge v_2 \wedge v_4) \cdot v_3 \\
&& \hspace{1cm} 
-  \zeta(v_3 \wedge v_4 \wedge v_1) \cdot v_2 + \zeta(v_3 \wedge v_4 \wedge v_2) \cdot v_1 \big \} \\
&=& 4 \big( \mathrm{det} (v_1 , v_2 , v_3 , v_4) - \mathrm{det} (v_3 , v_4 , v_1 , v_2) \big) = 0. 
\end{eqnarray*}
Hence $\mathrm{im}({\mathcal R}) \subset {\mathbb S}_{[2]}$. Furthermore $\mathrm{im}({\mathcal R}) \not = \{0\}$. Indeed, 
$$ {\mathcal R} \big( (e_1 \wedge e_2) \wedge (e_3 \wedge e_4) \big) = - (e_1 \vee e_1 + e_2 \vee e_2) + e_3 \vee e_3 + e_4 \vee e_4 \not = 0, $$
where $(e_i)_{i=1}^4$ is the canonical basis of $V = {\mathbb C}^4$. Since ${\mathbb S}_{[2]}$ is irreducible and $\mathrm{im}({\mathcal R})$ is a non-trivial subrepresentation of ${\mathbb S}_{[2]}$, we have $\mathrm{im}({\mathcal R}) = {\mathbb S}_{[2]}$. We have $\mathrm{dim} \, \Lambda^2(\Lambda^2(V)) = 15$, $\mathrm{dim} \, {\mathbb S}_{[2]} = 9$, hence by the rank nullity theorem, $\mathrm{dim} (\mathrm{ker} \, {\mathcal R}) = 6 = \mathrm{dim} \, \Lambda^2(V)$. Thus, $\mathrm{ker} \, {\mathcal R} = Z$ (indeed, $\mathrm{ker} \, {\mathcal R}$ has to be a subrepresentation of $\Lambda^2(\Lambda^2(V))$ and $Z \cong \Lambda^2(V)$ is the only such representation which has the right dimension). Consequently, ${\mathcal R}$ is an isomorphism from the complement $W$ of $Z$ in $\Lambda^2(\Lambda^2(V))$ onto ${\mathbb S}_{[2]}$ and is zero when restricted to $Z$. This shows that  ${\mathcal R}$ is the map that needed to be identified. By Schur's lemma,  there is a constant $C_6 \in {\mathbb C}$ such that $\tilde B_a = C_6 {\mathcal R}$. 

We have 
$$ \zeta (e_i \wedge e_j \wedge e_k) = \sum_{m=1}^4 \varepsilon_{ijkm} e_m, $$
where $\varepsilon_{ijkm}$ is equal to zero if two or more indices $i,j,k,m$ are equal and  equal to the signature of the permutation
$$ \left( \begin{array}{cccc} 1&2&3&4 \\i&j&k&m \end{array} \right), $$
otherwise. Then, 
$$ {\mathcal R}\big( (e_i \wedge e_j) \wedge (e_k \wedge e_\ell) \big) = \sum_{m=1}^4 \big[ 
\varepsilon_{ijkm} e_m \vee e_\ell - \varepsilon_{ij\ell m} e_m \vee e_k - \varepsilon_{k \ell im} e_m \vee e_j + \varepsilon_{k \ell jm} e_m \vee e_i  \big]. 
$$
It follows that 
$$ B_a(F_{ij},F_{ik})_{i \ell} = C_6 {\mathcal R}\big( (e_i \wedge e_j) \wedge (e_i \wedge e_k) \big)_{i \ell} = 2 C_6 \varepsilon_{ijk\ell}. $$
Thus, 
$$ \sum_{i,j,k, \ell} \epsilon_{i j k \ell} B_a(F_{ij},F_{ik})_{i \ell} = 2 C_6 \sum_{i,j,k, \ell} \epsilon_{i j k \ell} \varepsilon_{ijk\ell} = 2 C_6 \mathrm{Card}(\mathfrak{S}_4) =  48 C_6. $$
On the other hand, 
\begin{eqnarray*}
&&\hspace{-1cm}
\sum_{i,j,k, \ell} \epsilon_{i j k \ell} B_a(F_{ij},F_{ik})_{i \ell} = \frac{1}{2} \int_{\mathrm{SO}_4} \sum_{i,j,k, \ell} \epsilon_{i j k \ell} \Big[ \Big( \frac{A-A^T}{2} \Big)_{ij} \mu(A)_{ik} \\
&&\hspace{4cm}
 - \mu(A)_{ij} \Big( \frac{A-A^T}{2} \Big)_{ik} \Big] \Big( \frac{A+A^T}{2} \Big)_{i\ell} \, M(A) \, dA \\
&&\hspace{0cm}
= - \int_{\mathrm{SO}_4} \sum_{j,k, \ell, i} \epsilon_{j k \ell i} \Big( \frac{A-A^T}{2} e_i \Big)_j \big(\mu(A) e_i \big)_k \Big( \frac{A+A^T}{2} e_i \Big)_\ell \, M(A) \, dA \\
&&\hspace{0cm}
= - \int_{\mathrm{SO}_4} \sum_i  \zeta \Big(  \frac{A-A^T}{2} e_i \, \wedge \, \mu(A) e_i \, \wedge \,  \frac{A+A^T}{2} e_i  \Big) \cdot e_i  \, \, M(A) \, dA \\
&&\hspace{0cm}
= - \int_{\mathrm{SO}_4} \sum_i  \mathrm{det} \Big(  \frac{A-A^T}{2} e_i,  \,  \mu(A) e_i, \,  \frac{A+A^T}{2} e_i, \,   e_i  \Big) \, M(A) \, dA, 
\end{eqnarray*}
where we have used that 
$$ \zeta(u \wedge v \wedge w) = \sum_{i,j,k,\ell} \epsilon_{ijk\ell} u_i v_j w_k e_\ell, $$
with $u_i$ the $i$-th coordinates of $u$ in the basis $(e_i)_{i=1}^n$ and similarly for $v_j$ and $w_k$. Now, we use \eqref{eq:GCIId_varform_invar_prf1} and get 
\begin{eqnarray}
&&\hspace{-1cm}
{\mathcal D}_i =: \int_{\mathrm{SO}_4}  \mathrm{det} \Big(  \frac{A-A^T}{2} e_i,  \,  \mu(A) e_i, \,  \frac{A+A^T}{2} e_i, \,   e_i  \Big) \, M(A) \, dA \nonumber \\
&&\hspace{-1cm}
= \int_{{\mathcal T}} \int_{\mathrm{SO}_4}  \mathrm{det} \Big(  g \frac{A_\Theta-A_\Theta^T}{2} g^T e_i,  \,  g \mu(A_\Theta) g^T e_i, \, g \frac{A_\Theta+A_\Theta^T}{2} g^T e_i, \,   e_i  \Big) \, dg \, m(\Theta) \, d \Theta \nonumber \\
&&\hspace{-1cm}
= \int_{{\mathcal T}}  \int_{\mathrm{SO}_4}  \mathrm{det} \Big( \frac{A_\Theta-A_\Theta^T}{2} g^T e_i,  \,  \mu(A_\Theta) g^T e_i, \, \frac{A_\Theta+A_\Theta^T}{2} g^T e_i, \,  g^T e_i  \Big) \, dg \, m(\Theta) \, d \Theta \nonumber \\
&&\hspace{-1cm}
= \int_{{\mathcal T}}  \int_{\mathrm{SO}_4}  \sum_{j,k,\ell,m} g_{ij} \, g_{ik} \, g_{i\ell} \, g_{im} \,  \nonumber \\
&&\hspace{2.6cm} \mathrm{det} \Big( \frac{A_\Theta-A_\Theta^T}{2} e_j,  \, \mu(A_\Theta) e_k, \,  \frac{A_\Theta+A_\Theta^T}{2} e_\ell, \,  e_m  \Big)  \, m(\Theta) \, d \Theta\label{eq:Sjklmdet}
\end{eqnarray}
Now, thanks to \eqref{eq:GCIId_tildmuk_equation_prf2}, \eqref{eq:GCIId:mu_formula}, \eqref{eq:Gamma_L(P)_B(P,Q)_express_prf26}, we notice that the matrices $\frac{A_\Theta-A_\Theta^T}{2}$, $\mu(A_\Theta)$, $\frac{A_\Theta+A_\Theta^T}{2}$ are sparse. Hence the determinant in \eqref{eq:Sjklmdet} is equal to zero for many values of the quadruple $(j,k,\ell,m) \in \{1, \ldots, 4\}^4$. The non-zero values of this determinant are given in Table \ref{tab:determinant_value}. 
\begin{table}[h!]
\centering
\begin{tabular}{||c|c|c|c|c||c|c|c|c|c||} 
\hline \hline
 $j$ & $k$ & $\ell$ & $m$ & $\textrm{det}$ & $j$ & $k$ & $\ell$ & $m$ & $\textrm{det}$ \\  
 \hline\hline
 1 & 2 & 3 & 4 & $- \alpha_1 s_1 c_2$ & 3 & 1 & 1 & 3 & $- \alpha_1 s_2 c_1$\\ 
 1 & 2 & 4 & 3 & $  \alpha_1 s_1 c_2$ & 3 & 1 & 3 & 1 & $  \alpha_1 s_2 c_2$\\ 
\hline
 1 & 3 & 1 & 3 & $  \alpha_2 s_1 c_1$ & 3 & 2 & 2 & 3 & $- \alpha_1 s_2 c_1$\\ 
 1 & 3 & 3 & 1 & $- \alpha_2 s_1 c_2$ & 3 & 2 & 3 & 2 & $  \alpha_1 s_2 c_2$\\
\hline 
 1 & 4 & 1 & 4 & $  \alpha_2 s_1 c_1$ & 3 & 4 & 1 & 2 & $  \alpha_2 s_2 c_1$\\ 
 1 & 4 & 4 & 1 & $- \alpha_2 s_1 c_2$ & 3 & 4 & 2 & 1 & $- \alpha_2 s_2 c_1$\\ 
\hline \hline
 2 & 1 & 3 & 4 & $  \alpha_1 s_1 c_2$ & 4 & 1 & 1 & 4 & $- \alpha_1 s_2 c_1$\\ 
 2 & 1 & 4 & 3 & $- \alpha_1 s_1 c_2$ & 4 & 1 & 4 & 1 & $  \alpha_1 s_2 c_2$\\ 
\hline
 2 & 3 & 2 & 3 & $  \alpha_2 s_1 c_1$ & 4 & 2 & 2 & 4 & $- \alpha_1 s_2 c_1$\\ 
 2 & 3 & 3 & 2 & $- \alpha_2 s_1 c_2$ & 4 & 2 & 4 & 2 & $  \alpha_1 s_2 c_2$\\ 
\hline
 2 & 4 & 2 & 4 & $  \alpha_2 s_1 c_1$ & 4 & 3 & 1 & 2 & $  \alpha_2 s_2 c_1$\\ 
 2 & 4 & 4 & 2 & $- \alpha_2 s_1 c_2$ & 4 & 3 & 2 & 1 & $- \alpha_2 s_2 c_1$\\ 
 \hline \hline
\end{tabular}
\caption{Table of the non-zero values of the determinant in \eqref{eq:Sjklmdet} (called ``det'' in the table) as a function of $(j,k,\ell,m)$. The quantities $\alpha_q$, $c_q$ and $s_q$ for $q = 1, \, 2$ refer to $\alpha(\theta_q)$, $\cos(\theta_q)$, $\sin(\theta_q)$.}
\label{tab:determinant_value}
\end{table}
It results that 
\begin{eqnarray}
&&\hspace{-1cm}
{\mathcal D}_i = \int_{{\mathcal T}} \big( \alpha_2(\Theta) \sin \theta_1 - \alpha_1(\Theta) \sin \theta_2 \big) \big( \cos \theta_1 - \cos \theta_2) \, m(\Theta) \, d \Theta \nonumber \\
&&\hspace{7cm}
 \times \int_{\mathrm{SO}_4} ( g_{i1}^2 + g_{i2}^2 ) ( g_{i3}^2 + g_{i4}^2 ) \, dg, \label{eq:calDi}
\end{eqnarray}
and the first integral in \eqref{eq:calDi} is equal to $0$ by the symmetry relations \eqref{eq:sym_al1_al2}. \endproof

\setcounter{equation}{0}
\section{Conclusion}
\label{sec_conclu}

In this paper, we have derived a fluid model for a system of stochastic differential equations modelling rigid bodies interacting through body-attitude alignment in arbitrary dimensions. This follows earlier work where this derivation was done in dimension $3$ only on the one hand, or for simpler jump processes on the other hand. This extension was far from being straightforward. The main output of this work is to highlight the importance of concepts from Lie-group theory such as maximal torus, Cartan subalgebra and Weyl group in this derivation. We may anticipate that these concepts (which were hidden although obviously present in earlier works) may be key to more general collective dynamics models in which the control variables of the agents, i.e. the variable that determines their trajectory, belongs to a Lie group or to a homogeneous space (which can be regarded as the quotient of a Lie group by one of its subgroups). At least when these Lie groups or homogeneous spaces are compact, we may expect that similar concepts as those developed in this paper are at play. Obviously, extensions to non-compact Lie-groups or homogeneous spaces may be even more delicate.

\newpage
\vspace{1cm}
\appendix

\noindent
\textbf{{\Large Appendix}}

\setcounter{equation}{0}
\section{Direct derivation of strong form of equations satisfied by $(\alpha_k)_{k=1}^p$}
\label{appsec_direct_strong_form}

Beforehand, we need to give an expression of the radial Laplacian, defined by \eqref{eq:radlap_def}. 
\begin{eqnarray}
&&\hspace{-1cm}
L \varphi =  \sum_{j=1}^p \Big(\frac{\partial^2 \varphi}{\partial \theta_j^2} + \epsilon_n \frac{\sin \theta_j}{1 - \cos \theta_j} \, \frac{\partial \varphi}{\partial \theta_j} \Big) \nonumber \\
&&\hspace{1cm}
+ \sum_{1 \leq j < k \leq p} \frac{2}{\cos \theta_k - \cos \theta_j} \, \Big( \big( \sin \theta_j \, \frac{\partial \varphi}{\partial \theta_j} - \sin \theta_k \, \frac{\partial \varphi}{\partial \theta_k}) \Big) 
\label{eq:radlap_2p+1} \\
&&\hspace{-0.8cm}
= \sum_{j=1}^p \Big[ \frac{\partial^2 \varphi}{\partial \theta_j^2} + \Big( \sum_{k \not = j} \frac{2}{\cos \theta_k - \cos \theta_j} + \frac{\epsilon_n}{1 - \cos \theta_j} \Big) \sin \theta_j \frac{\partial \varphi}{\partial \theta_j} \Big]. \label{eq:radlap_equiv} \\
&&\hspace{-1cm}
=  \frac{1}{u_n} \nabla_\Theta \cdot \big( u_n \nabla_\Theta \varphi \big), \label{eq:radlap_conserv} 
\end{eqnarray}
with $\epsilon_n$ given by \eqref{eq:Gamma_eps_def}. 

In this section, we give a direct derivation of the strong form of the equations satisfied by $(\alpha_k)_{k=1}^p$. For this, we use a strategy based on \cite[Section 8.3]{faraud2008Analysis} (See also \cite{degond2023radial}). It relies on the following formula. 
Let $f$ be a function $\mathrm{SO}_n \to V$, where $V$ is a finite-dimensional vector space over ${\mathbb R}$. Then, we have
\begin{equation}
\varrho \big( \mathrm{Ad}(A^{-1}) X -X \big)^2 f(A) - \varrho \big( [\mathrm{Ad}(A^{-1}) X, X ] \big) f(A) = \frac{d^2}{dt^2} \big( f( e^{tX} A e^{-tX}) \big) \big|_{t=0}. 
\label{eq:fonda}
\end{equation}
for all $X \in \mathfrak{so}_n$ and all $A \in \mathrm{SO}_n$, where $\varrho$ is defined at \eqref{eq:rot_def_rho}. We note in passing that $\varrho$ is a Lie algebra representation of $\mathfrak{so}_n$ into $C^\infty(\mathrm{SO}_n,V)$, i.e. it is a linear map $\mathfrak{so}_n \to {\mathcal L}(C^\infty(\mathrm{SO}_n,V))$ (where ${\mathcal L}(C^\infty(\mathrm{SO}_n,V))$ is the space of linear maps of $C^\infty(\mathrm{SO}_n,V)$ into itself) which satisfies
$$ \varrho([X,Y]) f = [\varrho(X), \varrho(Y)] f , \quad \forall X, Y \in \mathfrak{so}_n, \quad \forall f \in C^\infty(\mathrm{SO}_n,V). $$
In the previous formula, the bracket on the left is the usual Lie bracket in $\mathfrak{so}_n$ while the bracket on the right is the commutator of two elements of ${\mathcal L}(C^\infty(\mathrm{SO}_n,V))$.  

The following proposition gives the equation satisfied by $(\alpha_k)_{k=1}^p$ in strong form.  

\begin{proposition}
(i) The functions $(\alpha_k)_{k=1}^p$ defined by \eqref{eq:GCIId:mu_generic} satisfy the following system of partial differential equations: 
\begin{eqnarray}
&&\hspace{-1cm}
L \alpha_\ell - \kappa \Big( \sum_{k=1}^p \sin \theta_k \frac{\partial}{\partial \theta_k} \Big) \alpha_\ell - \sum_{k \not = \ell} \Big( \frac{\alpha_\ell - \alpha_k}{1 - \cos(\theta_\ell - \theta_k)} +  \frac{\alpha_\ell + \alpha_k}{1 - \cos(\theta_\ell + \theta_k)} \Big)  \nonumber \\
&&\hspace{1cm}
- \epsilon_n \frac{\alpha_\ell}{1 - \cos \theta_\ell} 
 + \sin \theta_\ell = 0, \quad \forall \ell = 1, \ldots, p 
\label{eq:GCIId_tildmuk_equation_2p+1}
\end{eqnarray}

\smallskip
\noindent
(ii) System \eqref{eq:GCIId_tildmuk_equation_2p+1} is identical with \eqref{eq:GCIId_strongform_alpha}. 
\label{prop:GCIId_tildmuk_equation}
\end{proposition}

\noindent
\textbf{Proof.} (i) Eq. \eqref{eq:GCIId_equa_mu} can be equivalently written 
\begin{equation}
\Delta \mu(A) + (\nabla \log M \cdot \nabla \mu)(A) = \frac{A-A^T}{2}. 
\label{eq:GCIId_tildmuk_equation_prf1}
\end{equation}
We evaluate \eqref{eq:GCIId_tildmuk_equation_prf1} at $A = A_\Theta$ for $\Theta \in {\mathcal T}$. We recall the expression \eqref{eq:GCIId_tildmuk_equation_prf2} of $\frac{A_\Theta-A_\Theta^T}{2}$. 
Besides, with \eqref{eq:equi_VM} and \eqref{eq:equi_Q_express_prf1}, we get 
$$ \nabla \log M = \kappa \nabla (A \cdot \mathrm{I}) = \kappa P_{T_A} \mathrm{I} = \kappa A \frac{A^T - A}{2}, $$
which, thanks to \eqref{eq:GCIId_tildmuk_equation_prf2} and \eqref{eq:rot_operator_rho}, leads to 
\begin{eqnarray}
(\nabla \log M \cdot \nabla \mu)(A_\Theta) &=& \kappa \sum_{k=1}^p \sin \theta_k \, (\nabla \mu)(A_\Theta) \cdot (A_\Theta F_{2k-1 \, 2k}) \nonumber \\
&=& \kappa \sum_{k=1}^p \sin \theta_k \, \big( \varrho(F_{2k-1 \, 2k}) (\mu) \big)(A_\Theta). 
\label{eq:GCIId_tildmuk_equation_prf3}
\end{eqnarray}
We recall \eqref{eq:GCIId_tildmuk_equation_prf4}. The following identity is proved in the same manner: 
\begin{equation}
\big( \varrho(F_{2k-1 \, 2k})^2 (\mu) \big)(A_\Theta) = \sum_{\ell=1}^p \frac{\partial^2 \alpha_\ell}{\partial \theta_k^2}(A_\Theta) \, F_{2\ell-1 \, 2_\ell}. 
\label{eq:GCIId_tildmuk_equation_prf5}
\end{equation}
Inserting \eqref{eq:GCIId_tildmuk_equation_prf4} into \eqref{eq:GCIId_tildmuk_equation_prf3}, we eventually get 
\begin{equation}
(\nabla \log M \cdot \nabla \mu)(A_\Theta) = \sum_{\ell=1}^p \Big( - \kappa \sum_{k=1}^p \sin \theta_k \frac{\partial \alpha_\ell}{\partial \theta_k}(A_\Theta) \Big) \, F_{2\ell-1 \, 2_\ell}
\label{eq:GCIId_tildmuk_equation_prf6}
\end{equation}

It remains to find an expression of $\Delta \mu(A_\Theta)$. We use Formula \eqref{eq:rot_Laplacian} for $\Delta$. For $A=A_\Theta$, we see that $\varrho(F_{2k-1 \, 2k})^2 (\mu)$ for $k=1, \ldots, p$ is explicit thanks to \eqref{eq:GCIId_tildmuk_equation_prf5}. On the other hand, $\varrho(F_{ij})^2(\mu)$ for $(i,j) \not \in \{ (2k-1,2k), \, k=1, \ldots, p \}$ is not. To compute them, we apply the same strategy as in \cite[Section 8.3]{faraud2008Analysis} and in \cite{degond2023radial} based on \eqref{eq:fonda}. Given \eqref{eq:GCIId_conjug_invar_mu_0}, we get, after a few lines of computations
$$ \frac{d^2}{dt^2} \big( \mu ( e^{tX} A e^{-tX}) \big) \big|_{t=0} = \frac{d^2}{dt^2} \big( e^{tX} \mu (A) e^{-tX} \big) \big|_{t=0} = \big[X, [X,\mu(A)] \big], $$
and so, \eqref{eq:fonda} leads to 
\begin{equation}
\varrho \big( \mathrm{Ad}(A^{-1}) X -X \big)^2 \mu (A) - \varrho \big( [\mathrm{Ad}(A^{-1}) X, X ] \big) \mu (A) = \big[X, [X,\mu(A)] \big]. 
\label{eq:fonda_mu}
\end{equation}
We adopt the same proof outline as in \cite{degond2023radial} for the determination of the radial Laplacian. As in the proof of Prop. \ref{prop:GCIId_regularity_tau}, we successively treat the cases of $\mathrm{SO}_3$, $\mathrm{SO}_4$, $\mathrm{SO}_{2p}$, $\mathrm{SO}_{2p+1}$. We refer to the proof of Prop. \ref{prop:GCIId_regularity_tau} for the notations. 

\medskip
\paragraph{Case of $\mathrm{SO}_3$.} 
We use \eqref{eq:fonda_mu} with $A=A_\Theta$ and with $X=G^+$ and $X=G^-$ successively and we add up the resulting equations. The details of the computations of what comes out of the left-hand side of \eqref{eq:fonda_mu} can be found in \cite{degond2023radial}. On the other hand, after easy computations using \eqref{eq:commutF}, the right-hand side gives
$$ \big[G^+, [G^+,\mu(A_\Theta)] \big] + \big[G^-, [G^-,\mu(A_\Theta)] \big] = - 2 \alpha_1(\theta_1) \, F_{12}. $$  
Thus, we get: 
$$ 2 (1-\cos \theta) \Big( \big(\varrho(G^+)^2 +\varrho(G^-)^2 \big) \mu \Big)(A_\Theta) + 2 \sin \theta \big( \varrho(F_{12}) \mu \big) (A_\Theta) = - 2 \alpha_1(\theta_1) \, F_{12}.$$
With \eqref{eq:GCIId_tildmuk_equation_prf4} and \eqref{eq:GCIId_tildmuk_equation_prf5}, this yields 
\begin{eqnarray*}
\Delta \mu(A_\Theta) &=& \Big( \frac{\partial^2 \alpha_1}{\partial \theta_1^2}(\theta_1) + \frac{\sin \theta_1}{1 - \cos \theta_1} \frac{\partial \alpha_1}{\partial \theta_1}(\theta_1) - \frac{1}{1 - \cos \theta_1} \alpha_1(\theta_1) \Big) \, F_{12} \\
&=& \Big( (L \alpha_1)(\theta_1) - \frac{1}{1 - \cos \theta_1} \alpha_1(\theta_1) \Big) F_{12}. 
\end{eqnarray*}

\medskip
\paragraph{Case of $\mathrm{SO}_4$.} 
We use \eqref{eq:fonda_mu} with $A=A_\Theta$ and with $X=H^+$ and $X=K^-$ successively and we add up the resulting equations. We do similarly with $X = H^-$ and $X = K^+$. Again, what results from the left-hand sides of \eqref{eq:fonda_mu} can be found in \cite{degond2023radial}, while the right-hand sides, using \eqref{eq:commutF}, give: 
\begin{eqnarray*} 
\big[H^+, [H^+,\mu(A_\Theta)] \big] + \big[K^-, [K^-,\mu(A_\Theta)] \big] &=& 2 \big( - \alpha_1 + \alpha_2 \big) (\Theta) \, \big( F_{12} - F_{34} \big), \\
  \big[H^-, [H^-,\mu(A_\Theta)] \big] + \big[K^+, [K^+,\mu(A_\Theta)] \big] &=& - 2 \big( \alpha_1 + \alpha_2 \big) (\Theta) \, \big( F_{12} + F_{34} \big). 
\end{eqnarray*}
We get
\begin{eqnarray*}
&&\hspace{-1cm}
2 \big( 1-\cos (\theta_2 - \theta_1) \big) \Big( \big(\varrho(H^+)^2 +\varrho(K^-)^2 \big) \mu \Big)(A_\Theta) \\
&&\hspace{0.5cm}
- 2 \sin (\theta_2-\theta_1) \Big( \big( \varrho(F_{12}) - \varrho(F_{34}) \big) \mu \Big) (A_\Theta) = 2 \big( - \alpha_1 + \alpha_2 \big) (\Theta) \, \big( F_{12} - F_{34} \big), \\
&&\hspace{-1cm}
2 \big( 1-\cos (\theta_1 + \theta_2) \big) \Big( \big(\varrho(H^-)^2 +\varrho(K^+)^2 \big) \mu \Big)(A_\Theta) \\
&&\hspace{0.5cm}
+ 2 \sin (\theta_1+\theta_2) \Big( \big( \varrho(F_{12}) + \varrho(F_{34}) \big) \mu \Big) (A_\Theta) = - 2 \big(\alpha_1 + \alpha_2 \big) (\Theta) \, \big( F_{12} + F_{34} \big), 
\end{eqnarray*}
Collecting the results and using \eqref{eq:GCIId_tildmuk_equation_prf4} and \eqref{eq:GCIId_tildmuk_equation_prf5}, we get 
\begin{eqnarray*}
\Delta \mu(A_\Theta) &=& \Big\{ (L \alpha_1)(\Theta) - \Big( \frac{(\alpha_1 - \alpha_2)(\Theta)}{1 - \cos (\theta_1 - \theta_2)} + \frac{(\alpha_1 + \alpha_2)(\Theta)}{1 - \cos (\theta_1 + \theta_2)} \Big) \Big\} F_{12} \\
&&+ \Big\{ (L \alpha_2)(\Theta) - \Big( \frac{(\alpha_2 - \alpha_1)(\Theta)}{1 - \cos (\theta_2 - \theta_1)} + \frac{(\alpha_2 + \alpha_1)(\Theta)}{1 - \cos (\theta_2 + \theta_1)} \Big)  \Big\} F_{34}. 
\end{eqnarray*}

\medskip
\paragraph{Case of $\mathrm{SO}_{2p}$.} 
The computations are straighforward extensions of those done in the case of $\mathrm{SO}_{4}{\mathbb R}$ and lead to 
\begin{equation}
\Delta \mu(A_\Theta) = \sum_{\ell = 1}^p \Big\{ L \alpha_\ell (\Theta) - \sum_{k \not = \ell} \Big( \frac{(\alpha_\ell - \alpha_k)(\Theta)}{1 - \cos(\theta_\ell - \theta_k)} +  \frac{(\alpha_\ell + \alpha_k)(\Theta)}{1 - \cos(\theta_\ell + \theta_k)} \Big)  \Big\} F_{2 \ell -1 \, 2 \ell}. 
\label{eq:GCIId_tildmuk_equation_prf7}
\end{equation}

\medskip
\paragraph{Case of $\mathrm{SO}_{2p+1}$.} In this case, we combine the computations done for 
the cases $\mathrm{SO}_{2p}$ and $\mathrm{SO}_3$. They lead to 
\begin{eqnarray}
\hspace{-1cm}
\Delta \mu(A_\Theta) &=& \sum_{\ell = 1}^p \Big\{ L \alpha_\ell (\Theta) - \sum_{k \not = \ell} \Big( \frac{(\alpha_\ell - \alpha_k)(\Theta)}{1 - \cos(\theta_\ell - \theta_k)} +  \frac{(\alpha_\ell + \alpha_k)(\Theta)}{1 - \cos(\theta_\ell + \theta_k)} \Big) \nonumber \\
&& \hspace{6.5cm}  - \epsilon_n \frac{\alpha_\ell(\Theta)}{1 - \cos \theta_\ell}  \Big\} F_{2 \ell -1 \, 2 \ell}. 
\label{eq:GCIId_tildmuk_equation_prf9}
\end{eqnarray}

\medskip
Now, collecting \eqref{eq:GCIId_tildmuk_equation_prf2}, \eqref{eq:GCIId_tildmuk_equation_prf6} and \eqref{eq:GCIId_tildmuk_equation_prf7} or \eqref{eq:GCIId_tildmuk_equation_prf9} (according to the parity of $n$) and inserting them into \eqref{eq:GCIId_tildmuk_equation_prf1} gives a matrix identity which is decomposed on the basis vectors $(F_{2 \ell -1 \, 2 \ell})_{\ell = 1}^p$ of $\mathfrak{h}$. Hence it must be satisfied componentwise, which leads to \eqref{eq:GCIId_tildmuk_equation_2p+1} and ends the proof of Point (i).

\smallskip
\noindent
(ii) We have 
\begin{equation} 
m^{-1} \nabla_\Theta \cdot \big( m \nabla_\Theta \alpha_\ell \big) = \Delta_\Theta \alpha_\ell + \nabla_\Theta  \log m \cdot \nabla_\Theta \alpha_\ell, 
\label{eq:_GCIId_conservative_prf1}
\end{equation}
where $\Delta_\Theta$ is the Laplacian with respect to~$\Theta$ (i.e. $\Delta_\Theta \varphi = \nabla_\Theta \cdot (\nabla_\Theta \varphi)$ for any smooth function $\varphi \in {\mathcal T}$). Then, from \cite[Eq. (4.6) \& (4.7)]{degond2023radial}, we have 
$$ \frac{\partial \log m}{\partial \theta_j} = - \kappa \sin \theta_j + \frac{\partial \log u_n}{\partial \theta_j} = \sin \theta_j \Big( 2 \sum_{k \not = j} \frac{\displaystyle 1}{\displaystyle \cos \theta_k - \cos \theta_j} + \frac{\displaystyle \epsilon_n}{\displaystyle 1 - \cos \theta_\ell} - \kappa \Big). $$
Inserting this into \eqref{eq:_GCIId_conservative_prf1} and using \eqref{eq:radlap_equiv}, we find that the first term at the left-hand side of \eqref{eq:GCIId_strongform_alpha} corresponds to the first two terms of \eqref{eq:GCIId_tildmuk_equation_2p+1}. The other terms of \eqref{eq:GCIId_strongform_alpha} have exact correspondence with terms of \eqref{eq:GCIId_tildmuk_equation_2p+1}, which shows the identity of these equations. \endproof



\bigskip
\begin{thebibliography}{10}

\bibitem{aceves2019hydrodynamic}
P.~Aceves-S{\'a}nchez, M.~Bostan, J.-A. Carrillo, and P.~Degond.
\newblock Hydrodynamic limits for kinetic flocking models of {C}ucker-{S}male
  type.
\newblock {\em Math. Biosci. Eng.}, 16:7883--7910, 2019.

\bibitem{aoki1982simulation}
I.~Aoki.
\newblock A simulation study on the schooling mechanism in fish.
\newblock {\em Bulletin of the Japanese Society of Scientific Fisheries},
  48(8):1081--1088, 1982.

\bibitem{barbaro2016phase}
A.~B. Barbaro, J.~A. Canizo, J.~A. Carrillo, and P.~Degond.
\newblock Phase transitions in a kinetic flocking model of {C}ucker--{S}male
  type.
\newblock {\em Multiscale Model. Simul.}, 14(3):1063--1088, 2016.

\bibitem{barbaro2012phase}
A.~B. Barbaro and P.~Degond.
\newblock Phase transition and diffusion among socially interacting
  self-propelled agents.
\newblock {\em Discrete Contin. Dyn. Syst. Ser. B}, 19(3):1249--1278, 2014.

\bibitem{bazazi2008collective}
S.~Bazazi, J.~Buhl, J.~J. Hale, M.~L. Anstey, G.~A. Sword, S.~J. Simpson, and
  I.~D. Couzin.
\newblock Collective motion and cannibalism in locust migratory bands.
\newblock {\em Current Biology}, 18(10):735--739, 2008.

\bibitem{be2019statistical}
A.~Be\'er and G.~Ariel.
\newblock A statistical physics view of swarming bacteria.
\newblock {\em Movement Ecology}, 7(1):1--17, 2019.

\bibitem{bertin2006boltzmann}
E.~Bertin, M.~Droz, and G.~Gr{\'e}goire.
\newblock Boltzmann and hydrodynamic description for self-propelled particles.
\newblock {\em Phys. Rev. E}, 74(2):022101, 2006.

\bibitem{bertin2009hydrodynamic}
E.~Bertin, M.~Droz, and G.~Gr{\'e}goire.
\newblock Hydrodynamic equations for self-propelled particles: microscopic
  derivation and stability analysis.
\newblock {\em J. Phys. A}, 42(44):445001, 2009.

\bibitem{bertozzi2015ring}
A.~L. Bertozzi, T.~Kolokolnikov, H.~Sun, D.~Uminsky, and J.~Von~Brecht.
\newblock Ring patterns and their bifurcations in a nonlocal model of
  biological swarms.
\newblock {\em Commun. Math. Sci.}, 13(4):955--985, 2015.

\bibitem{bolley2012mean}
F.~Bolley, J.~A. Ca{\~n}izo, and J.~A. Carrillo.
\newblock Mean-field limit for the stochastic {V}icsek model.
\newblock {\em Appl. Math. Lett.}, 25(3):339--343, 2012.

\bibitem{briant2022cauchy}
M.~Briant, A.~Diez, and S.~Merino-Aceituno.
\newblock Cauchy theory for general kinetic {V}icsek models in collective
  dynamics and mean-field limit approximations.
\newblock {\em SIAM J. Math. Anal.}, 54(1):1131--1168, 2022.

\bibitem{cao2020asymptotic}
F.~Cao, S.~Motsch, A.~Reamy, and R.~Theisen.
\newblock Asymptotic flocking for the three-zone model.
\newblock {\em Math. Biosci. Eng.}, 17(6):7692--7707, 2020.

\bibitem{carrillo2010asymptotic}
J.~A. Carrillo, M.~Fornasier, J.~Rosado, and G.~Toscani.
\newblock Asymptotic flocking dynamics for the kinetic {C}ucker--{S}male model.
\newblock {\em SIAM J. Math. Anal.}, 42(1):218--236, 2010.

\bibitem{castellani2005spin}
T.~Castellani and A.~Cavagna.
\newblock Spin-glass theory for pedestrians.
\newblock {\em Journal of Statistical Mechanics: Theory and Experiment},
  2005(05):P05012, 2005.

\bibitem{chate2008collective}
H.~Chat{\'e}, F.~Ginelli, G.~Gr{\'e}goire, and F.~Raynaud.
\newblock Collective motion of self-propelled particles interacting without
  cohesion.
\newblock {\em Phys. Rev. E}, 77(4):046113, 2008.

\bibitem{cho2023continuum}
H.~Cho, S.-Y. Ha, and M.~Kang.
\newblock Continuum limit of the lattice {L}ohe group model and emergent
  dynamics.
\newblock {\em Math. Methods Appl. Sci.}, 2023.

\bibitem{cucker2007emergent}
F.~Cucker and S.~Smale.
\newblock Emergent behavior in flocks.
\newblock {\em IEEE Trans. Automat. Control}, 52(5):852--862, 2007.

\bibitem{degond2023radial}
P.~Degond.
\newblock Radial laplacian on rotation groups.
\newblock {\em Arxiv preprint}, 2302.00749, 2023.

\bibitem{degond2021body}
P.~Degond, A.~Diez, and A.~Frouvelle.
\newblock Body-attitude coordination in arbitrary dimension.
\newblock {\em arXiv preprint arXiv:2111.05614}, 2021.

\bibitem{Degond_eal_JNLS20}
P.~Degond, A.~Diez, A.~Frouvelle, and S.~Merino-Aceituno.
\newblock Phase transitions and macroscopic limits in a {B}{G}{K} model of
  body-attitude coordination.
\newblock {\em J. Nonlinear Sci.}, 30:2671--2736, 2020.

\bibitem{degond2021bulk}
P.~Degond, A.~Diez, and M.~Na.
\newblock Bulk topological states in a new collective dynamics model.
\newblock {\em SIAM J. Appl. Dyn. Syst.}, 21(2):1455--1494, 2022.

\bibitem{degond2013macroscopic}
P.~Degond, A.~Frouvelle, and J.-G. Liu.
\newblock Macroscopic limits and phase transition in a system of self-propelled
  particles.
\newblock {\em J. Nonlinear Sci.}, 23(3):427--456, 2013.

\bibitem{degond2015phase}
P.~Degond, A.~Frouvelle, and J.-G. Liu.
\newblock Phase transitions, hysteresis, and hyperbolicity for self-organized
  alignment dynamics.
\newblock {\em Arch. Ration. Mech. Anal.}, 216(1):63--115, 2015.

\bibitem{degond2017new}
P.~Degond, A.~Frouvelle, and S.~Merino-Aceituno.
\newblock A new flocking model through body attitude coordination.
\newblock {\em Math. Models Methods Appl. Sci.}, 27(06):1005--1049, 2017.

\bibitem{Degond_etal_proc19}
P.~Degond, A.~Frouvelle, S.~Merino-Aceituno, and A.~Trescases.
\newblock Alignment of self-propelled rigid bodies: from particle systems to
  macroscopic equations.
\newblock In {\em International workshop on Stochastic Dynamics out of
  Equilibrium}, pages 28--66. Springer, 2017.

\bibitem{Degond_etal_MMS18}
P.~Degond, A.~Frouvelle, S.~Merino-Aceituno, and A.~Trescases.
\newblock Quaternions in collective dynamics.
\newblock {\em Multiscale Model. Simul.}, 16(1):28--77, 2018.

\bibitem{degond2022hyperbolicity}
P.~Degond, A.~Frouvelle, S.~Merino-Aceituno, and A.~Trescases.
\newblock Hyperbolicity and non-conservativity of a hydrodynamic model of
  swarming rigid bodies.
\newblock {\em Quart. Appl. Math. (published online)}, 2023.

\bibitem{degond2008continuum}
P.~Degond and S.~Motsch.
\newblock Continuum limit of self-driven particles with orientation
  interaction.
\newblock {\em Math. Models Methods Appl. Sci.}, 18(supp01):1193--1215, 2008.

\bibitem{diez2020propagation}
A.~Diez.
\newblock Propagation of chaos and moderate interaction for a piecewise
  deterministic system of geometrically enriched particles.
\newblock {\em Electronic Journal of Probability}, 25:1--38, 2020.

\bibitem{faraud2008Analysis}
J.~Faraut.
\newblock {\em Analysis on Lie Groups, an introduction}.
\newblock Cambridge University Press, 2008.

\bibitem{fetecau2022emergent}
R.~C. Fetecau, S.-Y. Ha, and H.~Park.
\newblock Emergent behaviors of rotation matrix flocks.
\newblock {\em SIAM J. Appl. Dyn. Syst.}, 21(2):1382--1425, 2022.

\bibitem{figalli2018global}
A.~Figalli, M.-J. Kang, and J.~Morales.
\newblock Global well-posedness of the spatially homogeneous
  {K}olmogorov--{V}icsek model as a gradient flow.
\newblock {\em Arch. Ration. Mech. Anal.}, 227(3):869--896, 2018.

\bibitem{frouvelle2012continuum}
A.~Frouvelle.
\newblock A continuum model for alignment of self-propelled particles with
  anisotropy and density-dependent parameters.
\newblock {\em Math. Models Methods Appl. Sci.}, 22(07):1250011, 2012.

\bibitem{frouvelle2021body}
A.~Frouvelle.
\newblock Body-attitude alignment: first order phase transition, link with
  rodlike polymers through quaternions, and stability.
\newblock In {\em Recent Advances in Kinetic Equations and Applications}, pages
  147--181. Springer, 2021.

\bibitem{frouvelle2012dynamics}
A.~Frouvelle and J.-G. Liu.
\newblock Dynamics in a kinetic model of oriented particles with phase
  transition.
\newblock {\em SIAM J. Math. Anal.}, 44(2):791--826, 2012.

\bibitem{fulton2013representation}
W.~Fulton and J.~Harris.
\newblock {\em Representation theory: a first course}.
\newblock Springer, 2013.

\bibitem{gamba2016global}
I.~M. Gamba and M.-J. Kang.
\newblock Global weak solutions for {K}olmogorov--{V}icsek type equations with
  orientational interactions.
\newblock {\em Arch. Ration. Mech. Anal.}, 222(1):317--342, 2016.

\bibitem{golse2019mean}
F.~Golse and S.-Y. Ha.
\newblock A mean-field limit of the {L}ohe matrix model and emergent dynamics.
\newblock {\em Arch. Ration. Mech. Anal.}, 234(3):1445--1491, 2019.

\bibitem{griette2019kinetic}
Q.~Griette and S.~Motsch.
\newblock Kinetic equations and self-organized band formations.
\newblock In {\em Active Particles, Volume 2}, pages 173--199. Springer, 2019.

\bibitem{ha2017emergent}
S.-Y. Ha, D.~Ko, and S.~W. Ryoo.
\newblock Emergent dynamics of a generalized {L}ohe model on some class of
  {L}ie groups.
\newblock {\em J. Stat. Phys.}, 168(1):171--207, 2017.

\bibitem{ha2009simple}
S.-Y. Ha and J.-G. Liu.
\newblock A simple proof of the {C}ucker-{S}male flocking dynamics and
  mean-field limit.
\newblock {\em Commun. Math. Sci.}, 7(2):297--325, 2009.

\bibitem{hildenbrandt2010self}
H.~Hildenbrandt, C.~Carere, and C.~K. Hemelrijk.
\newblock Self-organized aerial displays of thousands of starlings: a model.
\newblock {\em Behavioral Ecology}, 21(6):1349--1359, 2010.

\bibitem{hsu2002stochastic}
E.~P. Hsu.
\newblock {\em Stochastic analysis on manifolds}.
\newblock Number~38. American Mathematical Soc., 2002.

\bibitem{jiang2016hydrodynamic}
N.~Jiang, L.~Xiong, and T.-F. Zhang.
\newblock Hydrodynamic limits of the kinetic self-organized models.
\newblock {\em SIAM J. Math. Anal.}, 48(5):3383--3411, 2016.

\bibitem{lopez2012behavioural}
U.~Lopez, J.~Gautrais, I.~D. Couzin, and G.~Theraulaz.
\newblock From behavioural analyses to models of collective motion in fish
  schools.
\newblock {\em Interface focus}, 2(6):693--707, 2012.

\bibitem{motsch2011new}
S.~Motsch and E.~Tadmor.
\newblock A new model for self-organized dynamics and its flocking behavior.
\newblock {\em J. Stat. Phys.}, 144(5):923--947, 2011.

\bibitem{parisi2020theory}
G.~Parisi, P.~Urbani, and F.~Zamponi.
\newblock {\em Theory of simple glasses: exact solutions in infinite
  dimensions}.
\newblock Cambridge University Press, 2020.

\bibitem{sarlette2010coordinated}
A.~Sarlette, S.~Bonnabel, and R.~Sepulchre.
\newblock Coordinated motion design on lie groups.
\newblock {\em IEEE Trans. Automat. Control}, 55(5):1047--1058, 2010.

\bibitem{sarlette2009autonomous}
A.~Sarlette, R.~Sepulchre, and N.~E. Leonard.
\newblock Autonomous rigid body attitude synchronization.
\newblock {\em Automatica J. IFAC}, 45(2):572--577, 2009.

\bibitem{sepulchre2010consensus}
R.~Sepulchre, A.~Sarlette, and P.~Rouchon.
\newblock Consensus in non-commutative spaces.
\newblock In {\em 49th IEEE conference on decision and control (CDC)}, pages
  6596--6601. IEEE, 2010.

\bibitem{Simon}
B.~Simon.
\newblock {\em Representations of finite and compact groups}.
\newblock Number~10 in Graduate Studies in Mathematics. American Mathematical
  Soc., 1996.

\bibitem{toner1998flocks}
J.~Toner and Y.~Tu.
\newblock Flocks, herds, and schools: A quantitative theory of flocking.
\newblock {\em Phys. Rev. E}, 58(4):4828, 1998.

\bibitem{vicsek1995novel}
T.~Vicsek, A.~Czir{\'o}k, E.~Ben-Jacob, I.~Cohen, and O.~Shochet.
\newblock Novel type of phase transition in a system of self-driven particles.
\newblock {\em Phys. Rev. Lett.}, 75(6):1226, 1995.

\bibitem{vicsek2012collective}
T.~Vicsek and A.~Zafeiris.
\newblock Collective motion.
\newblock {\em Phys. Rep.}, 517(3-4):71--140, 2012.

\end{thebibliography}
\end{document}